\documentclass[a4paper,fleqn]{cas-sc}

\usepackage[authoryear]{natbib}
\usepackage[]{listings}
\usepackage{algorithmicx}
\usepackage{algpseudocode}
\usepackage[exercise]{amsthm}
\usepackage{caption}  
\usepackage{amsmath}  

\def\tsc#1{\csdef{#1}{\textsc{\lowercase{#1}}\xspace}} \tsc{WGM} \tsc{QE}
\tsc{EP} \tsc{PMS} \tsc{BEC} \tsc{DE}

\newproof{pf}{Proof}
\newproof{pot}{Proof of Theorem \ref{thm2}}

\theoremstyle{plain}

\theoremstyle{definition}
  \newtheorem{definition}{Definition}[subsection]

  \newtheorem{conjecture}{Conjecture}[section]
  \newtheorem{example}{Example}[subsection]
  
  \newtheorem{algorithm}{Algorithm}[]
  
  \newtheorem{data}{Data}[subsection]

  \newtheorem{property}{Property}[subsection]

\theoremstyle{remark}
  \newtheorem*{remark}{Remark}
  \newtheorem*{note}{Note}

\usepackage{inconsolata} 
\usepackage{listings}

\lstdefinelanguage{julia}
{
  keywordsprefix=\@,
  morekeywords={
    exit,whos,edit,load,is,isa,isequal,typeof,tuple,ntuple,uid,hash,finalizer,convert,promote,
    subtype,typemin,typemax,realmin,realmax,sizeof,eps,promote_type,method_exists,applicable,
    invoke,dlopen,dlsym,system,error,throw,assert,new,Inf,Nan,pi,im,begin,while,for,in,return,
    break,continue,macro,quote,let,if,elseif,else,try,catch,end,bitstype,ccall,do,using,module,
    import,export,importall,baremodule,immutable,local,global,const,Bool,Int,Int8,Int16,Int32,
    Int64,Uint,Uint8,Uint16,Uint32,Uint64,Float32,Float64,Complex64,Complex128,Any,Nothing,None,
    function,type,typealias,abstract
  },
  sensitive=true,
  morecomment=[l]{\#},
  morestring=[b]',
  morestring=[b]" 
}

\begin{document} 

\let\WriteBookmarks\relax 
\def\floatpagepagefraction{1}
\def\textpagefraction{.001} 
\shorttitle{Finite Boolean Algebras for Solid Geometry using Julia's Sparse Arrays} 

\shortauthors{Alberto Paoluzzi et~al.}
%
\title [mode = title]{Finite Boolean Algebras for Solid Geometry\\ using Julia's
Sparse Arrays} 
\tnotemark[1,2] 

\tnotetext[1]{This work was partially supported from Sogei S.p.A. -- the ICT company of the Italian Ministry of Economy and Finance, by grant 2016-17.}
\tnotetext[2]{V.S. was supported in part by National Science Foundation grant CMMI-1344205 and National Institute of Standards and Technology.}

\author[1]{Alberto Paoluzzi}[type=editor, auid=000,bioid=1, orcid=0000-0002-3958-8089] 


\address[1]{Roma Tre University, Rome, RM, Italy}
\address[2]{University of Wisconsin, Madison \& International Computer Science Institute (ICSI), Berkeley, California, USA}
\address[3]{Scientific Computing and Imaging Institute (SCI), Salt Lake City, Utah, USA}
\address[4]{CECAM-IT-SIMUL Node, Rome, Italy}

\author[2,]{Vadim Shapiro}[]
\author[4] {Antonio DiCarlo}  
\author[3] {Giorgio Scorzelli}  
\author[1] {Elia Onofri}  



\def\E{\mathbb{E}}
\def\I{\mathbb{I}}
\def\R{\mathbb{R}}
\def\Z{\mathbb{Z}}
\def\N{\mathbb{N}}
\def\P{\mathbb{P}}



\cortext[cor1]{Corresponding author}

\begin{abstract} The goal of this paper is to introduce a new method in computer-aided 
geometry of solid modeling. We put forth a novel algebraic technique
to evaluate any variadic expression between polyhedral $d$-solids ($d=2,3$) with
regularized operators of union, intersection, and difference, {i.e.},~any CSG tree.
The result is obtained in three steps: first, by computing an independent set of
generators for the $d$-space partition induced by the input; then, by reducing
the solid expression to an equivalent logical formula between Boolean terms, made by
zeros and ones; and, finally, by evaluating this expression using bitwise native
operators. This method is implemented in Julia using sparse arrays.
The computational evaluation of every possible solid expression, usually denoted as CSG
(Constructive Solid Geometry), 
is reduced to an equivalent logical expression of a finite set algebra over the cells
of a space partition, and solved by native bitwise operators.
\end{abstract}

\begin{graphicalabstract} 
	\includegraphics[width=0.2\linewidth]{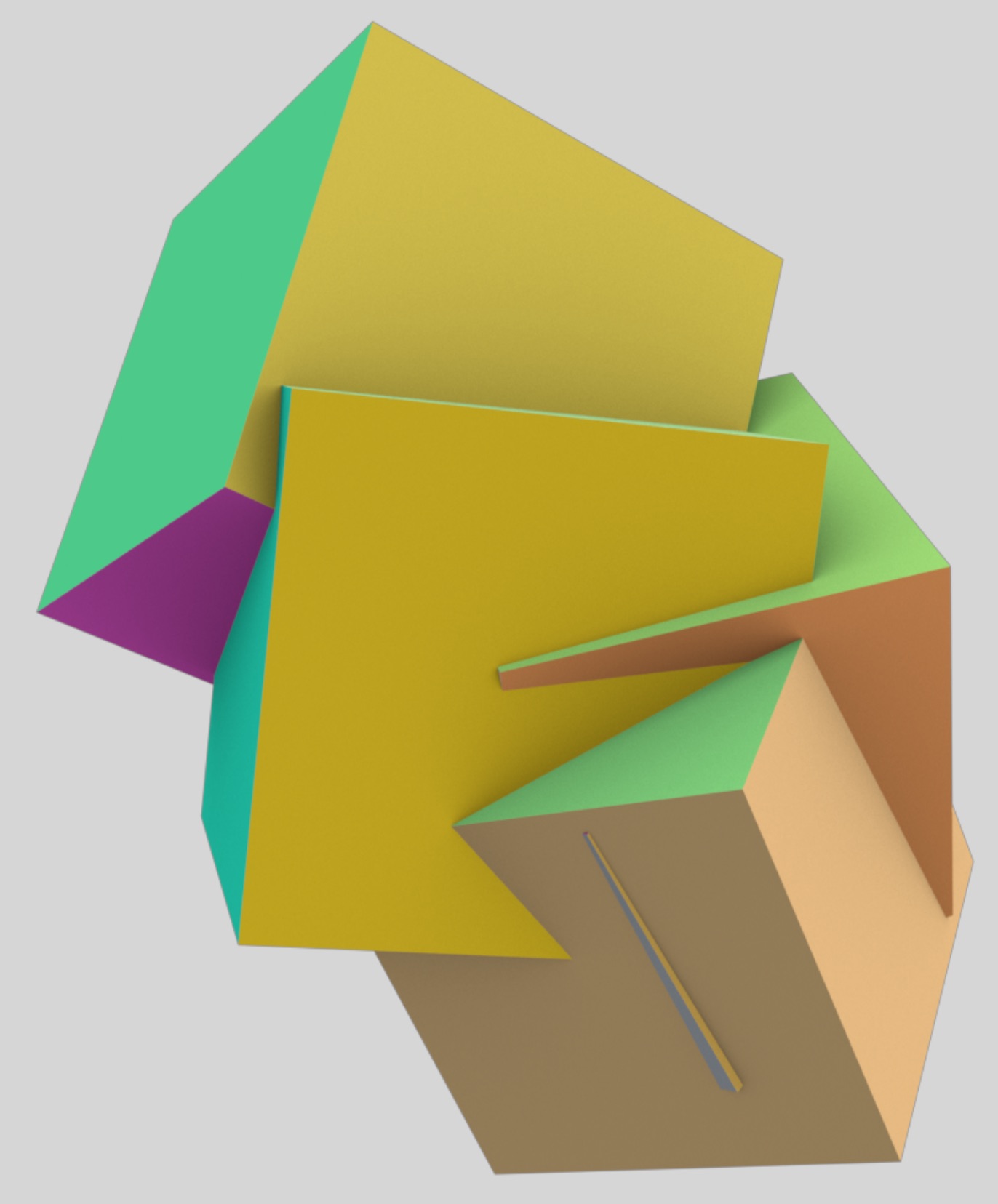}
	\includegraphics[width=0.3\linewidth]{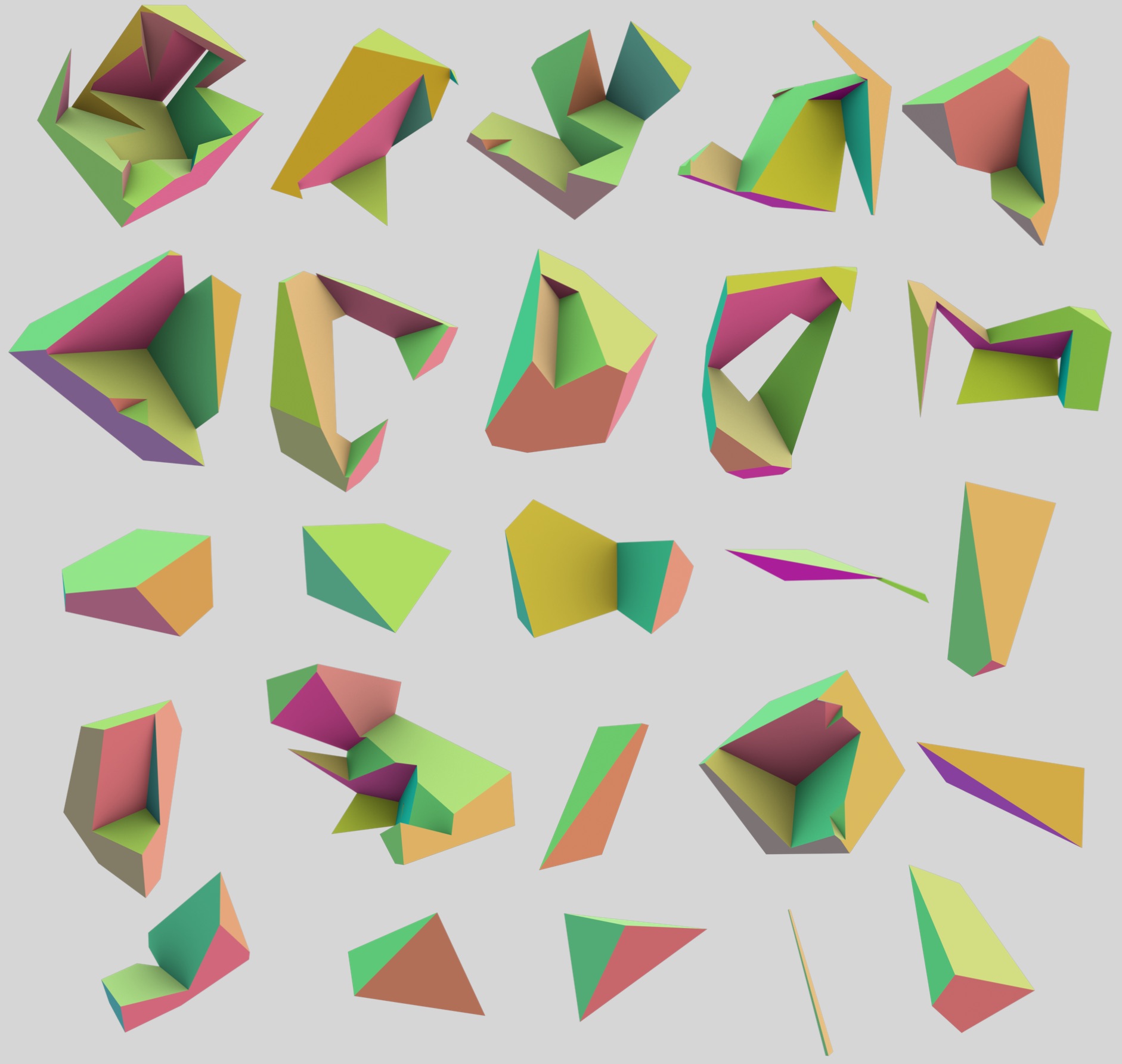}

	\vspace{5mm}\small
	{The set of join-irreducible atoms of the Boolean Algebra\\ 
	generated by a partition of $\E^3$, composing a basis of 3-chains.}
\end{graphicalabstract}

\begin{highlights} 

\item An algebraic approach to computation of Boolean operations between solid models is introduced in this paper. 

\item Any Boolean form with solid models is evaluated by assembling atoms of the finite algebra generated by the arrangement of Euclidean space produced by solid terms of the form. 

\item The atoms of this algebra are the basis of the linear chain space associated with this partition of Euclidean space. 

\item Basic tools of linear algebra and algebraic topology are used, namely (sparse) matrices of linear operators and matrix multiplication and transposition. Interval-trees and $k$d-trees are used for acceleration. 

\item Validity of topology computations is guaranteed, since operators satisfy by construction  the (graded) constraints $\partial^2=0$ and $\delta^2=0$.

\item CSG expressions of arbitrary complexity are evaluated in a novel way.
Input solid objects give a collection of boundary 2-cells, each operated \emph{independently}, generating local topologies merged by boundary congruence. 
 
\item  Once the space partition is generated, the atoms of its algebra are classified w.r.t.~all input solids. Then, {all CSG expressions}---of any complexity---may be evaluated by bitwise vectorized logical operations. 

\item Distinction is removed between manifolds and non-manifolds, allowing for mixing meshes, cellular decompositions, and 3D  grids, using Julia's sparse arrays. 

\item This approach can be extended to general dimensions and/or implemented on highly parallel computational engines, using standard GPU kernels.

\end{highlights}

\begin{keywords} 
Computational Topology  
\sep Chain Complex  
\sep Cellular Complex  
\sep Solid Modeling  
\sep Constractive Solid Geometry (CSG)
\sep Linear Algebraic Representation (LAR)  
\sep Arrangement  
\sep Boolean Algebra
\end{keywords}

\maketitle


\section{Introduction}\label{sec:introduction}
%
We introduce a novel method for evaluation of Boolean 
algebraic expressions including polyhedral solid objects, often called constructive solid
geometries, by reduction to bitwise evaluation of coordinate
representations within a linear space of chains of cells. This approach merely 
uses  basic algebraic topology and basic linear algebra, in accord with current
trends in data science. 
We show that any Boolean form with solid models may be evaluated assembling atoms of the finite algebra 
generated by the arrangement of Euclidean space produced by solid terms of the form. \footnote{
An open-source implementation is available. The source codes of all examples are on Github, in
\href{https://github.com/cvdlab/LinearAlgebraicRepresentation.jl/blob/master/CAGD.jl/paper/examples/all_examples.jl}{CAGD.jl/paper/}.
}

\subsection{Motivation}\label{sec:motivation}

From the very beginning, solid modeling suffered from a dichotomy between
``boundary'' and ``cellular'' representations, that induced practitioners to separate the
computation of the object's surface from that of its interior, and/or to
introduce the so-called ``non-manifold'' representations, often requiring
special and/or very complex data structures and algorithms.
Conversely, our approach relies on standard mathematical methods, {i.e.},~on basic
linear algebra and algebraic topology, and allows for an unified
evaluation of variadic Boolean expressions with solid models, by using cell
decompositions of both the interior and the boundary. In particular, we use
graded linear spaces of (co)chains of cells, as well as graded linear (co)boundary  
operators, and compute the operator matrices between such spaces.

The evaluation of a solid expression  is therefore reduced to the computation of
the \emph{chain complex} of the $\E^d$ partition (arrangement) induced by the
input, followed by the translation of the solid geometry formula to an equivalent
binary form of a finite algebra over the set of generators of the
$d$-chain space. Finally, the evaluation of this binary expression,
using native bitwise operators, is directly
performed by the compiler. The output is the coordinate representation of the
resulting $d$-chain in chain space $C_d$, {i.e.}, the binary vector representing
the solid result, the boundary set of which is optionally produced as the product of the
boundary matrix times this binary column vector.

Applied algebraic topology and linear algebra require the use of matrices as fundamental data
structures, in accord with the current development of computational
methods~\citep{strang:2019}. Our (co)boundary matrices may be very large, but are
always very sparse, so yielding comparable space and time complexity with
previously known methods.

The advantages of formulating the evaluation of Boolean expressions between
solid models in terms of (co)chain complexes and operations on sparse matrices
are that: (1) the common and general algebraic topological nature of such
operations is revealed; (2) implementation-specific low-level details and
algorithms are hidden; (3) explicit connection to computing kernels (sparse
matrix-vector and matrix-matrix multiplication) and to sparse numerical linear
algebra systems on modern computational platforms is provided; (4) systematic
development of correct-by-construction algorithms is
supported (topological constraints $\partial^2=0$ are automatically satisfied); (5) the computational solution of every possible solid
expression with union, intersection and complement, denoted as CSG, is reduced to an equivalent logical expression of a
finite Boolean algebra, and natively solved by vectorized bitwise operators.

\subsection{Related work}\label{sec:previous-work}

An up-to-date extensive survey of past and current methods and
representations for solid modeling can be found in \citep{HDCG:2017}. 
We discuss in the following a much smaller set of
references, that either introduced  the ideas discussed in our work 
or are directly linked to them. Note that such concepts were mostly published
in the foundational decades of solid modeling technologies, {i.e.}~in the 70's
trough the 90's. Time is ripe to go beyond.

Some milestones are hence recalled in the following paragraphs, starting
from~\cite{Baumgart:1972:WEP:891970}, who introduced the data
structure of Winged Edge Polyhedra for manifold representations at Stanford, and
 included the first formulation of ``Euler
operators'' in the ``Euclid'' modeler, using primitive solids,  
operators for affine transforms, and imaging 
procedures for hidden surface removal. 
\cite{Braid:1975:SSB:360715.360727}, starting from primitives like cubes,
wedges, tetrahedrons, cylinders, sectors, and fillets, or from a planar
primitive of straight lines, illustrated how to synthesize solids
bounded by many faces, giving algorithms for addition (quasi-disjoint or
intersecting union) and subtraction of solids.

The foundational Production Automation Project (PAP) at Rochester in the
seventies described computational models of solid
objects~\citep{requicha1977mathematical} by using relevant results scattered
throughout the mathematical literature, placed them in a coherent framework and
presented them in a form accessible to engineers and computer scientists. 
\cite{RequichaVoelcker:77} provided also a mathematical foundation for
\emph{constructive solid geometry} by drawing on established results in modern
axiomatic geometry and point-set topology. The term ``constructive solid
geometry'' denotes a class of schemes for describing solid objects as
compositions of primitive solid ``building blocks''.

Weiler introduced at RPI the first non-manifold representation \citep{Weiler:86}, 
called radial-edge data structure, and boundary graph operations for non-manifold
geometric modeling topology. Since then, several similar data structures for
non-manifold boundaries and interior structures (solid meshes) were introduced and
implemented in commercial systems,
with similar operations and performances. Hoffmann, Hopcroft, and Karasick at 
Cornell~\citep{Hoffmann:1987:RSO:866286} provided a reliable method for regularized
intersection, union, difference, and complement of polyhedral solids described
using a boundary representation and local cross-sectional graphs of any two
intersecting surfaces. An \emph{algebra of polyhedra} was devised by Paoluzzi and his
students in Rome~\citep{Paoluzzi:1989:BAO:70248.70249}, using boundary triangulations,
together with very simple algorithms for union, intersection, difference and
complement. Their data structure, called winged-triangle, is space-optimal for
piecewise-linear representations of polyhedra with curved boundaries.

The Selective Geometric Complex (SGC) by~\cite{Rossignac:89} at IBM Research 
provided a common framework for representing cellular decompositions of
objects of mixed dimensions, having internal structures and possibly
incomplete boundaries. `Boundary-of' relations capture incidences between cells of
various dimensions. \cite{Shapiro:1991:RSS:124951} presented
a \emph{hierarchy of algebras} to define formally a
family of Finite Set-theoretic Representations (FSR) of semi-algebraic subsets
of $\E^d$, including many known representation schemes for solid and non-solid
objects, such as boundary representations, Constructive Solid Geometry, cell
decompositions, Selective Geometric Complexes, and others.  Exemplary
applications included B-rep $\to$ CSG and CSG $\to$ B-rep conversions.

Semi-dynamical algorithms for maintaining \emph{arrangements} of polygons on the plane and the sphere were given by~\cite{Goldwasser:1995:IMA:220279.220337}.
A recent work more related to the present paper is by~\cite{Zhou:2016:MAS:2897824.2925901}.  
They compute mesh \emph{arrangements} for \emph{solid
geometry}, taking as input any number of triangle meshes, iteratively resolving
triangle intersections with previously subdivided 3D cells, and assigning
winding number vectors to  cells, in order to evaluate variadic Boolean
expressions. Their approach applies only  to boundary triangulations and uses
standard geometric computing methods, while the present one applies to any
cellular decomposition, either of the interior or of the boundary, computes
intersections only between line segments in 2D, and transforms every Boolean solid
expression into a logical expression solved natively by the compiler with vectorized bitwise operations.

Our related work in geometrical and physical modeling  with \emph{chain and cochain
complexes} was introduced in \cite{DiCarlo:2009:DPU:1629255.1629273} and~\cite{ieee-tase}.
The Linear Algebraic Representation (LAR), using sparse matrices, and its
applications to the computation of (co)boundary matrices and other chain
adjacencies and incidences is discussed in~\cite{ Dicarlo:2014:TNL:2543138.2543294}. The
computational pipeline and the detailed algorithms to compute the space
decomposition induced by a collection of solid models is given
in~\cite{arXiv:2017}, \cite{TSAS:19}. An open-source prototype 
implementation is available at ~\href{https://github.com/cvdlab/LinearAlgebraicRepresentation.jl}{\texttt{https://github.com/cvdlab/LinearAlgebraicRepresentation.jl}}.

\subsection{Overview}\label{sec:paper-contents}
%
In Section~\ref{sec:background} we provide a short introduction to the basic
algebraic-topological concepts and notations used in this paper, including graded
vector spaces, chain and cochain spaces, boundary and coboundary operators, cellular and chain 
complexes, arrangements,  and finite Boolean algebras.  Section~\ref{sec:pipeline}
discusses a computational pipeline introduced to build  
the algebras generated by decompositions of $\E^d$, 
within the frame of a representation theory for solid modeling.  In particular, we introduce 
a linear representation based on independent generators of chain spaces.
The main tasks are implemented by a sequence of novel 2D/3D geometric and topological algorithms, 
illustrated by simple examples scattered in the text. 
Section~\ref{sec:examples} explores the evaluation of some simple
Boolean formulas with solid objects, both in 2D and in 3D, 
including the computation of the Euler characteristic of the boundary of a triple Boolean union. 
In~Section~\ref{sec:discussion} the main results of our approach are summarized 
and compared with the current state of the art, 
including (a) solid modeling via sparse
matrices and linear algebra, (b) the computation of generators for
the column space of boundary matrices, and (c) our new method for CSG evaluation. 
The closing section~\ref{sec:conclusion}
presents a summary of contents, and outlines possible applications of the
ideas introduced. Small snippets of Julia code are inserted in the examples.

\section{Background}\label{sec:background}

In this section we provide a set of definition and examples for most of the basic concepts used in this paper. In particular, we will introduce complexes of \emph{cells} and \emph{chains}, the cellular decompositions of the ambient Euclidean space, called \emph{arrangements}, and the specific topic of geometric and solid modeling called \emph{Constructive Solid Geometry} (CSG), which is the subject matter of this paper. 

We restrict out attention to dimensions \emph{two} and \emph{three}, and to piecewise-linear (PL) objects, i.e. to triangulable polyhedra. For the sake of space we give mostly 2D examples in this section.  Everywhere we will privilege simplicity and readability to  exactness and consequentiality. 

In addition, we discuss in this paper several small readable scripts of \emph{Julia}, the novel language for numeric and scientific computing, that we chose as our implementation platform, after we previously developed our first experiments in Python. There are many reasons to prefer Julia over other languages, but
Julia's main innovation is the very combination of productivity and performance. For a discussion of this point, the most important  reference is~\cite{BEKS14}. Another important point is the near automatic optimization and translation to vectorized, parallel and distributed computation, when resources are available, starting from readable sequential prototypes. 

\subsection{Complexes}\label{sec:complexes}

A \emph{complex} is a graded set $S = \{ S_i \}_{i\in I}$ i.e.~a family of sets, indexed in this paper over $I = \{0,1,2,3\}$.
We use two different but intertwined types of {complexes}, and specifically complexes of \emph{cells} and complexes of \emph{chains}. 
Their definitions and some related concepts are given in this section. Greek letters are used for the {cells} of a space partition, and roman letters for {chains} of cells, coded as either (un)signed integers or sparse arrays of (un)signed integers.

\begin{definition}[$d$-Manifold]
A \emph{manifold} is a topological space that resembles a flat space locally, i.e., near every point. 
Each point of a $d$-dimensional manifold has a neighborhood that is homeomorphic to $\E^d$, the Euclidean space of dimension $d$. Hence, this geometric object is often referred to as $d$-manifold.
\end{definition}

\begin{definition}[Cell]
A \emph{$p$-cell} $\sigma$ is a $p$-manifold ($0\leq p\leq d$) which is 
piecewise-linear, connected, possibly non convex, and not necessarily contractible\footnote{This definition refers to cellular complexes used in this paper}$^,$\footnote{In our representation, cells may contain internal holes; cells of CW-complexes~\citep{hatcher:2002} are, conversely,  contractible to a point. }. 
\end{definition}

\begin{remark}
We deal here with Piecewise-Linear (PL) cells of dimension 0, 1, 2, and 3, respectively. It should be noted that 2- and 3-cells may contain holes, while remaining connected.  In other words, the cells are $p$-polyhedra, i.e.~segments, polygons and polyhedrons embedded in two- or three-dimensional space. Even if they  are often convex, cells in a polyhedral decomposition of a space are not necessarily convex.
\end{remark}

\begin{definition}[Cellular complex]
A \emph{cellular $p$-complex} is a finite set of cells that have at most dimension $p$, together with all their $r$-dimensional faces ($0\leq r\leq p$). A \emph{face} is an element of the PL boundary of a cell, that satisfy a \emph{boundary compatibility} condition. Two $p$-cells $\alpha, \beta$ are boundary-compatible when their point-set intersection contains the same $r$-faces ($0\leq r\leq p$) of $\alpha$ and $\beta$. A cellular $p$-complex is \emph{regular} when each $r$-cell ($0\leq r\leq p$) is face of a $p$-cell. 
\end{definition}

\begin{definition}[Skeleton]
The $s$-skeleton of a $p$-complex $\Lambda_p$ ($s\leq p$) is the set $\Lambda_s\subseteq\Lambda_p$ of all $r$-cells ($r\leq s$) of $\Lambda_p$. Every skeleton of a regular complex is a regular subcomplex.
\end{definition}

\begin{definition}[Support space]
The support space $|\Lambda|$ of a cellular complex is the point-set union of its cells. 
\end{definition}

\begin{remark}[Geometric Representation]
The LAR representation\footnote{
The linear algebraic representation (LAR), by~\cite{Dicarlo:2014:TNL:2543138.2543294}, started using sparse binary arrays to compute and represent the topology (linear spaces of chains, and linear (co)boundary operators) of cellular complexes.
}
 of a PL complex $\Lambda_p$ in the Euclidean space $\E^d$ ($p\leq d$) is given by 
an embedding map $\mu: \Lambda_0\to\E^d$, and by the discrete sets $U_0:=\chi(\Lambda_0)$, and $U_r:=\chi(\Lambda_r - \Lambda_{r-1}) $, with $1\leq r\leq p$, where $\chi: \Lambda\to\mathscr{P}(\Lambda_0)$ is the \emph{characteristic function}\footnote{
Given a subset $S$ of a larger set $A$, the characteristic function $\chi_A(S)$, sometimes also called the \emph{indicator function}, is the function defined to be identically one on $S$, and zero elsewhere. \citep{Wolfram}. 
}
 from cells to subsets of 0-cells, that links every cell to the subset of its ``vertices''.
\footnote{
A constrained Delaunay triangulation (CDT) is the triangulation of a set of vertices and edges in the plane such that: (1) the  edges are included in the triangulation, and (2) it is as close as possible to the Delaunay triangulation. It can be shown that the CDT can be built in optimal ?(n log n) time \citep{10.1145/41958.41981}.
}$^,$\footnote{
It is possible to show that for a given CDT of 2-cells, the PL affine functions that map each $p$-face to $\E^d$ can be produccd in a unique way,  by properly combining  $\mu$ and CDT$(U_p)$
}
\end{remark}

\begin{data}[Cellular complex] \label{ex:data}
Input data to generate the 2D cellular complex in (figure of) Data~\ref{fig:simplest} follows:

\begin{minipage}{0.96\textwidth}
\small 
\begin{lstlisting}[mathescape]
V = $\ $[0.0  1.5  3.0  1.0  1.5  2.0  1.0  1.5  2.0  0.0  1.5  3.0 ; $\qquad\qquad\qquad\qquad\qquad\qquad\ $# $\gamma : \Lambda_0 \to \E^2$
      $\,\,$0.0  0.0  0.0  1.0  1.0  1.0  2.0  2.0  2.0  3.0  3.0  3.0 ] 
EV = [[1,2],[2,3],[4,5],[5,6],[7,8],[8,9],[10,11],[11,12], $\qquad\ \ \ $# $\chi_{U_0}(U_1)$ as array of array
      $\,\,$[1,10],[4,7],[6,9],[3,12],[2,5],[8,11]] 
FV = [[1,2,4,5,7,8,10,11],[2,3,5,6,8,9,11,12],[4,5,6,7,8,9]] $\qquad\ $# $\chi_{U_0}(U_2)$ as array of array
\end{lstlisting} \label{data-1}
\end{minipage} 
\end{data}

\begin{remark}[Minimal representation]
The cellular representation given above is actually redundant, since the map \lstinline{FV} (``faces-by-vertices'') can be generated from \lstinline{V} and \lstinline{EV} (``edges-by-vertices'') using the methods given in~\cite{TSAS:19}, 
that will be summarized in the first half of Section~\ref{sec:pipeline}. In other words, given the PL embedding of a planar graph in a plane, the cellular 2-complex---i.e., the graph complement---of its plane partition is completely specified.
\end{remark}

\begin{remark}[Characteristic matrices]
Given the geometric representation of a cellular $p$-complex, the topology of ordered $s$-cells ($0\leq s\leq p$), denoted $\lambda_i\in U_s$, is fully represented by the sparse binary matrices $K_s = [\chi(\lambda_i)]$, where each binary row gives the image of the characteristic function of the $s$-cell $\lambda_i$ with respect to the set of 0-cells.
\end{remark}

\begin{algorithm}[Characteristic matrices] 
Are used to denote the cells of a cellular complex as binary vectors, i.e., as rows of binary matrices.
The code snippet below computes the characteristic matrix for a set \texttt{CV} of cells, represented by-vertices.
The function \lstinline{K} returns the Julia matrix of type \lstinline{SparseArrays}  providing the characteristic matrix \lstinline{K}$_r$ given as input an array \lstinline{CV} specifying each $r$-cell as array of vertex indices. 
The embedding of cells, i.e., the affine map that locate them in $\E^d$, will be specified by a $d\times n$ array \lstinline{V}, where $n$ is the number of $0$-cells. See \texttt{V} in Data~\ref{ex:data}.

\begin{minipage}{0.96\textwidth}
\small 
\begin{lstlisting}[mathescape]
function K( CV )
	I = vcat( [ [k for h in CV[k]] for k=1:length(CV) ]...) 
	J = vcat( CV...)
	X = Int8[1 for k=1:length(I)]
	return SparseArrays.sparse(I,J,X)
end
\end{lstlisting}
\end{minipage}
\end{algorithm}

\begin{definition}[CSC sparse matrix]
The compressed sparse column (CSC) format represents a sparse matrix \lstinline{M} by three (one-dimensional) arrays.
In Julia, CSC is the preferred (actually unique, by now) storage format for sparse matrices. The sparse \lstinline{M} is stored as a \lstinline{struct} containing: Number of rows, Number of columns, Column pointers, Row indices of stored values, and Stored values, typically nonzeros.
\end{definition}

\begin{data}[Characteristic matrices] \label{fig:simplest}
The sparse binary matrices \lstinline{K1 = K(EV)} and \lstinline{K2 = K(FV)} are generated from data of 2-complex~\ref{data-1}. Here we show the corresponding dense matrices for sake of readibility. Of course, the storage space of a sparse matrix is linear with the total number $nz$ of non-zero elements, and the sparsity, defined as $0\leq 1-nz/n^2 \leq 1$, grows  with the number $n$ of cells, i.e.~the \lstinline{EV} array length.

\begin{minipage}{0.325\textwidth}
\centering
\begin{lstlisting} 
julia> Matrix(K(EV))
14x12 Array{Int8,2}:
 1  1  0  0  0  0  0  0  0  0  0  0
 0  1  1  0  0  0  0  0  0  0  0  0
 0  0  0  1  1  0  0  0  0  0  0  0
 0  0  0  0  1  1  0  0  0  0  0  0
 0  0  0  0  0  0  1  1  0  0  0  0
 0  0  0  0  0  0  0  1  1  0  0  0
 0  0  0  0  0  0  0  0  0  1  1  0
 0  0  0  0  0  0  0  0  0  0  1  1
 1  0  0  0  0  0  0  0  0  1  0  0
 0  0  0  1  0  0  1  0  0  0  0  0
 0  0  0  0  0  1  0  0  1  0  0  0
 0  0  1  0  0  0  0  0  0  0  0  1
 0  1  0  0  1  0  0  0  0  0  0  0
 0  0  0  0  0  0  0  1  0  0  1  0
\end{lstlisting} 
\end{minipage}
\begin{minipage}{0.275\textwidth}
\centering
\begin{lstlisting} 
julia> Matrix(K(FV)')
12x3 Array{Int8,2}:
 1  0  0
 1  1  0
 0  1  0
 1  0  1
 1  1  1
 0  1  1
 1  0  1
 1  1  1
 0  1  1
 1  0  0
 1  1  0
 0  1  0
\end{lstlisting} 
\end{minipage}
\begin{minipage}{0.5\textwidth}
\flushleft
   \includegraphics[width=0.7\linewidth]{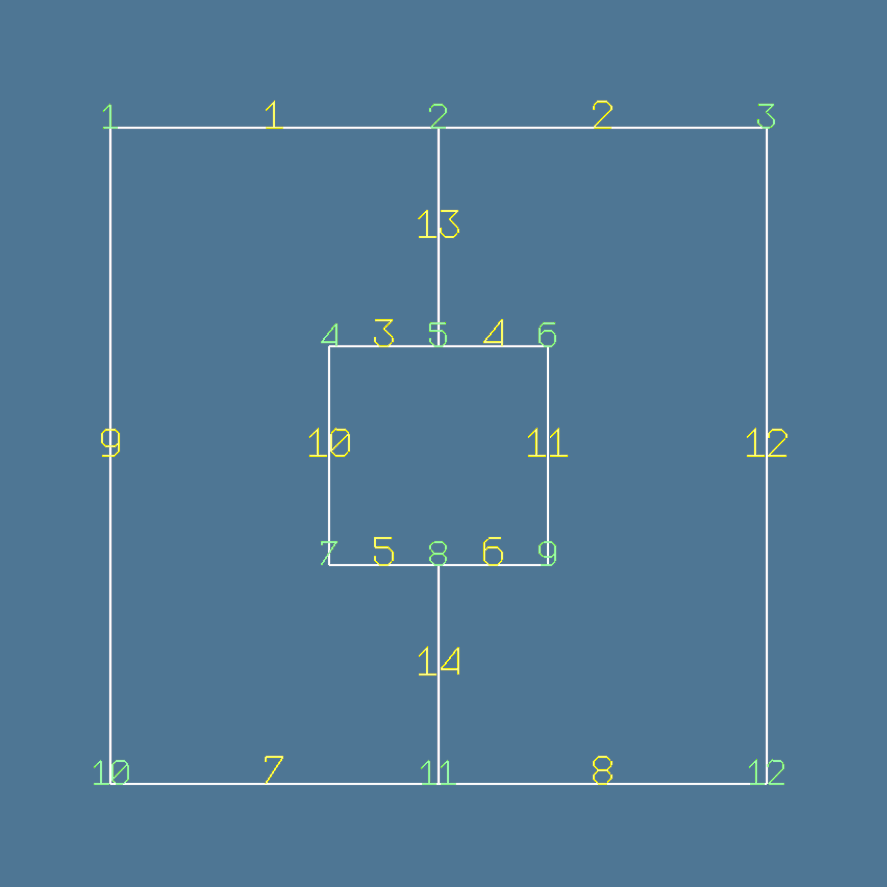} 
\end{minipage}
\end{data}

\begin{definition}[Chains] 
A \emph{$p$-chain} can be seen, with some abuse of language, as a collection of $p$-cells. In this flavor we may write $C_p = \mathscr{P}(U_p)$ for the space of $p$-chains and $U_p = \Lambda_p - \Lambda_{p-1}$ for the set of unit $p$-chains ($0\leq p\leq d$).
\end{definition}

\begin{definition}[Chain space]
The set $C=\oplus\ C_p$, direct sum of chain spaces, can be given the structure of a graded vector space (see Definition~\ref{def:graded} and~\citealp[pgs.~11--12]{Arnold:2018}) by
defining sums of chains with the same dimension, and products times scalars in a
field, with the usual properties.
\end{definition}

\begin{definition}[Chain space bases]
As a linear space, each set $C_p$ contains a natural set of irreducibles generators.
The natural \emph{basis} $U_p \subset C_p$  is the set of \emph{independent} (or \emph{elementary}) chains $u_p \in C_p$, given
by singleton elements. Consequently, every chain $c\in C_p$ can be written as 
a linear combination of the basis with field elements, and is uniquely generated. Once  the  basis is fixed, the unsigned coordinate
representation of each $\{\lambda_k\} = u_k \in C_p$ is unique, and is a binary array with just one element non-zero in position $k$, and all other elements 0. The ordered
sequence of scalars may be drawn either from $\{0,1\}$ (unsigned representation) or from
$\{-1,0,+1\}$ (signed representation). With abuse of language, we
often call $p$-cells the independent generators of $C_p$, {i.e.}~the elements of
$U_p$.
\end{definition}

\begin{definition}[Graded vector space]\label{def:graded}
A \emph{graded vector space} is a vector space $V$ expressed as a direct sum  of
spaces $V_p$ indexed by integers in $[0,d] := \{p\in\N \ |\ 0\leq p\leq d\}$:
\begin{equation} 
V = \oplus_{p = 0}^d V_p.
\end{equation} 
A linear map $f:V\to W$ between graded vector spaces is called a \emph{graded
map} of degree $k\ $ if $f(V_k) \subset W_{p+k}$.
\end{definition}

\begin{definition}[Chain complex]
A \emph{chain complex} is a graded vector  space $V$ furnished with a graded
linear map $\partial : V \to V$ of degree $-1$ called
\emph{boundary operator}, which satisfies $\partial^2 = 0$. In other words, a chain complex
is a sequence of vector spaces $C_p$ and linear maps $\partial_p : C_p \to C_{p-1}$,
such that $\partial_{p-1} \circ\ \partial_{p} = 0$.
\end{definition}

\begin{definition}[Cochain complex]
A \emph{cochain complex} is a graded vector space $V$ furnished with a graded
linear map $\delta : V \to V$ of degree $+1$ 
called \emph{coboundary operator},  which satisfies $\delta^2 = 0$. That is to say, a cochain complex is a
sequence of vector spaces $C^p$ and linear maps $\delta^p : C^p \to C^{p+1}$,
such that $\delta^{p+1} \circ\ \delta^{p} = 0$.
\end{definition}

\begin{property}[Duality of chain complexes] 
Any {chain} space, being linear, is associated with a unique dual {cochain} space. A linear map $L: V\to W$ between linear spaces induces a dual map $L^*:  W^* \to V^*$ between their dual spaces. If (and only if) the primal chain space has finite dimension, as in our case, then its dual cochain space has the same dimension, and is therefore linearly isomorphic to the primal. Moreover, the coboundary operator $\delta^{k}$  is the dual of the boundary operator $\partial_{k+1}$ :
\begin{equation}
(\delta^k \omega) g = \omega (\partial_{k+1} g), \qquad \mbox{for every\ } \omega \in C^k, g \in C_{k+1}.
\label{eq:duality1}
\end{equation}
There exist infinitely many linear isomorphisms between a finite-dimensional linear space and its dual (and selecting one of them is tantamount to endowing the primal space with a metric structure). 
Each such isomorphism is produced by identifying elementwise a basis of the primal space with a basis of its dual. 
Since we are only interested in topological properties, we adopt the most straightforward choice, identifying elementwise the natural bases of the corresponding chain and cochain spaces. 
Under the selected identification, we have that the matrix $[\delta_{p-1}]$, representing $\delta_{p-1}$ in the natural bases of $C_{p-1}$ and $C_p$, equals the transpose of the matrix $[\partial_p]$, representing $\partial_p$ in the natural bases of $C_p$ and $C_{p-1}$, so that $[\delta_{p-1}] = [\partial_p]^t$, and represent identification and duality in the diagram below, where identified (co)chain spaces are graded with lower indices: 
\begin{equation}
C_\bullet = (C_p, \partial_p) := 
C_3 \ 
\substack{
\delta_2 \\
\longleftarrow \\[-1mm]
\longrightarrow \\
\partial_3 
}
\ C_2 \
\substack{
\delta_1 \\
\longleftarrow \\[-1mm]
\longrightarrow \\
\partial_2 
}
\ C_1 \ 
\substack{
\delta_0 \\
\longleftarrow \\[-1mm]
\longrightarrow \\
\partial_1 
}
\ C_0 , 
\qquad\mbox{\rm where\ }\qquad
\partial C_{p-1} \circ\,\partial C_{p}
=
\delta C_{p+1} \circ\,\delta C_{p}
=
0
\label{eq:duality2}
\end{equation}
\end{property}


\begin{definition}[Operator matrices]
Once fixed the bases, i.e.~ordered the sets of $p$-cells, the matrices of boundary and coboundary operators, that we see in Example~\ref{ex:example-3}, are uniquely determined.
Such matrices are very
sparse, with sparsity growing linearly with the number $m$ of cells (rows). Sparsity may be
defined as one minus the ratio between non-zeros and the number of matrix
elements. It~is fair to consider that the non-zeros per row are bounded by a small constant in
topological matrices, hence the number of non-zero elements grows linearly
with $m$. With common data structures~\citep{coosparse} for sparse matrices, 
the storage cost $O(m)$ is linear with the number of cells, with $O(1)$ small cost per cell  that depends on the storage scheme.
\end{definition}

\begin{example}[Boundary matrix] 
\label{ex:example-3}
In this small example we construct the sparse matrix $[\partial_2]$ from multiplication of characteristic matrices \lstinline{EV} and the transposed \lstinline{FV} (for details see~\cite{Dicarlo:2014:TNL:2543138.2543294}). Some matrices are written transposed here for sake of space.  We show also how it is easy to extract either the boundary or any cycles of a cellular complex, using the semiring\footnote{A \emph{semiring} is an algebraic structure with sum and product, but without the requirement that each element must have an additive inverse.}~\citep{DBLP:journals/corr/KepnerABBFGHKLM16} matrix multiplication\footnote{GraphBLAS, resounding the BLAS (Basic Linear Algebra Subprograms) standard, is a new paradigm to define normalized building blocks for graph algorithms in the language of linear algebra, and provides a powerful and expressive framework for creating graph algorithms based on the elegant mathematics of sparse matrix operations on a semiring.}. Currently, the algebraic multiplication of matrices is used in our algorithms in the form of a standard (sparse) matrix multiplication, followed by a filtering of resulting matrix. A  porting to Julia of the \texttt{SuiteSparse:GraphBLAS}  library by Tim~\cite{10.1145/3322125} is  currently being developed by our group.

\small
\begin{minipage}{0.435\textwidth}
\centering
\begin{lstlisting}[mathescape=true]
julia> Matrix( K(EV) * K(FV)')'
 2  1  2  1  2  1  2  1  2  2  0  0  2  2
 1  2  1  2  1  2  1  2  0  0  2  2  2  2
 0  0  2  2  2  2  0  0  0  2  2  0  1  1

julia> EF = (K(EV) * K(FV)') .$\div$ 2;

julia> Matrix(EF')
3x14 Array{Int64,2}:
 1  0  1  0  1  0  1  0  1  1  0  0  1  1
 0  1  0  1  0  1  0  1  0  0  1  1  1  1
 0  0  1  1  1  1  0  0  0  1  1  0  0  0

julia>  b1 = (EF * [1 1 1]') .% 2
1x14 Array{Int64,2}:
 1  1  0  0  0  0  1  1  1  0  0  1  0  0

julia> b2 = (EF * [1 1 0]') .% 2
1x14 Array{Int64,2}:
 1  1  1  1  1  1  1  1  1  1  1  1  0  0
 
julia> SparseArrays.findnz(b1)[2]
[1, 2, 7, 8, 9, 12]

julia> SparseArrays.findnz(b2)[2]
[1, 2, 3, 4, 5, 6, 7, 8, 9, 10, 11, 12]
\end{lstlisting} 
\end{minipage}
\begin{minipage}{0.265\textwidth}

\texttt{b1} and \texttt{b2} must be read over \newline the graph of Example~\ref{fig:simplest}.

\begin{lstlisting} 
julia> EV[findnz(b1)[2]]
 [1, 2]  
 [2, 3]  
 [10, 11]
 [11, 12]
 [1, 10] 
 [3, 12] 

julia> EV[findnz(b2)[2]]
 [1, 2]  
 [2, 3]  
 [4, 5]  
 [5, 6]  
 [7, 8]  
 [8, 9]  
 [10, 11]
 [11, 12]
 [1, 10] 
 [4, 7]  
 [6, 9]  
 [3, 12] 
\end{lstlisting} 
\end{minipage}
\begin{minipage}{0.42\textwidth}
\flushleft
   \includegraphics[width=0.625\linewidth]{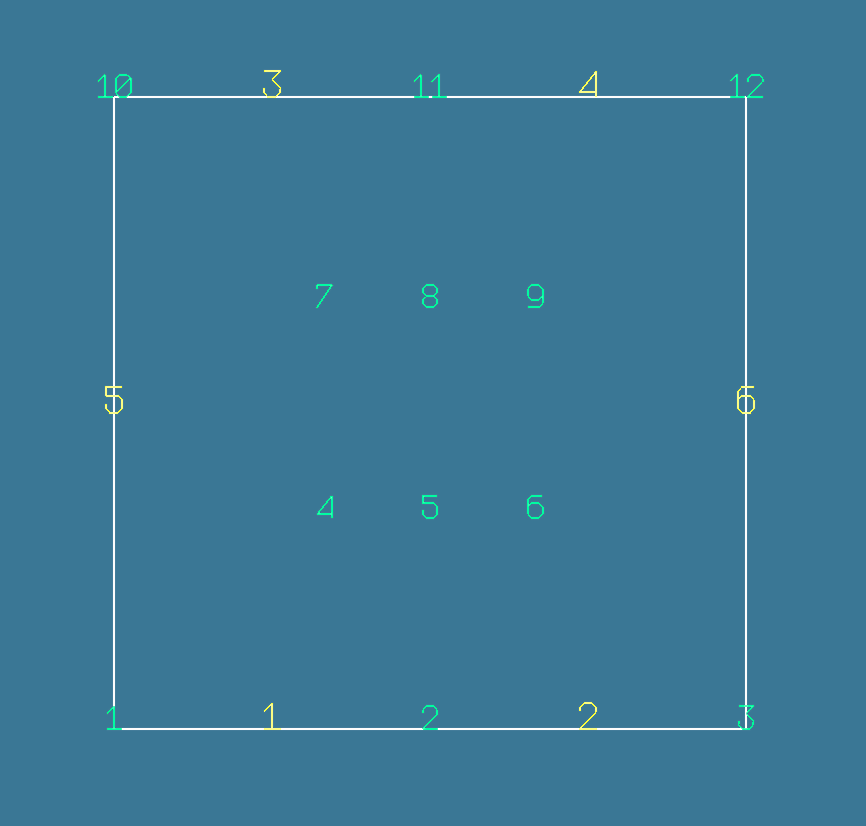} 

   \includegraphics[width=0.625\linewidth]{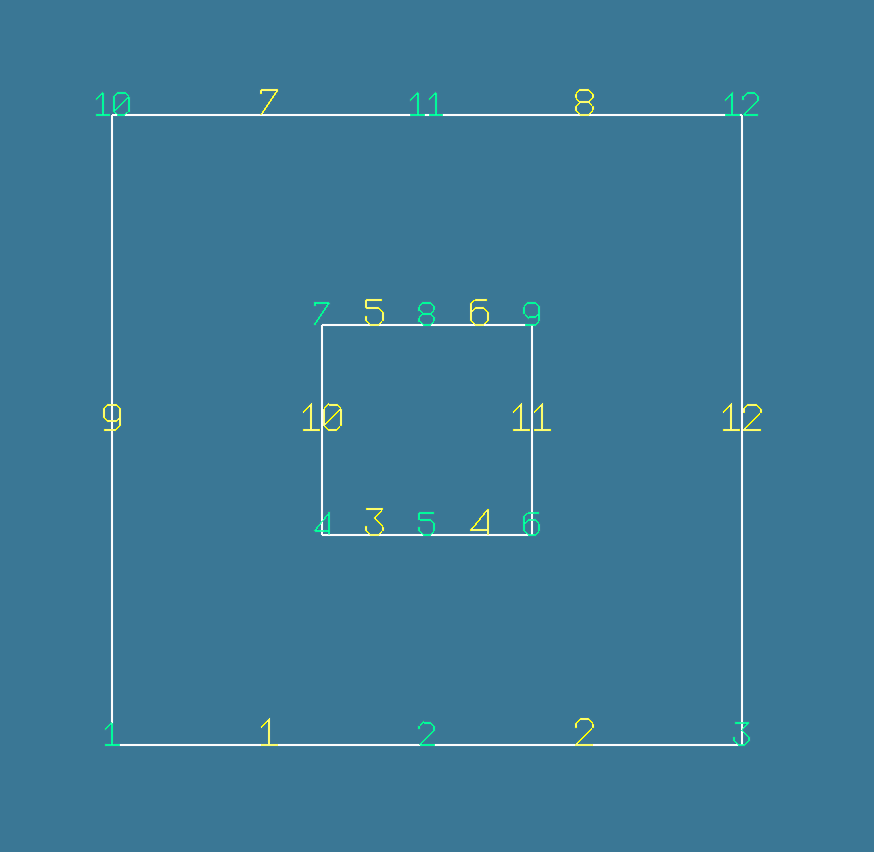} 
   \label{fig:example2}
\end{minipage}
\end{example}
\noindent
Julia names made by two characters from \lstinline{V,E,F,C} (for 0-, 1-, 2-, 3-cells) will by often used instead of mathematical simbols $\partial_p$ and $\delta_q$. In this case we have always: \lstinline{YX} : \lstinline{X} $\to$ \lstinline{Y}, and $\texttt{YX'} = \texttt{XY}$. 

\begin{remark}
We remark that \lstinline{b1} and \lstinline{b2} are the 1-cycles in Figures (a) and (b), generated by right product of 2-boundary operator matrix $[\partial_2]\equiv{}$\lstinline{EF}, times a 2-chain (\lstinline{[1 1 1]'} and \lstinline{[1 1 0]', respectively}). Note that the  \lstinline{b2} cycle is disconnected, i.e.~reducible to the sum of two cycles. Note also that in the product matrix \lstinline{A = E}$\,$\lstinline{F} of characteristic matrices, like in row 1 of left snippet, a term $a_{ij} = e^i \times f_j$ denotes the \emph{number} of 0-cells shared by chains $e^i$ and $f_j$.
\end{remark}

\begin{definition}[Natural (co)chain bases] In topological algorithms, see Algorithm~\ref{alg:tgw}, we make large use of  coboundary maps between cochain spaces: $\delta^p: C^p\to C^{p+1}$. By identification of cochain and chain spaces---see diagram~(\ref{eq:duality2}), the maps $\delta_p: C_p\to C_{p+1}$ go between the same natural bases than $\partial_p: C_p\to C_{p-1}$, i.e., from/to graded cells.
\end{definition}

\subsection{Space Arrangement}\label{sec:arrangement}

The word \emph{arrangement} is used in combinatorial geometry and computational geometry and topology as a synonym of space partition. Construction of arrangements of lines~\citep{Dimca:2017}, segments, planes and other geometrical objects is discussed in~\cite{fhktww-a-07}, with a description of CGAL software~\citep{Fabri:2000:DCC:358668.358687}, implementing 2D/3D arrangements with Nef polyhedra~\citep{bieri:95} by~\cite{Hachenberger:2007:BOS:1247750.1248141}. A review of papers and algorithms concerning construction and counting of cells may be found in the chapter on Arrangements in the ``Handbook of Discrete and Computational Geometry''~\citep{Goodman:2017:HDC:285869}. 
Arrangements of polytopes, hyperplanes and $d$-circles are discussed in~\cite{Ziegler:92}.

\begin{definition}[Space Arrangement]
Given a finite collection $\mathcal{S}$ of geometric objects in $\E^d$, the arrangement $\mathcal{A}(\mathcal{S})$ is the decomposition of $\E^d$ into connected open cells of dimensions $0, 1, \ldots , d$ induced by $\mathcal{S}$~\citep{Halperin:2017}. We are interested in the Euclidean space partition induced by a collection of PL cellular complexes.
\end{definition}

\begin{example}[Space Arrangement] The irreducible 2-cells of the $\E^2$ partition generated by random rectangles and by random polygonal approximations of the 2D circle. Take notice of the fact that 2-cells may be non convex and/or non contractible. 

\begin{minipage}{0.18\textwidth}\small\flushleft
(a) 2D partition generated by rectangles of random size, position and orientation;

(b) exploded view;

(c) random circles approximations by polygons.

\end{minipage}\hfill
\begin{minipage}{0.25\textwidth}
   \centering
   \includegraphics[height=0.9\linewidth,width=\linewidth]{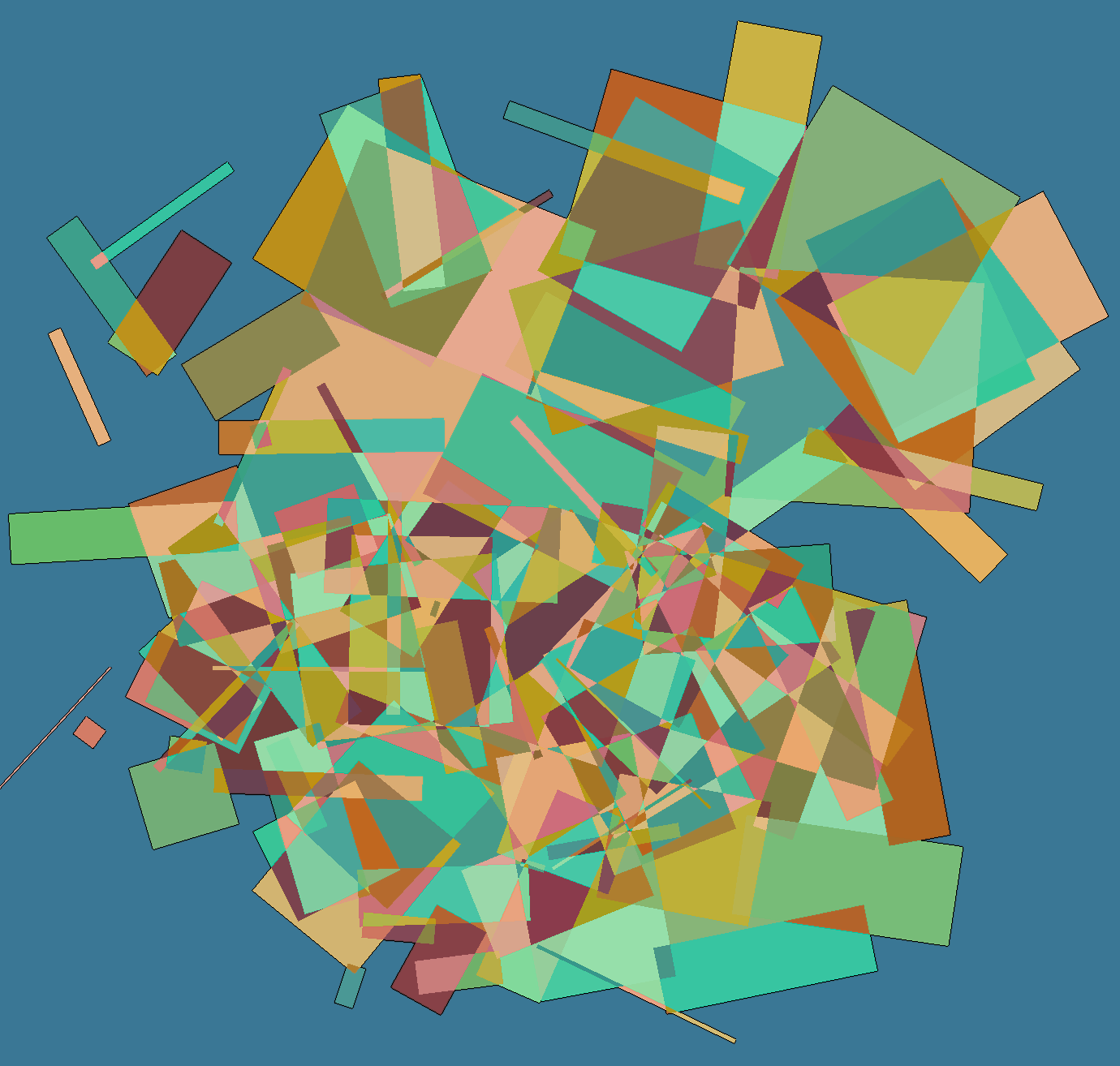} 
   \label{fig:example}
\end{minipage}
\begin{minipage}{0.25\textwidth}
   \centering
   \includegraphics[height=0.9\linewidth,width=\linewidth]{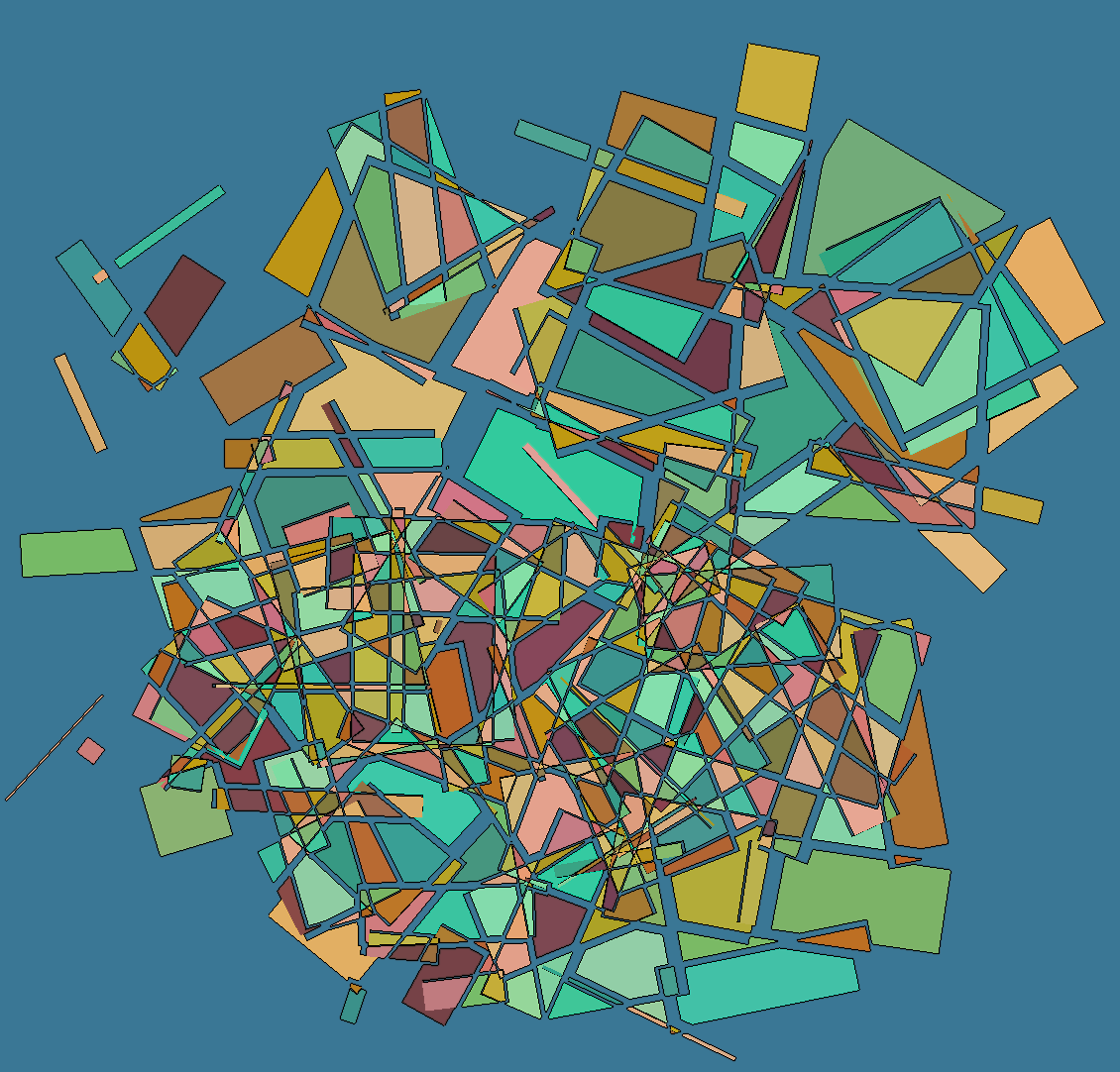} 
   \label{fig:example}
\end{minipage}
\begin{minipage}{0.25\textwidth}
   \centering
   \includegraphics[height=0.9\linewidth,width=\linewidth]{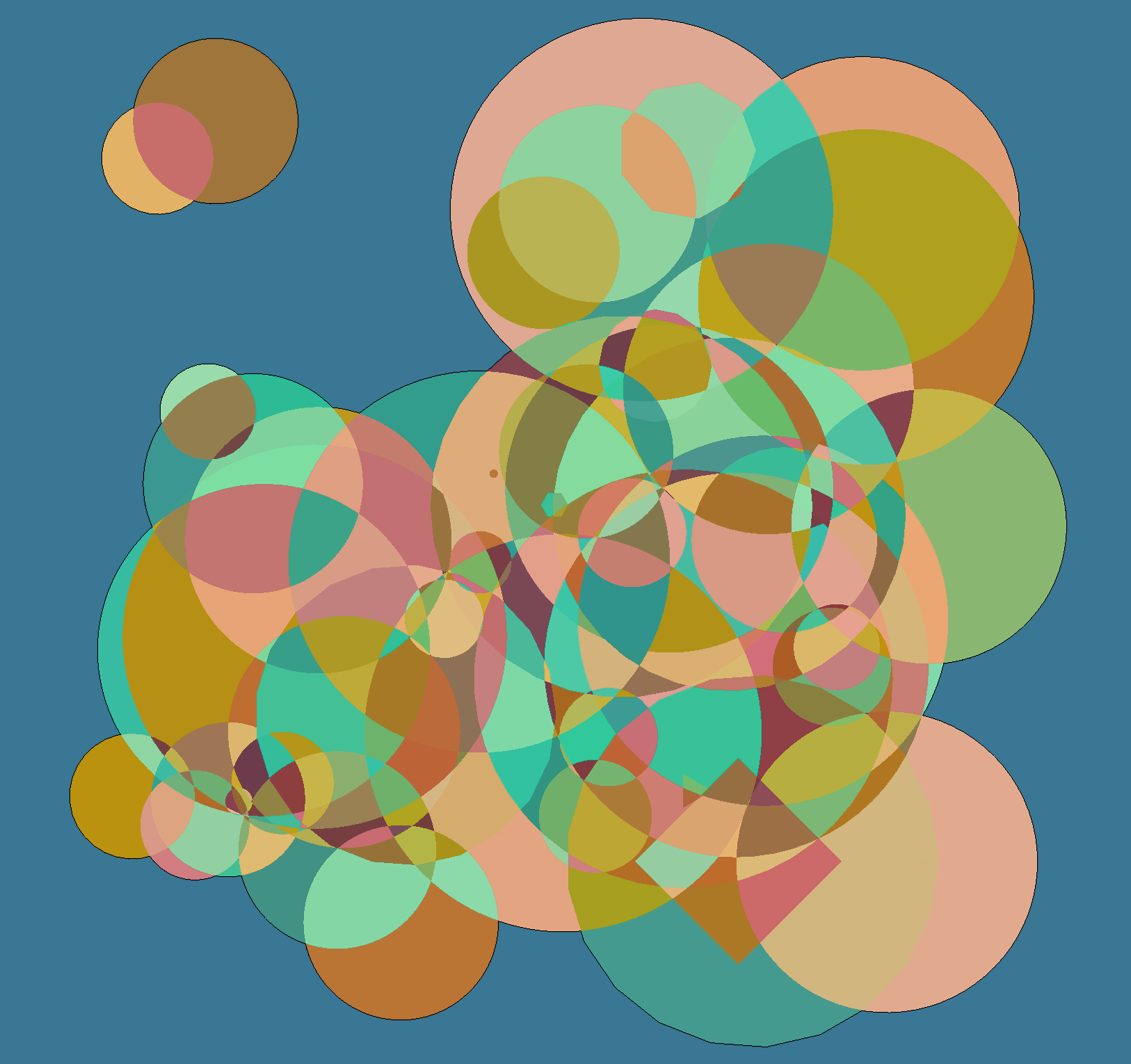} 
   \label{fig:example}
\end{minipage}
\vspace{-5mm}
{\footnotesize ~\flushleft\hspace{5.8cm}(a)\hspace{3.8cm}(b)\hspace{3.9cm}(c)\hspace{3cm}~}
\end{example}

\begin{example}[Random segments in 2D] \label{ex:lines} Regularized plane arrangement generated by random line segments.
\vspace{1mm}

\begin{minipage}{0.25\textwidth}
   \centering
   \includegraphics[height=0.96\linewidth,width=0.96\linewidth]{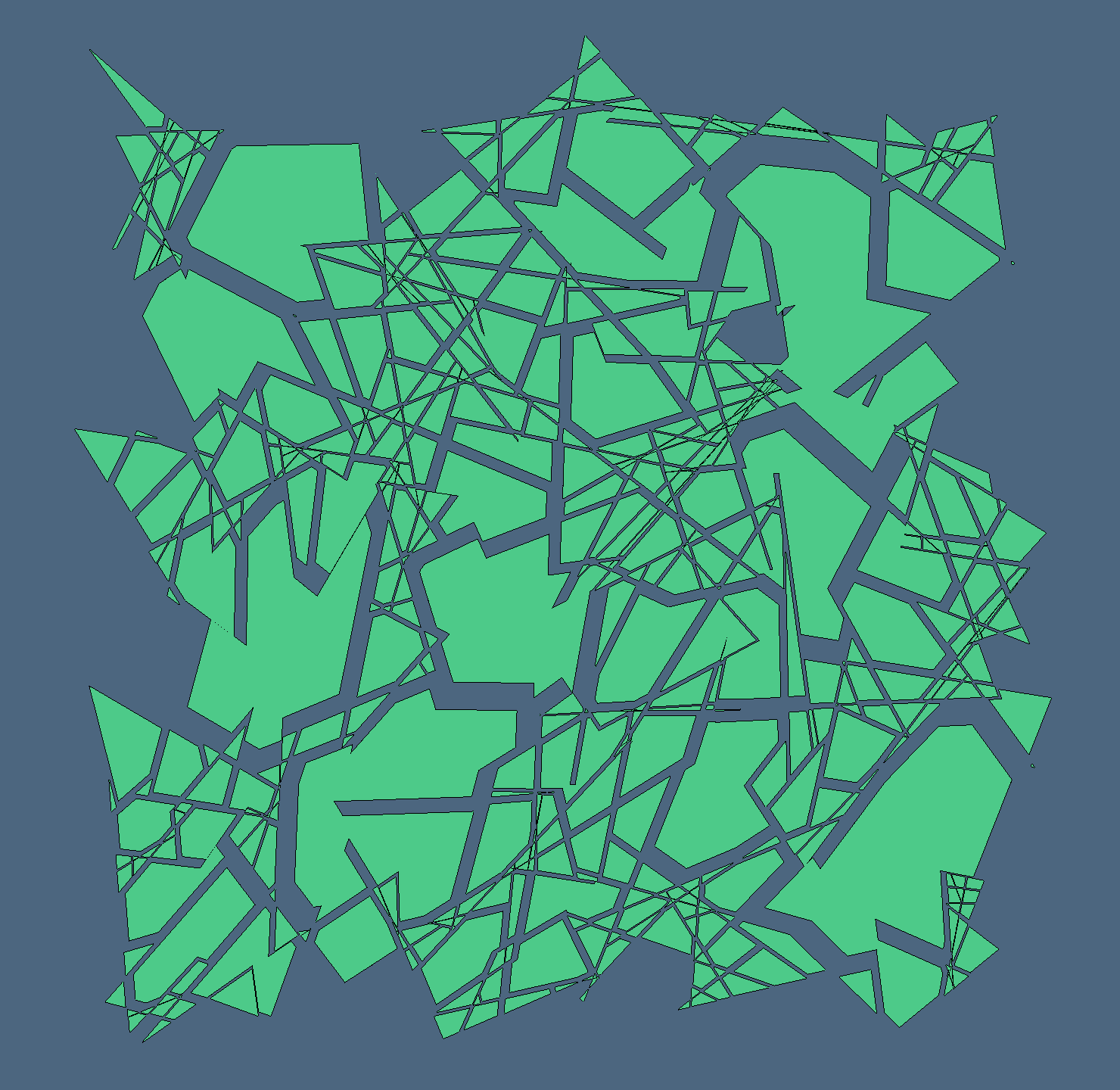}%
   \centering
   \includegraphics[height=0.96\linewidth,width=0.96\linewidth]{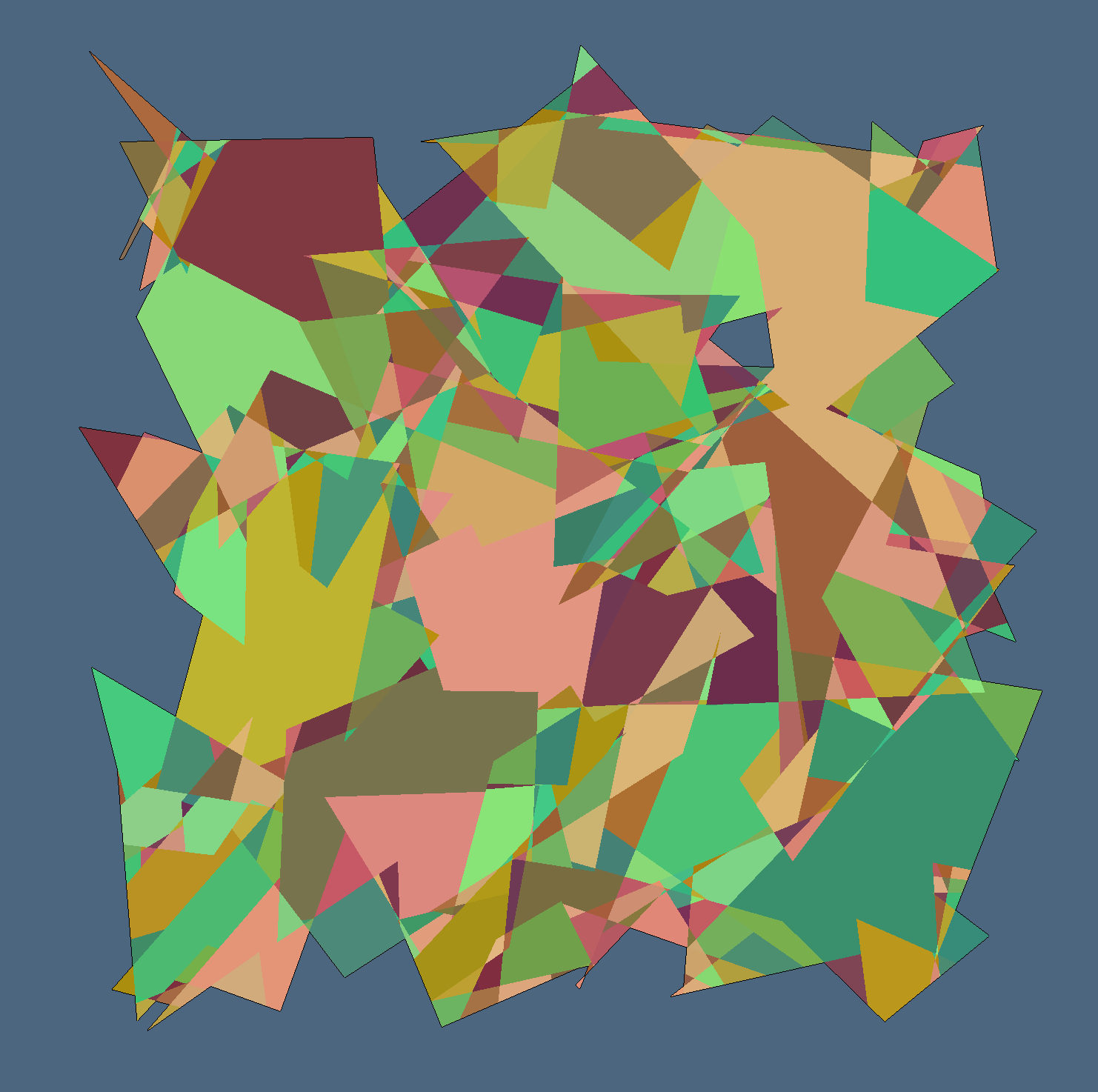}%
   \centering
   \includegraphics[height=0.96\linewidth,width=0.96\linewidth]{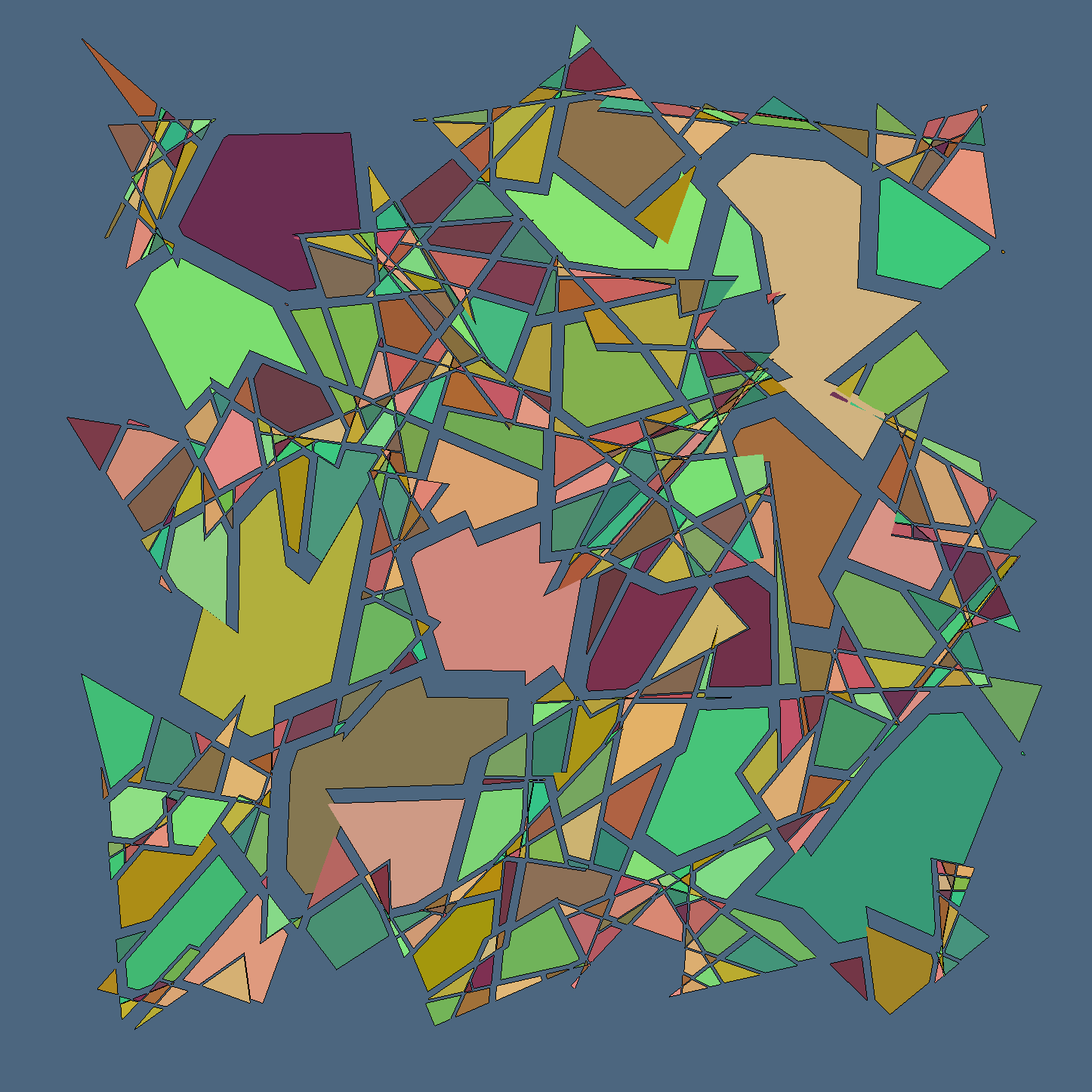}%
   \centering
   \includegraphics[height=0.96\linewidth,width=0.96\linewidth]{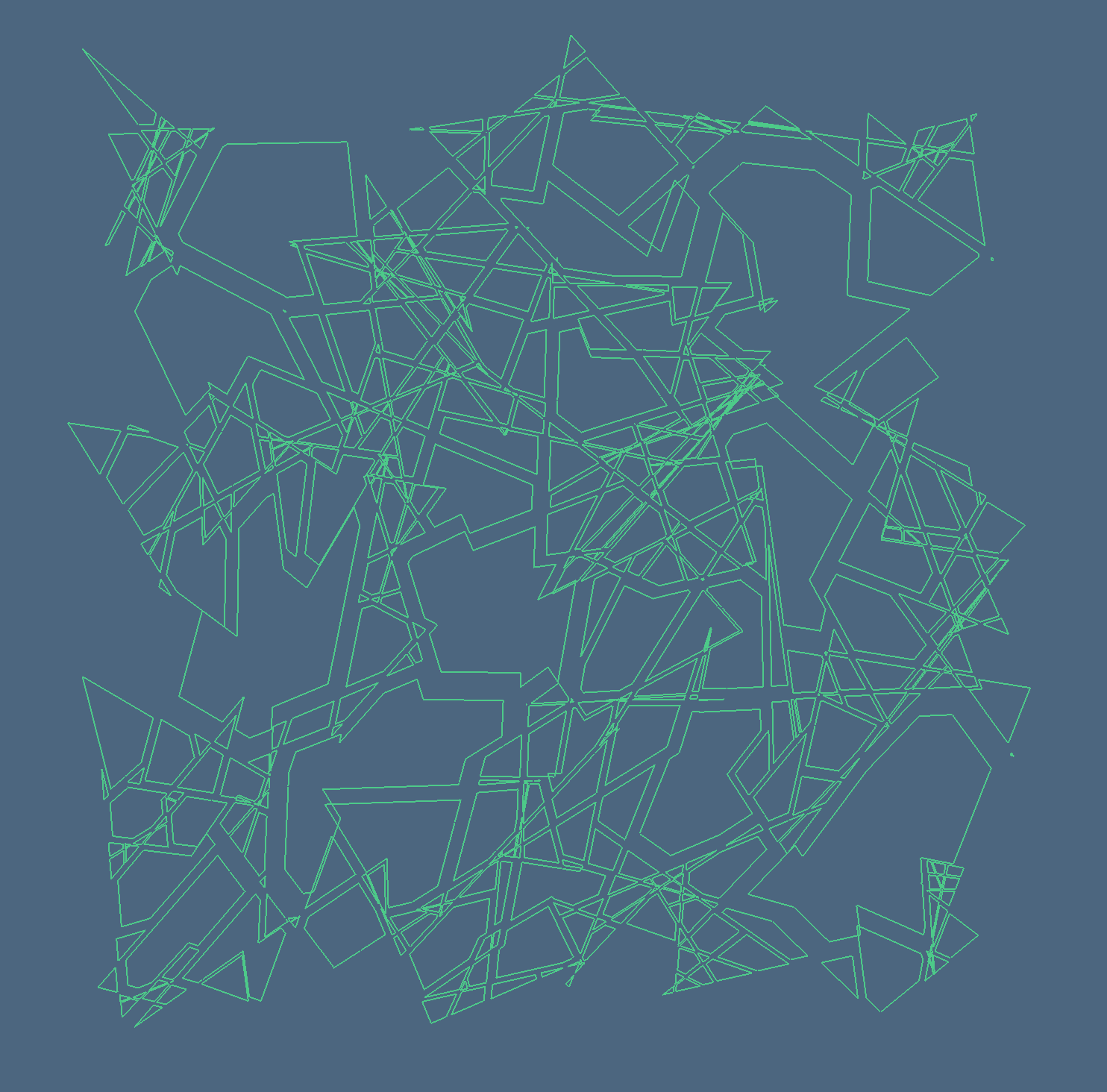}%
\end{minipage}
\end{example}

\subsection{Constructive Solid Geometry}\label{sec:csg}
The term \emph{Constructive Solid Geometry} (CSG) was coined by~\cite{RequichaVoelcker:77} in one the foundamental reports of Production Automation Project (PAP) at Rochester University. It is used 
to create a complex object by using Boolean operators to combine a few primitive objects. Sometimes it is used to procedurally describe the sequence of operations performed by a designer during the initial phases of shape creation. Today it is more used to characterize the data structures underlying the \lstinline{UNDO} operation within an interactive design shell, than to specify the generative description of a complex shape, i.e.~as a \emph{representation scheme}~\citep{Requicha:80}. Representation conversion algorithms between CSG and Boundary representations (B-reps) were given by~\cite{10.1145/169728.169723}.

\begin{definition}
Constructive solid geometry is a representation scheme for modeling rigid solid objects as set-theoretical compositions of primitive solid "building blocks"~\citep{RequichaVoelcker:77}.
\end{definition}

\begin{note}
CSG can be described as the combination of volumes occupied by overlapping 3D objects, using Boolean set operations. Typical primitive are defined as a combination of half-spaces delimited by general quadric surfaces (parallelepipeds, cylinders, spheres, cones, tori, closed splines, wedges, swept solids). Operations are union, intersection, difference, and complement.
\end{note}

\begin{remark}
Two more formal definitions of CSG are the following: (a) non complete binary tree with primitive solids in local coordinates on the leafs and either affine maps or Boolean operations on the non-leaf nodes; (b) complete binary tree with Booleans on the non-leafs and primitive solids in world coordinates (the coordinates of the root) on the leafs. In other words: the solid is represented as union, intersection, and difference of primitive solids that are positioned in space by rigid-body transformations~\citep{HDCG:2017}.
\end{remark}

\begin{algorithm}[Pairwise CSG evaluation]\label{alg:pairwise}
Pairwise CSG evaluation methods require a post-order DFS traversal of the expression tree, and the pairwise sequential computation of each Boolean operation encountered on each node with the current partial result, until the root operation is evaluated. See for example the $n$-ary computation of Booleans by~\cite{Zhou:2016:MAS:2897824.2925901}. Other types of traversal do not change the number and complexity of operations.
\end{algorithm}

\begin{remark}
It is well known that Boolean operations between B-reps of two solids $A$ and $B$ have $O(n^2)$ worst case time complexity, since every 2-face of $A$ may intersect with every face of $B$. Therefore it is possible to show that a flat subdivision is much efficient than a hierarchical one. See Property~\ref{prop:complexity}.
%
\end{remark}

\begin{note}
Standard (pairwise) n-ary CSG computational processes also lack in robustness since would accumulate numerical errors, that ultimately modify the topology of partial result, and make the applications easily stop in error. Hence, intermediate boundary representations need to be carefully curated, before continuing the traversal.  
\end{note}

\begin{algorithm}[Flat CSG evaluation]
In this paper we are going to give a novel organization of the computational process for evaluating CSG expressions of any depth and complexity. 
The evaluation of a CSG expression of any complexity is done here with a different computational approach. The tree itself is only used to apply affine transformations to solid primitives, in order to scale, rotate and translate them in their final (``world'') positions and attitudes. All their 2-cells are thus accumulated in a single collection, and each of them is efficiently fragmented against the others, so generating a collection of independent local topologies that are merged by boundary congruence, using a \emph{single} round-off  identification of close vertices. The global space partition is hence generated, and all initial solids are classified with respect to such 3-cells, with a single point-set containment test. Finally,  \emph{any Boolean form} of arbitrary complexity, using the same solid components, can be immediately evaluated by bitwise vectorized logical operations.
\end{algorithm}

\begin{property}[Complexity of evaluation]\label{prop:complexity}
The worst-case complexity of flat approach is better than the hierarchical one, as seen in an extremely simplified plot (see Remark after Algorithm~\ref{alg:pairwise}), supposing that on every pair of faces a single cut is executed in constant time,   In case of $m$ solid objects, each with $n$ boundary 2-cells, in the flat hypothesis that each 2-cell is intersected by all the others $mn-1$, we have a polynomial number $O(m^2n^2)$ of possible cuts. Conversely, in the hierarchical case, it happens that each 2-cell intersects all the ones \emph{previously split}, requiring an exponential number $O(2^{mn})$ of cuts. Even for $m=2$ and $n=6$, it is $m^2  n^2=144$ and $2^{mn} =4096$. Of course, this extreme hypothesis that each face cuts all the others, will never apply in practice, but the comparison seems anyway useful. 
\end{property}

\begin{algorithm}[CSG by ray-casting]
A typical \emph{approximate algorithm} for \emph{evaluation} of a CSG tree for graphical display is by way of ray-casting, and intersects for each ray the appropriate limit surfaces, to be computed by \emph{set-membership classification} (SMC) function.
This discrete algorithm requires a proper Boolean combination of the ray subdivision into line segments~\citep{tilove:78}.  
\end{algorithm}

\begin{example}[CSG example --- 1/2] \label{ex:explode}

An example of \emph{Constructive Solid Geometry} is shown below (from Wikipedia). The Boolean formula is built using three primitive solids (cylinder, cube, sphere). The expression tree is shown with operators on non-leaves, and primitive solids on the leaves. In our approach the assembly \texttt{Struct}, with parameterized generating functions and affine transformations (scale, rotation, translation) is used for the 3D positioning of the 5 terms \texttt{X$_1$,X$_2$,X$_3$,Y,Z}, respectively. 
Our prefix Julia translation is: \texttt{ @CSG (-,(*,Y,Z),(+,X$_1$,X$_2$,X$_3$))}. \vspace{3mm}

\begin{minipage}{0.28\textwidth}
   \centering\flushleft{\small
First (a) A standard CSG model is shown from Wikipedia, where it is defined as the \emph{binary tree} of the Boolean expression; then in (b)  is displayed the exploded set of all the atoms (randomly colored) of  $\E^3$ arrangement $\mathcal{A}(\mathcal{S})$ induced by the flat \emph{input collection} of five cellular $(d-1)$-complexes $\mathcal{S} = \{\texttt{X\_1,X\_2,X\_3, Y, Z} \}$, already mapped in world coords.}
   
\end{minipage}
\begin{minipage}{0.35\textwidth}
   \centering
   \includegraphics[height=0.9\linewidth,width=\linewidth]{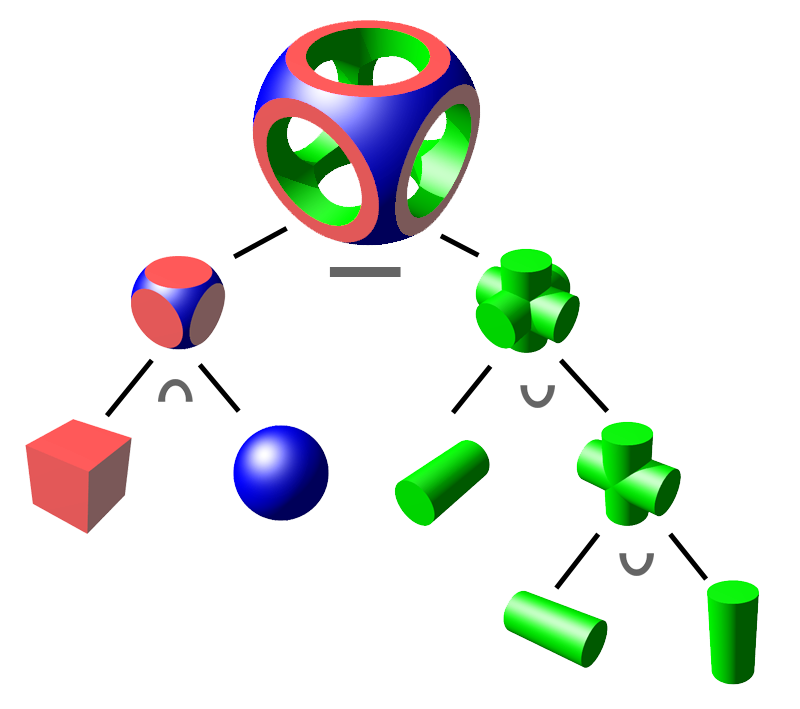} 
\end{minipage}
\begin{minipage}{0.33\textwidth}
   \centering
   \includegraphics[height=0.9\linewidth,width=\linewidth]{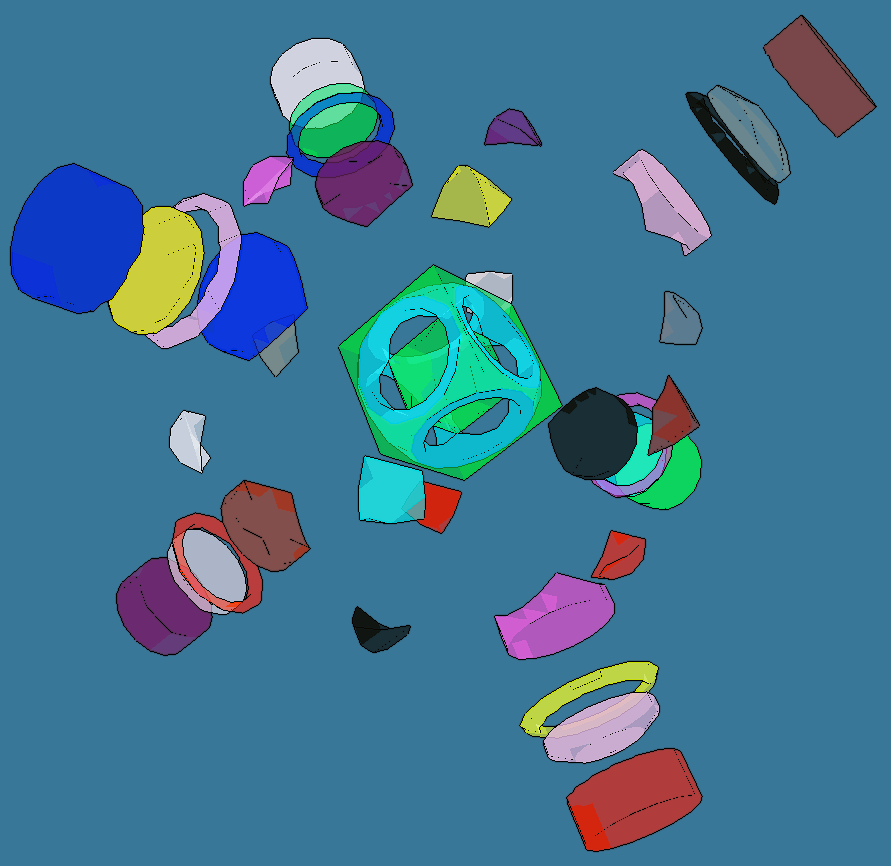} 
\end{minipage}
\end{example}

\subsection{Boolean Algebra}\label{sec:algebra}
In mathematics and mathematical logic, \emph{Boolean algebra} is the type of algebra in which the values of the variables are the truth values \emph{true} and \emph{false}, usually denoted 1 and 0, respectively.
In this paper we represent and implement for $d=2,3$ the solid Boolean algebras of piecewise-linear CSG with closed regular cells, generated by the arrangement of $\E^d$  induced by a collection of cellular complexes with polyhedral cells of dimension $d-1$~\citep{TSAS:19}. In particular, we show that the \emph{structure} of each term of this algebra is characterized by a discrete set of points, each one computed once and for all in the interior of each atom. Set-membership classifications (SMC) with respect to such single internal points of atoms, computes the \emph{structure} of any algebra element, and in particular transforms each solid term (each input) of every Boolean formula, into a \emph{logical array} of length $m$, equal to the number of atoms.

\begin{definition}[Finite algebra]
Let $\mathcal{A}$ be a nonempty set, and operations $\otimes_i : \mathcal{A}^{n_i} \to \mathcal{A}$ be functions of $n_i$ arguments. The system $\langle \mathcal{A}; \otimes_1, \ldots, \otimes_k \rangle$ is called an \emph{algebra}.
Alternatively we say that $\mathcal{A}$ is a set with operations $\otimes_1, \ldots, \otimes_k$.
By definition, an algebra $\mathcal{A}$  is closed under its operations. If $\mathcal{A}$  has a finite number of elements it is called a \emph{finite} algebra.
\end{definition}

\begin{definition}[Generators]
A set $\mathcal{H}$ generates the algebra $\mathcal{A}$ (under some operations) if $\mathcal{A}$ is the smallest set closed w.r.t the operations and containing $\mathcal{H}$. The elements $h_i \in \mathcal{H}$ are called \emph{generators} or \emph{atoms} of the algebra $\mathcal{A}$.
\end{definition}

\begin{definition}[Boolean Algebra]
We may think of Boolean algebras $\mathscr{B}$ as a set which is isomorphic to the power set $\mathscr{P}(X)$ of some finite set $X$. The power set is naturally equipped with complement, union and intersection operations, which corresponds to 
$-, \vee, \wedge$ operations in Boolean algebra.
\end{definition}

\begin{property}[Boolean Algebra]
Any algebra $\mathscr{B}$ with $n$ generators is isomorphic to the Boolean algebra $\mathscr{P}(X)$, with set $X$ containing $n$ elements. Therefore $\mathscr{B}$  can be mapped-one-to-one with the set $\chi_X(\mathscr{P}(X))$ of the images of characteristic functions of $\mathscr{B}$ elements with respect to $X$, i.e.~with the set of strings of $n$ elements in $\{0,1\}$.
\end{property}

\begin{definition}[Atom]
An \emph{atom} is something which cannot be decomposed into two proper subsets, like a singleton which cannot be written as a union of two strictly smaller subsets.
An atom is a minimal non-zero element;
$a$ is an atom iff for every $b$, either $b \wedge a = a$ or $b \wedge a = 0$.  In the first case we say that $a$ belongs to the \emph{structure} of $b$.
\end{definition}

\begin{definition}[Structure of algebra elements]
We call \emph{structure} of  $b \in \mathscr{P}(X)$  the atom subset $S$ such that $b$ is irreducible union of $S$. By extension, we also call structure of $b$ the \emph{binary string} associated to the ordered sequence of its atoms (elements of $X$). In other words, the structure $S(b)$ is the image of the characteristic function $\chi_X(S)$
\end{definition}

\begin{table}
\caption{Truth table associating 2-cells $c_i\in U_2$ ($1\leq i\leq 4$) to rows, and solid variables $A, B$ and $\Omega=\overline{A\cup B}$ to columns. See Example~\ref{fig:one}.}
\begin{tabular}{|c||ccc|}
\hline $U_2$ & $\Omega$ & $A$ & $B$  \\ \hline \hline $c_{1}$  & 1 & 0 & 0  \\ $c_{2}$ 
& 0 & 1 & 0  \\ $c_{3}$  & 0 & 1 & 1  \\ $c_{4}$  & 0 & 0 & 1  \\ \hline
\end{tabular} 
\label{tab:one}
\end{table}

\begin{table}
\caption{Truth table providing the complete enumeration of elements $S$ of the finite Boolean algebra  $\mathcal{A}$ generated by two solid objects $A, B$ and four atoms $c_i$ in the basis $U_2$ (blue) of chain space $C_2$ associated with the arrangement $\mathcal{A}(\mathcal{S})$, with $\mathcal{S}=\{\partial(A), \partial(B)\}$.}
\begin{tabular}{|c||cccccccccccccccc|} 
\hline 
\rotatebox{90}{$U_2$} &
\rotatebox{90}{$X = \E^2\ $} & \rotatebox{90}{{\color{red} $A$}} &
\rotatebox{90}{{\color{red} $B$}} & \rotatebox{90}{$A\cup B$} &
\rotatebox{90}{$\overline{A\cup B}$} & \rotatebox{90}{$A \!\setminus\! B$} &
\rotatebox{90}{$A \cap B\,$} & \rotatebox{90}{$B\!\setminus\! A$} &
\rotatebox{90}{$A\oplus B$} & \rotatebox{90}{$\overline{A\!\setminus\! B}$} &
\rotatebox{90}{$\overline{B}$} & \rotatebox{90}{$\overline{B\!\setminus\! A}$} &
\rotatebox{90}{$\overline{A}$} & \rotatebox{90}{$\overline{A\oplus B}$} &
\rotatebox{90}{$\overline{A\cap B}\ $} & \rotatebox{90}{$\varnothing$}  
\\
\hline \hline  
$c_{1}$  & $1$ & {\color{red} $0$} & {\color{red} $0$} & $0$ & {\color{blue}$1$} & {\color{blue}$0$} & {\color{blue}$0$} & {\color{blue}$0$} & $0$ & $1$ & $1$ & $1$ & $1$ & $1$ & $1$ & $0$  
\\ 
$c_{2}$  & $1$ & {\color{red} $1$} & {\color{red} $0$} & $1$ & {\color{blue}$0$} & {\color{blue}$1$} & {\color{blue}$0$} & {\color{blue}$0$} & $1$ & $0$ & $1$ & $1$ & $0$ & $0$ & $1$ & $0$  
\\ 
$c_{3}$  & $1$ & {\color{red} $1$} & {\color{red} $1$} & $1$ & {\color{blue}$0$} & {\color{blue}$0$} & {\color{blue}$1$} & {\color{blue}$0$} & $0$ & $1$ & $0$ & $1$ & $0$ & $1$ & $0$ & $0$  
\\ 
$c_{4}$  & $1$ & {\color{red} $0$} & {\color{red} $1$} & $1$ & {\color{blue}$0$} & {\color{blue}$0$} & {\color{blue}$0$} & {\color{blue}$1$} & $1$ & $1$ & $0$ & $0$ & $1$
& $0$ & $1$ & $0$  
\\ 
\hline 
\end{tabular} 
\label{tab:table2}
\end{table}

\begin{example}
Let $\mathcal{P}$ be a set of $m$ hyperplanes that partition $\E^d$ into convex relatively open cells of dimension ranging from zero (points) to $d$. The collection of all such cells is called a \emph{linear arrangement} $\mathcal{A}(\mathcal{P})$ and has been studied extensively in computational geometry~\citep{Edelsbrunner:1987:ACG:28905}. The cells in $\mathcal{A}(\mathcal{P})$ are atoms of a Boolean algebra of subsets of $\E^d$ that can be formed by union of convex cells in the arrangement.
\end{example}

\begin{example}[2D space arrangement] 
generated in $\E^2$ by two overlapping 2-complexes. Figure (c) below shows a partition of $\E^2$ into four irreducible subsets: the red region A, the green region B, the overlapping region $A\cup B$ and the outer region $\overline{A\cup B}$, {i.e.,}~the rest of the plane.
The set of \emph{atoms} of Boolean algebra $\mathcal{B}$ is the same: the colored regions  are the four atoms of $\mathcal{B}$ algebra, and there are $2^4 = 16$ distinct elements $S \in \mathcal{B}$. The \emph{structure} of each element $S\in\mathcal{B}$ is a \emph{union} of $\mathcal{B}$ atoms;  as a chain in $C_2$, it is a \emph{sum} of basis elements.
\vspace{3mm}

\begin{minipage}{0.25\textwidth}
   \flushleft{\small
	The 4+4 line segments (1-cells) in
   the 1-skeletons, generating the \emph{four} 2-cells:  blue ($c_1$),
   red ($c_2$), white ($c_3$), and green ($c_4$). The first ($c_1$) is the
   \emph{outer} or \emph{exterior} cell $\Omega$, which is the complement of the
   union of the others. }
   \label{fig:one}
\end{minipage}
\begin{minipage}{0.235\textwidth}
   \centering
   \includegraphics[height=0.9\linewidth,width=\linewidth]{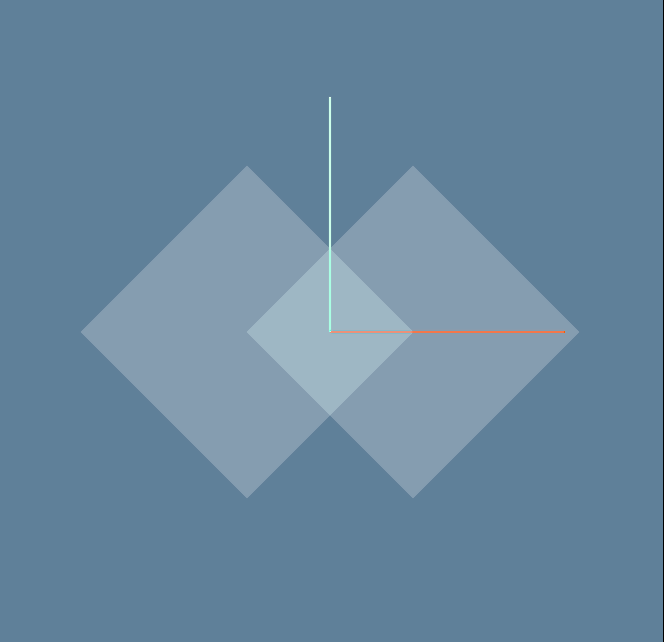} 
\end{minipage}
\begin{minipage}{0.235\textwidth}
   \centering
   \includegraphics[height=0.9\linewidth,width=\linewidth]{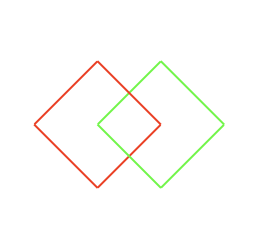} 
\end{minipage}
\begin{minipage}{0.235\textwidth}
   \centering
   \includegraphics[height=0.9\linewidth,width=\linewidth]{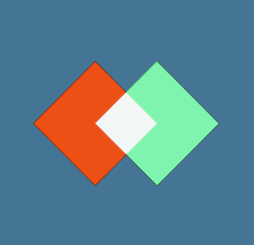} 
\end{minipage}
\end{example}

\begin{example}[Boolean algebra] 
Let consider the columns of Table~\ref{tab:one}. 
By columns, the table contains the coordinate representation of the 2-chains $A$, $B$, and
$\Omega = \overline{A\cup B}$ in chain space $C_2$. In algebraic-topology language:
\[
A = c_2 + c_3,\quad  B = c_3 + c_4,\quad  \Omega = c_1 = - (c_2 + c_3 + c_4).
\]

The Boolean algebra of sets is represented here by the power set $\mathscr{A} = \mathscr{P}(U_2)$, with $U_2 = \{c_1,c_2,c_3,c_4\}$, whose
$2^4 = 16$ binary terms of length $\#\,U_2 = 4$ are given  in Table~\ref{tab:table2}, together with their
semantic interpretation in set algebra. Of course, the coordinate vector representing the universal set, {i.e.}~the whole
topological space $X := \E^2$ generated by $[\partial_2]$ columns is given by $[X] =
(1,1,1,1)$, including the exterior cell. Let us remember that the
\emph{complement} of $A$, denoted $-A$ or $\overline{A}$, is defined as $X\!\setminus\! A$ and that the 
\emph{difference} operation $A \!\setminus\! B$ is defined as $A \cap \overline{B}$. 
\end{example}

\begin{example}[Solid algebra 3D]
 Each
\emph{3-cell is generated} by \emph{a column} of the \emph{sparse matrix} of
\emph{boundary map} $\partial_3: C_3\to C_2$,  with values in $\{-1,0,1\}$. Columns of $[\partial_3]$ are 2-cycles, i.e.~closed chains in $C_2$. In fact, closed chains have no boundary.
3-cells are the join-irreducible \emph{atoms} of the CSG algebra with closed regular cells. They 
may be non contractible to a point (when they contain a hole) and non convex. The exterior cell is the complement of their union. Any model from the Boolean algebra (out of $2^{25}$ -- see figure below) of the CSG generated by these 
five cubes is made by the same 25 atoms.

\begin{minipage}{0.25\textwidth}
   \flushleft{\small
   (a) A collection $\mathcal{S}$ of five random cubes in $\E^3$; \\[3mm]
   (b) the  display 
   of 3-cells of the generated $\E^3$ arrangement $\mathcal{A}(\mathcal{S})$.  
   3-cells are here not in scale, and are suitably rotated to better exhibit 
   their complex structure. Their quasi-disjoint union gives the five cubes.}
\end{minipage}
\begin{minipage}{0.33\textwidth}
   \centering
\includegraphics[height=\linewidth]{figs/image1.jpg}%
\end{minipage}
\begin{minipage}{0.33\textwidth}
   \centering
\includegraphics[height=\linewidth]{figs/image2.jpg}%
\end{minipage}
\label{fig:example} 

\end{example}

\subsection{Topological Algebras}\label{sec:algebra}
\cite{Shapiro:1991:RSS:124951} presented
a \emph{hierarchy of algebras} to define formally a
family of Finite Set-theoretic Representations (FSR) of semi-algebraic\footnote
{
A \emph{semialgebraic set} is a subset $S\subset R^n$  defined by a finite sequence of polynomial equations (of the form $P(x_1,...,x_n)=0$) and inequalities (of the form $Q(x_1,...,x_n) > 0$.
Finite unions, intersections, complement and projections of semialgebraic sets are still semialgebraic sets.
} 
subsets of $\E^d$, including many known representation schemes for solid and non-solid
objects, such as boundary representations, Constructive Solid Geometry, cellular 
decompositions, Selective Geometric Complexes, and others.  Exemplary
applications included B-rep $\to$ CSG and CSG $\to$ B-rep conversions.
In particular, a hierarchy  of algebras over semi-algebraic subsets
of $\E^d$ is discussed, where algebra elements can be constructed using a finite number of FSR operations (standard set ops, closure, and connected component).  

\begin{definition}[Chains as topological basis]
A \emph{basis} $B$ for a topological space $X$ with topology $T$ is a subset of open 
sets in $T$ such that all other open sets can be written as union of elements of $B$. 
A set is said to be \emph{closed} if its complement is \emph{open}.
When both a set and its complement are open, they are called \emph{clopen}. 
Let us consider both $d$-cells and $d$-chains as point-sets.
The chains in a linear space $C_d$ can be seen as open point-sets generated by
\emph{topological sum} of topological spaces. 
\end{definition}
%
\begin{property}[Chain space as discrete topology]
It can be proved that, if all connected components of a space are open, then a
set is clopen if and only if it is the union of pairwise disjoint
connected components. All chains in $C = \oplus_d C_d$ space are clopen. Even more, all
sets $C_p$ of all $p$-chains, including $C_p$ itself and $\varnothing$, has the
\emph{discrete topology}, since all elements of the power set  $X =
\mathscr{P}(U_d)$ are clopen, where $U_d$ is the basis.
\end{property}

\begin{property}[Chain spaces as Boolean algebras]
The set $X$ of subsets of independent
$d$-chains may represent a \emph{discrete topological space}. Using the union and
intersection as operations, the clopen subsets of a discrete  topological space
$X$ form a \emph{finite Boolean algebra}. 
It can be shown that every finite Boolean algebra is isomorphic to the Boolean algebra $\mathcal{B}$
of all subsets of a finite set~\citep{Halmos:1963}, and can be represented by the
$2^{N}$ binary terms in $\mathcal{B}$. In our case $N = \dim C_d = \#U_d$.
\end{property}

\section{Computational Pipeline}\label{sec:pipeline}
Every Boolean form of the solid PL algebra investigated in this paper is evaluated by the following computational pipeline: (a) computing the arrangement of $\E^d$ space induced by the input polyhedra, followed by (b) the computation of a sparse array of bits, denoting the atomic structure, for each input solid variable, and by (c) the native application of bitwise operators on bit strings,  according to the symbolic formula to be solved. A single $\E^d$ decomposition may be used to evaluate every functional form over the same algebra, including every subexpression between the same arguments (see Example~\ref{ex:forms}).

\subsection{Subdivision of 2-cells}\label{sec:subdivision}

Every $2$-cell $\sigma$ in the input set $\mathcal{S}_2$ is \emph{independently} intersected with the others of possible intersection, producing its own arrangement of~$\E^2$ space as a chain 2-complex $C_\bullet(\sigma)$. The 2-cells of possible intersection with $\sigma$ are those whose containment boxes intersect the box of $\sigma$ .

\begin{algorithm}[Splitting of 2-cells in 2D]\label{alg:splitting}
Each set $\mathcal{I}(\sigma)$ of $2$-cells of possible
intersection with $\sigma$, is efficiently computed by combinatorial intersection
of query results  on $d$ different (one for each coordinate)
interval-trees~\citep{Preparata:1985:CGI:4333}. Every
$\mathcal{I}(\sigma)$ set, for $\sigma\in\mathcal{S}_2$, is affinely mapped in $\E^3$, 
leading $\sigma$ to the $z=0$ subspace.  The arrangement
$\mathcal{A}(\sigma) = C_\bullet(\sigma)$  is firstly computed in $\E^2$, and then mapped back into
$\E^3$. 
\end{algorithm}

\begin{remark}
The actual intersection of 2-cells is computed between line segments in 2D, retaining only the non external part of the maximal 2-connected subgraph of results, so obtaining a regular 2-complex decomposition of 2-cell $\sigma$. See Figure in Example~\ref{fig:2D}.
This computation is executed \emph{independently} for each 2-cell $\sigma$ in the input 
cellular complexes. 
To strongly accelerate the computation, the splitting may be computed through Julia's \texttt{channels}, which are able to take advantage of 
both local and remote compute nodes, making use of this \emph{embarrassingly parallel} 
data-driven approach.
\end{remark}

\begin{property}[Time complexity]   
The  complexity of the 2-cell subdivision Algorithm~\ref{alg:splitting} can be regarded in the average case as linear in the total number of 2-cells, times a $\log n$ factor, needed to compute those of probable intersection, when the splitting of a single cell may be considered done in constant time. In fact, normally, each cell is small with respect to the scene, and is intersected by a small number of other cells. Under this hypotheses both the mapping to/from $z=0$ and each 1-cell intersection are done in constant time and are in small number. Furthermore, the 2D cell complex creation produces two small sparse matrices $\delta_0(\sigma)$ and $\delta_1(\sigma)$ per 2-cell $\sigma$. The numbers of their rows and columns are those of vertices, edges and faces of the split $\sigma$, i.e.~very small bounded numbers. The complexity of Algorithm~\ref{alg:splitting} is hence $O(\ell k\, (n\log n))$, with $\ell$ average number of produced component cells (depending on both size e incidence number of each $\sigma$) and $0\leq k\leq 1$, a distribution factor inversely proportional to the number of used cores. The initial construction of interval trees has the same complexity of (double) sorting of intervals.
\end{property}

\begin{example}[Subdivision of 2-cells]   \label{fig:2D} 
After computation of 2-cells $\mathcal{I}(\sigma)$ of possible intersection, for each input 2-cell $\sigma$, a (parallel) computation is executed  of the regularized arrangements  in $\E^2$ of sets of lines. Successive operations are displayed in the figure below, with their captions on the left.

\begin{minipage}{0.50\textwidth} 
\flushleft {\small
   \begin{description}
\item 
   (a) input to \emph{cell splitting} Algorithm~\ref{alg:splitting}: \emph{i.e.},~the 2-cell $\sigma$ (signed blue) and the line segments (solid lines), from intersection of $\mathcal{I}(\sigma)$ with the plane $z=0$ ; 
\item 
   (b) all pairwise intersections of such 1-cells in 2D, producing a linear graph; 
\item 
   (c)~removal of the 1-subgraph external to $\partial\sigma$; 
\item 
   (d) chain 2-complex (0-, 1-, and 2-cells) generated by $\sigma\cup\mathcal{I}(\sigma)$ via TGW in 2D (see Section~\ref{sec:2D-TGW}).
\end{description}
}
\end{minipage}
\begin{minipage}{0.47\textwidth}
   \centering
   \includegraphics[width=0.75\linewidth]{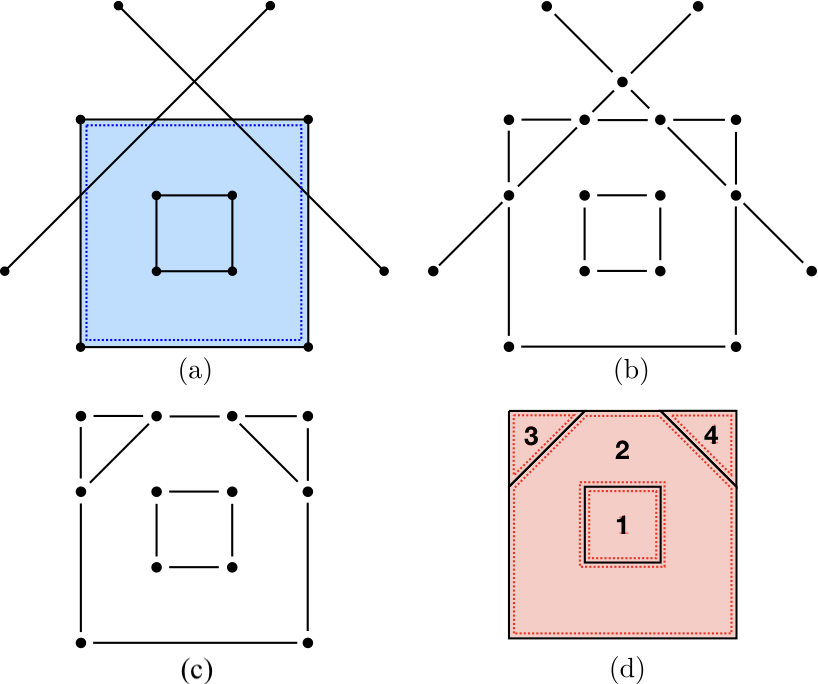} 
\end{minipage}
\end{example}

\subsection{Topological Gift Wrapping in 2D}\label{sec:2D-TGW}

To compute the 2-cells as 1-cycles and the corresponding $[\partial_2]$ matrices, 
the \emph{Topological Gift Wrapping} (TGW) algorithm~\citep{TSAS:19} is applied in 2D to the matrix $[\partial_1](\sigma)$, for each $\sigma$. A step-by step construction of a 1-cycle starting from a single 1-cell within a 1-complex, is shown in Example~\ref{fig:2D-full}. A complete description of this algorithm and the related pseudocode are given in~\cite{arXiv:2017} and~\cite{TSAS:19}. A summary description is given below.

\begin{remark}[Meaning of matrix columns]
First recall that the matrix of a linear transformation between two linear spaces is uniquely determined once the bases of the domain and target spaces are fixed. Then notice that the matrix columns are the coordinate representation of the basis element of domain, represented as linear combination of the basis of the target space. In particular, the columns contain the scalar coefficients of such linear combinations, in our case taken from $\{-1,0,+1\}$. Hence, a basis $d$-cell is represented by his boundary, a linear combination of basis ($d$-1)-cells.
\end{remark}

\begin{algorithm}[Topological Gift Wrapping]\label{alg:tgw}
We summarize here the TGW algorithm in $d$-space.
The input is the sparse matrix $[\partial_{d-1}]$; the output is the matrix $[\partial_{d}^+] : C_d \to Z_{d-1}$, from $d$-chains to ($d$-1)-cycles (see Section~\ref{sec:cycles}). 

{\small
\begin{enumerate}
\item Initialization: $m,n = [\partial_{d-1}].shape\ $; $marks = zeros(n)$; $[\partial_d^+] = []\ $; 

\item \textbf{while} sum(marks) < 2n \textbf{do}
\begin{enumerate}

\item select the $(d-1)$-cell seed of the column extraction

\item compute boundary $c_{d-2}$ of seed cell

\item \textbf{until} boundary $c_{d-2}$ becomes empty \textbf{do}
\begin{enumerate}

\item $corolla$ = []

\item \textbf{for each} ``stem'' cell $\tau\in c_{d-2}$ \textbf{do}
\begin{enumerate}

\item[a.] compute the $\tau$ coboundary

\item[b.] compute the new $petal$ cell

\item[c.] orient $petal$ and insert it in $corolla$

\end{enumerate}

\item insert $corolla$ cells in current $c_{d-1}$

\item compute again the boundary of $c_{d-1}$
\end{enumerate}

\item update the mark counters of used cells

\item append a new column to $[\partial_d^+]$
\end{enumerate}

\end{enumerate}
}
\end{algorithm}

\begin{example}[Topological Gift Wrapping] \label{fig:2D-full} A step by step example of computation of a 2-chain boundary as a 1-cycle is given below.

\begin{minipage}{0.6\textwidth} \flushleft
Figure on right side shows a fragment of a 1-complex $X=X_1$ in $\E_2$, with unit chains $u_0^k\in C_0$ and $u_1^h\in C_1$. Here we compute stepwise the 1-chain representation $c\in C_1$ of the central 2-cell of the unknown complex $X_2 = \mathcal{A}(X_1)$, using the Topological Gift Wrapping Algorithm~\ref{alg:tgw}. Refer to to the figure below to stepwise follow  the extraction of a 2-cell as 1-cycle. 
\end{minipage}
\begin{minipage}{0.33\textwidth}
	\centering
   \includegraphics[width=0.8\textwidth]{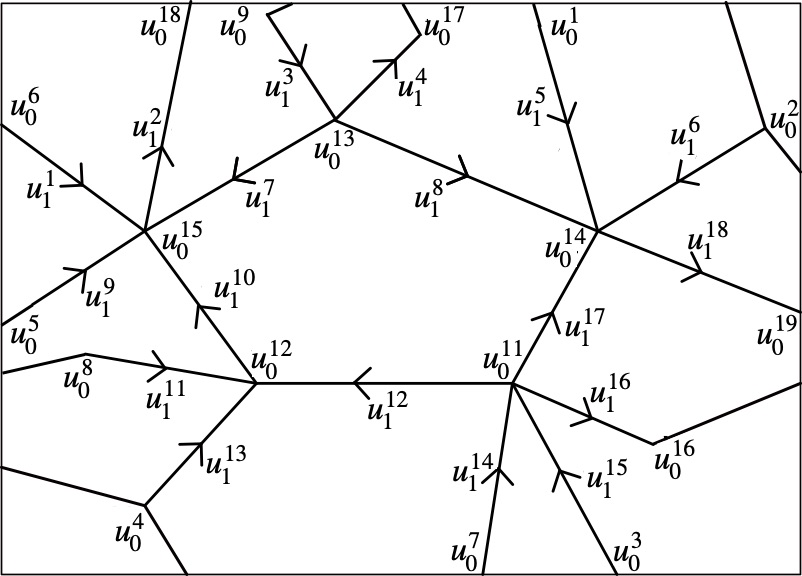} 
   
   {\small A portion of 1-complex in $\E^2$, with unit chains $u_0^h \in C_0$ and $u_1^k \in C_1$.}
\end{minipage}\vspace{2mm}

\includegraphics[height=0.17\textwidth]{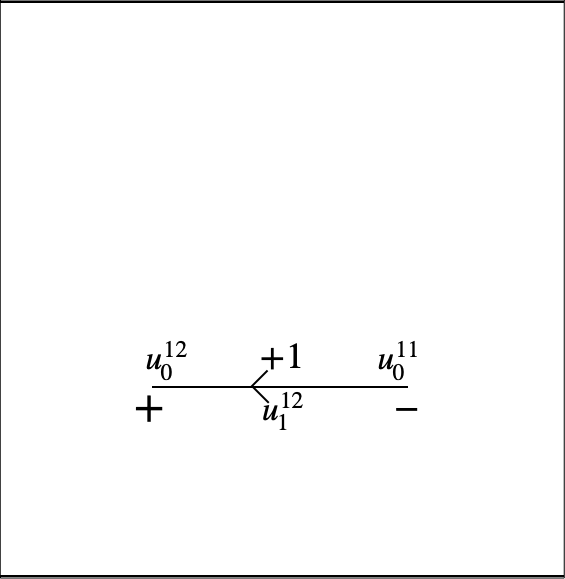}%
\includegraphics[height=0.17\textwidth]{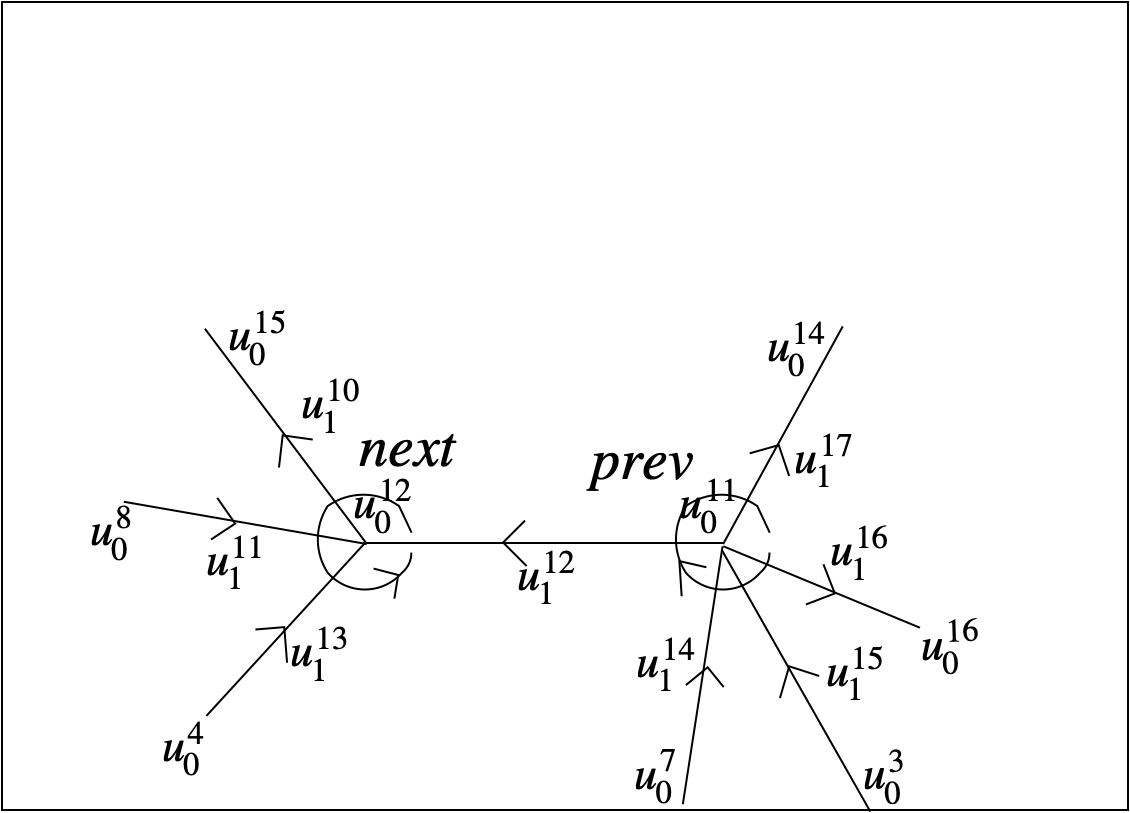}%
\includegraphics[height=0.17\textwidth]{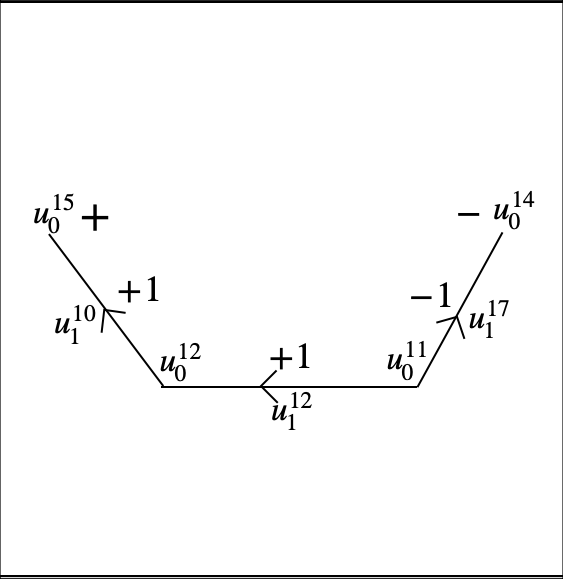}%
\includegraphics[height=0.17\textwidth]{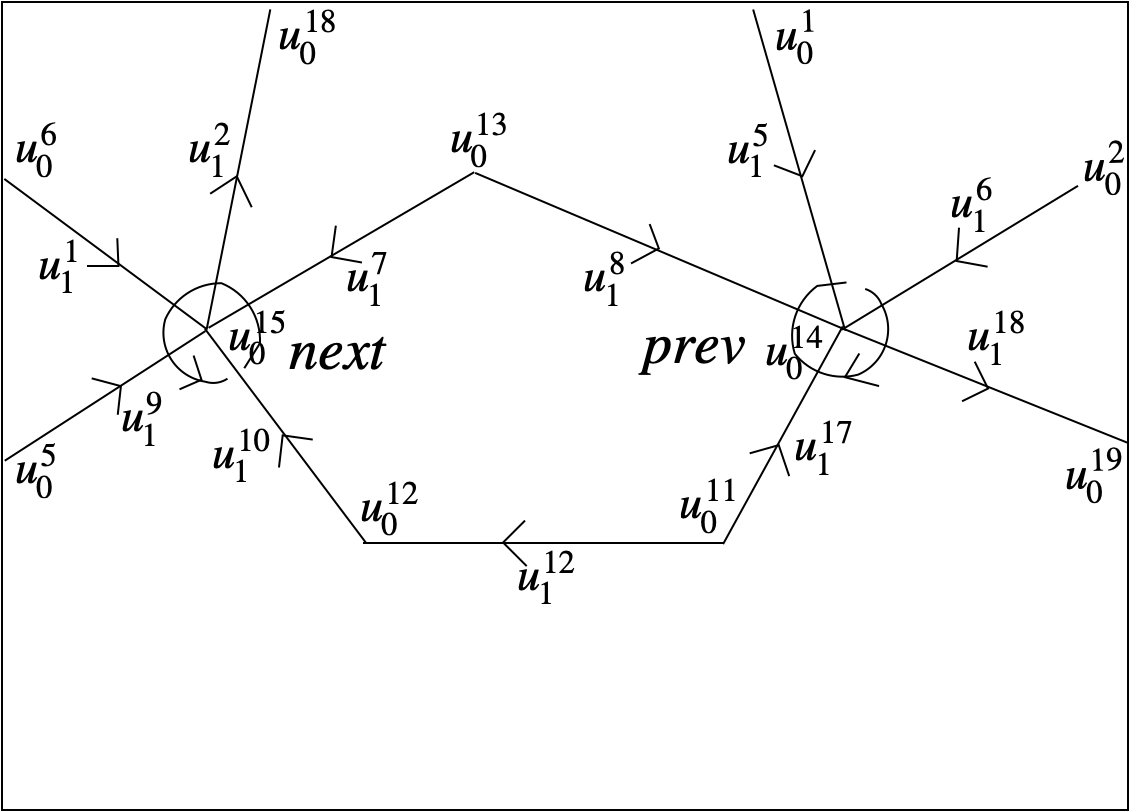}%
\includegraphics[height=0.17\textwidth]{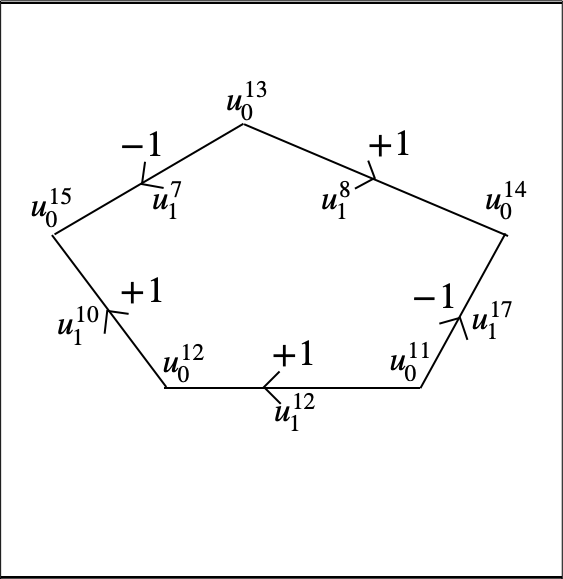}%

{\footnotesize\hspace{.09\textwidth}(a)\hfill(b)\hfill(c)\hfill(d)\hfill(e)\hspace{.09\textwidth}}

Extraction of an irreducible 1-cycle from the arrangement from $\mathcal{A}(X_1)$ of $\E^2$ generated by a cellular 1-complex $X_1$: (a) the initial value for $c\in C_1$ and the signs of its oriented boundary; (b) cyclic subgroups on $\delta\partial c$; (c) new (coherently oriented) value of $c$ and $\partial c$; (d) cyclic subgroups on $\delta\partial c$; (e) final value of $c$, with $\partial c = \emptyset$.
\end{example}

%

\subsection{Congruence Chain Complex}\label{sec:congruence}

We introduce in this section a simple way to glue together a set of local 
geometric and topological information
generated independently from each input 2-cell, by giving few definitions of 
nearness and congruence, which will allow to transform the union of \emph{local} topologies 
into a single \emph{global} ``quotient topology''. For a deeper discussion the reader is referred to~\cite{2020arXiv200400046D}.

\begin{definition}[\(\epsilon\)-nearness]
We say that two points \(u,v\in V\) are \emph{$\epsilon$-near}, and write
\(u\overset{\epsilon}{\sim} v\), when their Euclidean distance is
\(d(u,v) \leq 2\epsilon\). 
Let $S_i$ be the subset of points at distance less than $\epsilon$ from $v_i$.
The \(\epsilon\)-nearness \(\overset{\epsilon}{\sim}\) is an equivalence relation, since it is reflexive, symmetric, and transitive. In particular, it is transitive since
\emph{any} pair of points \(u,v \in S_i\)
are \(\epsilon\)-near, because both have a distance less than
\(\epsilon\) from \(v_i\), and hence have a distance no more than $2\epsilon$ from each other.
More formally, if \(u\overset{\epsilon}{\sim}v\) and \(v\overset{\epsilon}{\sim}w\), then \(u\overset{\epsilon}{\sim}w\), since the distance  from $v_i$  of every point ({e.g.}, $u,v,w$) in $S_i$ is less than $\epsilon$. 
\end{definition}

\begin{definition}[\(\epsilon\)-congruence]
We say that two \(p\)-cells $e,f$ are \emph{\(\epsilon\)-congruent}, and write $e\cong f$, when there exists a bijection $\mu$ 
between their 0-faces that pairwise maps vertices to \(\epsilon\)-near
vertices. 
\end{definition}

\begin{definition}[Quotient topology]
The quotient space of a topological space $X$ under a given equivalence relation is a new topological space constructed by endowing the quotient set of $X$ with the \emph{quotient topology}. This one is the finest topology that makes continuous the \emph{canonical projection} $\pi: x \mapsto [x]$, i.e., the function that maps points to their equivalence classes.
\end{definition}

\begin{note}[Quotient topology]
The \emph{quotient} or \emph{identification topology} gives a method of getting a
topology on \(X/\!\!\cong\) from a topology on \(X\). The quotient
topology $\pi(T)$ is exactly the one that makes the resulting space `look like'
the original one, with the identified elements glued together.
\end{note}

\begin{property}[Chain Complex Congruence]
The \(\epsilon\)-congruence between
elementary chains (\emph{aka} cells) \(c, d\in C_p\), denoted \(c\cong d\), is a
graded equivalence relation, so that a chain complex \((C_p,\delta_p)\)
may be represented by a much smaller one, that we call \emph{congruence chain complex}, with 
\(\pi(C_p,\delta_p)= (G_p, \delta'_p)\), where  \(G_p =  \pi(C_p) =  C_p/\!\!\cong\),
and where
\[\delta'_p = \delta_p\circ\pi := \pi(C_p)\to \pi(C_{p+1}) = \delta_p: G_p \to G_{p+1}.\]
\end{property}

\begin{data}[Block-diagonal accumulator matrix]
While computing the space arrangement generated by a collection of cellular complexes, we started from 
independent computation of the intersections of each single input 2-cell with the others.
The topology of these intersections is codified within a set of (0--2)-dimensional chain complexes,
stored within  \emph{accumulator matrices} $[\Delta_1]:C_1\to C_2$ and $[\Delta_0]:C_0\to C_1$. 
\vspace{2mm}

\begin{minipage}{0.66\textwidth} 
{\small\flushleft
Both \emph{accumulator matrices} $[\Delta_1]:C_1\to C_2$ and $[\Delta_0]:C_0\to C_1$ have a sparse block-diagonal structure, with two nested levels of (sparse) diagonal blocks. Each \emph{outer block} concerns one of the $m$ input geometric objects, and $n_k$ \emph{inner blocks}, $1\leq k\leq m$, store the matrices of each decomposed $\xi(\sigma)$ 2-cell. 
 We call $[\theta_h], 1\leq h \leq m$, the exterior blocks (light gray), and $[\epsilon_k], 1\leq k \leq m_h$, the interior blocks (dark gray), where $m_h$ is the number of 2-cells in $k$-th input geometric object. }
\end{minipage}
\begin{minipage}{0.3\textwidth}
\centering
   \includegraphics[width=0.7\textwidth]{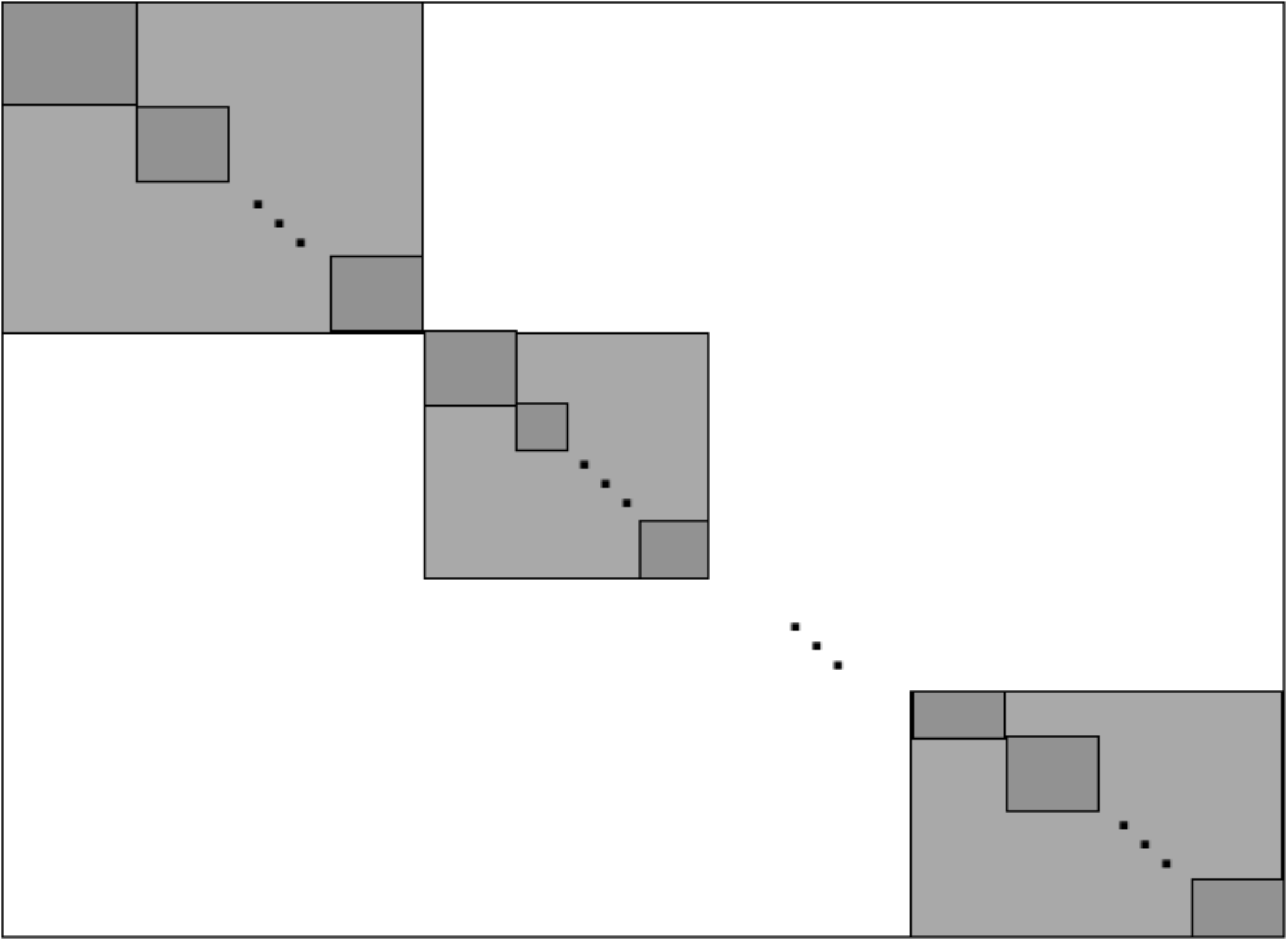} 
\end{minipage}

\vspace{2mm}\noindent 
In the algorithm specification below, 
a dense array  \texttt{W} and two sparse arrays \texttt{Delta\_0}, \texttt{Delta\_1} are respectively used for the input vertex coordinates and the sparse repositories of block-matrices $[\Delta_0]$ and $[\Delta_1]$.
\end{data}

\begin{algorithm}[Chain Complex Congruence (CCC)]\label{alg:ccc}
We have discussed  above the block diagonal marshaling $[\Delta_0]$ and $[\Delta_1]$ of local coboundary matrices. 
The target of the CCC algorithm is to merge the local chains by using the equivalence relations of $\epsilon$-congruence between 0-, 1-,  and 2-cells (elementary chains).

In particular, we reduce the block-diagonal coboundary matrices $[\Delta_0]$ and $[\Delta_1]$, used as matrix accumulators of the \emph{local} coboundary chains, to the \emph{global} matrices $[\delta_0]$ and $[\delta_1]$, representative of congruence topology, i.e., of congruence quotients between all 0-,1-,2-cells, via elementary algebraic operations on their columns. 

\begin{enumerate}

\item
We discover the \emph{$\epsilon$-nearness} of vertices by calling the function \texttt{CSG.vcongruence(V::Matrix;epsilon=1e-6)} on the $3\times n$ input \texttt{V} of 3D coordinates, which returns the new vertices \texttt{W} and the \texttt{vclasses} map of $\epsilon$-congruence.  The $3\times m$ matrix \texttt{W} holds the coordinates of class representatives, mapped to each class centroid.

\item
With the Julia function \texttt{CSG.cellcongruence} we replace each subset of columns of \texttt{Delta\_0} sparse matrix  corresponding to \emph{$\epsilon$-near vertices}, with their centroid. A new matrix is produced from the array of new vectors. Finally, equal rows of this new matrix, discovered via a dictionary, are substituted by a single representative. 

\item
The same function \texttt{CSG.cellcongruence} is also applied to $[\Delta_1]$, by summing each subset of columns corresponding to each class of \emph{congruent edges}, so generating a new Julia sparse matrix from the resulting set of columns. Then, we reduce every subset of equal rows, if any, to a single row representative of congruent faces. 

\item
Finally, a higher-level Julia function \texttt{CSG.chaincongruence} maps the input data \texttt{W}, \texttt{Delta\_0}, \texttt{Delta\_1}, into a compact representation \texttt{V}, \texttt{EV}, \texttt{FE} of the chain complex $\texttt{V}:C_0\to\E^3$, $\delta_0:C_0\to C_1$, and $\delta_1:C_1\to C_2$
\end{enumerate}

\end{algorithm}

\begin{property}[Topological robustness]
We would like to remark that decomposed 1-cells (i.e., input edges) and the representatives of their congruence classes (i.e., output edges) are associated one-to-one with the rows of $[\Delta_0]\to[\delta_0]$ and the columns of $[\Delta_1]\to[\delta_1]$, respectively, so satisfying the topological constraints $[\Delta_1][\Delta_0]=[\delta_1][\delta_0]=[0]$, that are checked as invariants in our tests (see Algorithm~\ref{alg:invariants}).
\end{property}

\begin{property}[Time complexity]
The algorithm that builds a balanced $k$d-tree to perform \emph{range-search} queries  has a worst-case complexity of $O(kn \log n)$. Step 1. of Algorithm~\ref{alg:ccc} requires a single range search query of a $k$d-tree, writh range size $\epsilon$, done in a single tree traversal, hence in $O(n)$, with tree construction in $O(dn\log n)$. Local centroid computations are done in expected constant time to output $m$ vertices, for a total time $O(n+m)$.
Each one of $d-1$ iterations of \texttt{CSG.cellcongruence} function, that performs the transformation $[\Delta_p]\to[\delta_p]$, must rewrite a smaller matrix from a bigger one, with scaling coefficients for non-zeros amount, on both rows and columns, going from 3 to 10 in average. It seems fair to estimate the average total time to $O($k$d-tree) + O(centroid) + O(out\_writing) = O(dn\log n+m+nnz)$, where $nnz$ (non-zeros) is equal to $k_1 m$ ($3\leq k_1\leq 10$), i.e., proportional to final  vertices.
\end{property}

\begin{algorithm}[Topological invariants] \label{alg:invariants}
Invariants are predicates (functions that return a Boolean value) that must be satisfied by current values of variables in specific points during the program evaluation process. In particular, topological invariants are evaluated dynamically during execution to catch common numerical errors. The TWG algorithm is particularly fragile with respect to topological errors of this type, so that few invariants are evaluated in various points of the pipeline. The more common invariant is the topological characteristic, or Euler number, both in 2D and in 3D. In fact we have $\texttt{V} - \texttt{E} + \texttt{F} = 2$ and $\texttt{V} - \texttt{E} + \texttt{F} - \texttt{C}=0$ on the $2$-sphere and the $3$-sphere, topologically equivalent to the plane and the space, respectively. Some invariants are therefore computed:
\begin{enumerate}
\item Before 2D splitting, for each input 2-face, we have a ``soup'' of line segments to mutually intersect, with $\texttt{E} \leq 2\texttt{V}$; 
\item After 2D splitting and TGW, for each simply connected 2-face component, it is necessarily $\texttt{V} - \texttt{E} + \texttt{F} = 2$.
\item Before chain complex congruence, when building $[\Delta_0]$ and $[\Delta_1]$ sparse container matrices, $[\Delta_1][\Delta_0]=0$
\item After CCC we must have, for the quotient topology, that $[\delta_1][\delta_0]=0$ holds, with $\ell=\texttt{V}$ columns of $[\delta_0]$;
\item Before TGW in 3D, with input $[\partial_2]_{m,n} = [\delta_1]^\top$ we have $m=\texttt{E}, n=\texttt{F}$, and $\ell-m+n=p$;
\item After TGW in 3D, the identity $\texttt{V} - \texttt{E} + \texttt{F} - \texttt{C}=0$ must hold, where $\texttt{C}=p$. 
\end{enumerate}
\end{algorithm}
\subsection{Topological Gift Wrapping in 3D}\label{sec:tgw}

The input in $\E^3$ is the sparse $[\partial_2]$ matrix; the output is the sparse $[\partial_3^+]$ matrix of the operator $\partial_3^+: Z_3\to Z_2$ from irreducible 3-cycles to 2-cycles (see Algorithm~\ref{alg:TGW3D}). 

\begin{property}[Complexity of Topological Gift Wrapping]
The time complexity of TGW algorithm~\ref{alg:TGW3D} is the one necessary to write down the cycle matrix $[\partial_3^+]$, i.e., to compute its $nnz$ (non-zero) terms. Looking to detailed pseudocode given in~\citep{TSAS:19}, one can see that: if $n$  is the number of $d$-cells and $m$ is the number of $(d-1)$-cells, the time complexity of this algorithm is $O(n m\log m)$ in the worst case of unbounded complexity of $d$-cells, and roughly $O(n k\log k)$ if their $(d-1)$-cycle complexity is bounded by a constant $k$.

\end{property}

\begin{example} \label{fig:3D}
Once again, we suggest the reader to refer to~\cite{arXiv:2017} for a full discussion of this multidimensional algorithm, summarized here in Section~\ref{sec:2D-TGW}. In the following we show a cartoon display of the 2-cycle boundary of an irreducible unit 3-chain in $U_3\subset C_3$.\vspace{2mm}

\noindent\includegraphics[height=0.15\textwidth,width=0.1428\textwidth]{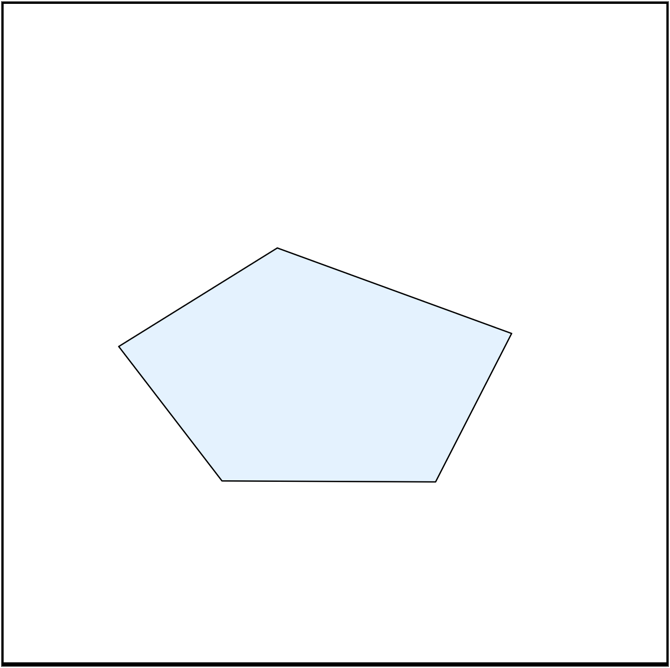}%
\includegraphics[height=0.15\textwidth,width=0.1428\textwidth]{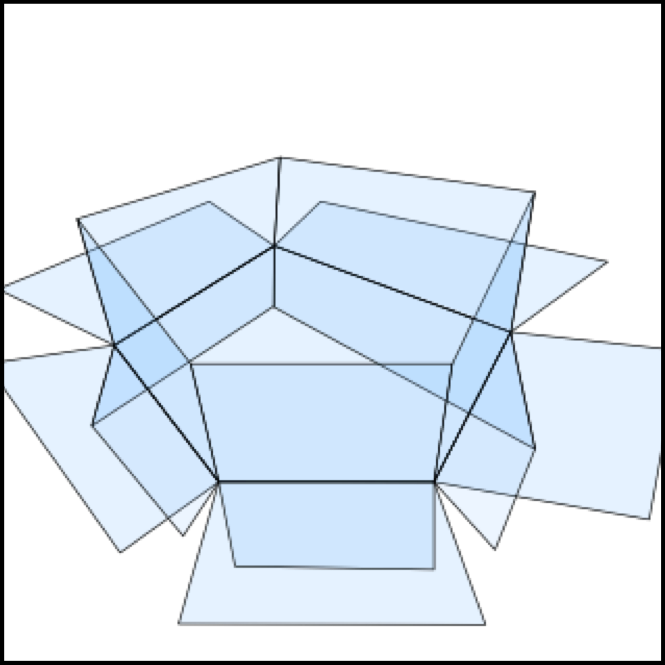}%
\includegraphics[height=0.15\textwidth,width=0.1428\textwidth]{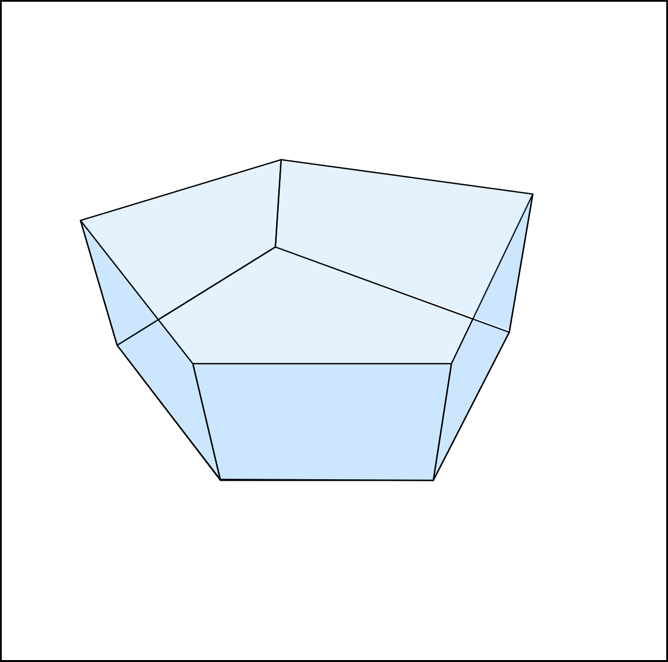}%
\includegraphics[height=0.15\textwidth,width=0.1428\textwidth]{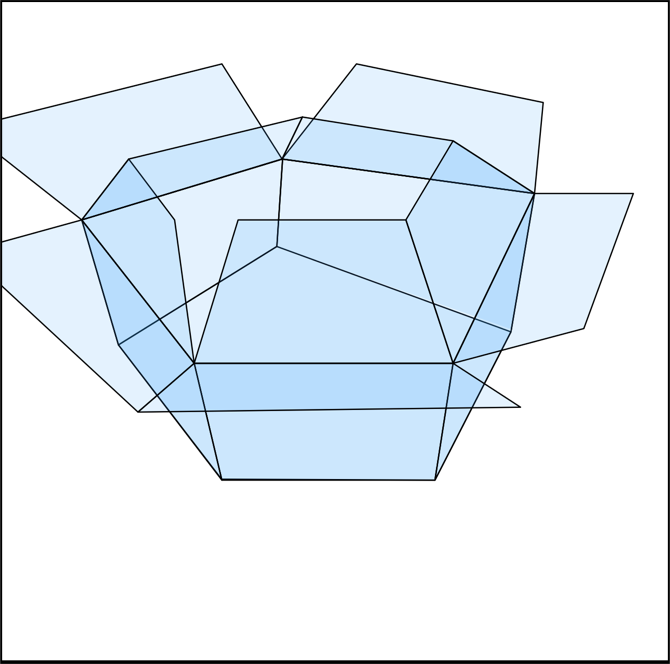}%
\includegraphics[height=0.15\textwidth,width=0.1428\textwidth]{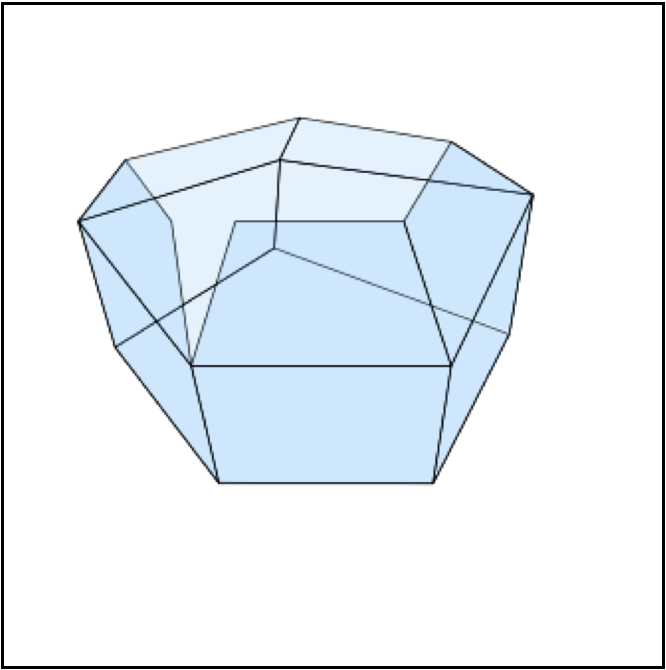}%
\includegraphics[height=0.15\textwidth,width=0.1428\textwidth]{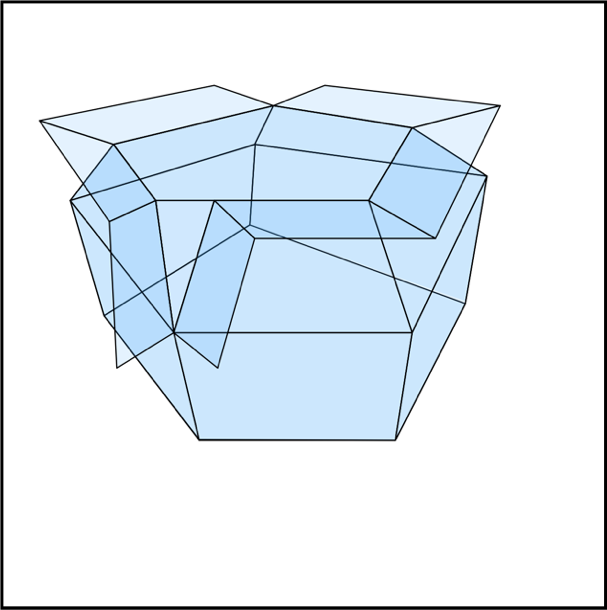}%
\includegraphics[height=0.15\textwidth,width=0.1428\textwidth]{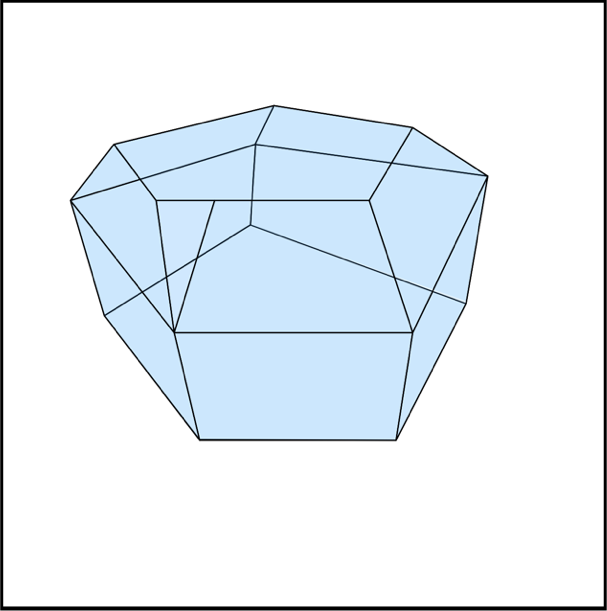}

{\noindent\footnotesize\hspace{.06\textwidth}(a)\hfill(b)\hfill(c)\hfill(d)\hfill(e)\hfill(f)\hfill(g)\hspace{.06\textwidth}}

\noindent
{\small Extraction of a minimal 2-cycle from $\mathcal{A}(X_2)$: (a) initial (0-th) value for $c\in C_2$; (b) cyclic subgroups on  $\delta\partial c$; (c)~1-st value of $c$; (d)~cyclic subgroups on $\delta\partial c$; (e) 2-nd value of $c$; (f) cyclic subgroups on $\delta\partial c$; (g) 3-rd value of $c$, such that $\partial c=0$, hence stop.}
 \end{example}

\begin{remark}[Stem $d-2$-cells]
Let us remark, with respect to Examples~\ref{fig:2D-full} and~\ref{fig:3D} and to~Algorithm~\ref{sec:2D-TGW}, that the ``stem'' cells are all the unit $d-2$-cells in the \emph{partial} $\partial c$, with $d\in C_{d-1}$. Of course, such stem cells are always two when $d=2$, and at least three when $d=3$. 
\end{remark}

\subsection{Cycles and Boundaries}\label{sec:cycles}

\begin{definition}[] A \emph{$p$-cycle} is defined as a 
$p$-chain without a boundary, hence it is an element of the kernel $Z_p$ of $\partial_p$. (The red sets in Figure below.) \vspace{2mm}

\begin{minipage}{0.30\textwidth}\flushleft
   {\small  A \emph{$p$-boundary} is a $p$-chain which is the boundary of a ($p$+1)-chain, so it is an element of the image $B_p$ of $\partial_{p+1}$. The set $B_p$ is a subset of the kernel of $\partial_p$, as~the boundary of a boundary is either~empty, or $\partial_{p}\partial_{p+1} = 0$.}
\end{minipage}
\begin{minipage}{0.64\textwidth}
   \flushleft
\includegraphics[width=\linewidth]{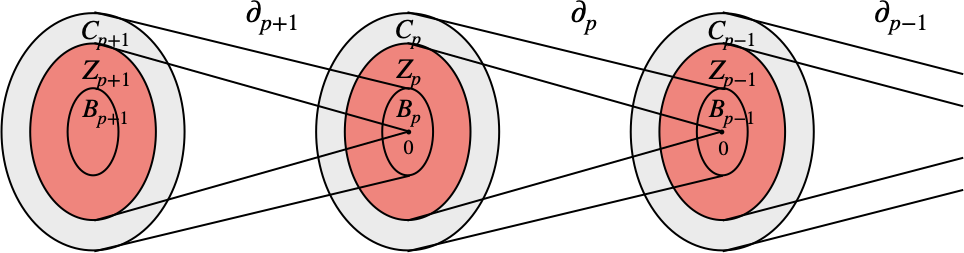}%
\end{minipage}
\label{fig:example} 
\end{definition}

\begin{property}[Columns of $\partial_3$ matrix are 2-cycles]
As we have seen in the previous section, the TGW algorithm in 3D produces the sparse matrix $[\partial_3]$, starting from the sparse matrix $[\partial_2]$ of the 2-skeleton of the $\E^3$ partition induced by the input data, i.e.~by a collection $\mathcal{S}_{d-1}$ of cellular complexes. Every column of $[\partial_3]$ matrix, say the $k$-th column $[u^k]$, \emph{is a 2-cycle} by construction, since its TGW building as a 2-chain stops when  $[\partial_2] [u^k] = [0]$ , i.e., when its boundary is empty.
\end{property}

\begin{property}[Rows of $\partial_3$ matrix sum to zero]
Each row of $\partial_3$ matrix corresponds to an irreducible element in the basis  $U_2\subset C_2$.
By construction in the TGW algorithm, each basis element $u\in U_2$ is used exactly  twice in 3-cell boundary building, with opposite coefficients $+1$ and $-1$, so proving the assertion. In other words, the set $Z_2$ of 2-cycles produced by TGW in 3D is not linearly independent. In particular, each one of them is generated by the topological sum of the others.
\end{property}

\begin{remark}[Cycles are non-intersecting]
Two remarks are very important for Algorithm~\ref{alg:TGW3D}, concerning transformations of cycle chains to boundary chains. The first is that, by construction, the 2-cycles corresponding to $[\partial_3]$ columns are (a) elementary (irreducible) and (b) non intersecting; the second one concern their numbers. It is well known that $Z_p \supseteq B_p$ or, in words, there may be cycles which are not boundaries. 
\end{remark}

\begin{example}[Boundary of concentric spheres]
Consider the 3D space partition generated by two 2-spheres $S_1$ and $S_2$ with the same center and different radiuses $r_1>r_2$. There are three solid cells:  (a) the outer cell, i.e., $\E^3$ minus the ball of radius $r_1$; (b) the solid intermediate ball with spherical hole inside, and thickness $r_1-r_2$; and (c) the solid inner ball of radius $r_2$. There are four closed irreducible 2-chains (cycles) generated as columns of $[\partial_3]$ by TGW in this complex, pairwise summing to zero and with opposite orientations. Let us denote orderly, from exterior to interior, as 
\[
[\partial_3]=[u_1\ u_2\ u_3\ u_4].
\]
If we denote the three ``solid'' basis elements in $U_3\subset C_3$ (3-chains) as $A, B, C$ and the whole space as $X$, we can express them as Boolean algebra expression:
\[
A = X-B-C;\quad  B=X-A-C;\quad C = X-A-B
\]
In terms of oriented 2-chains it is easy to see that 
\[
\partial A = u_1;\quad  \partial B = u_2 + u_3;\quad \partial C = u_4, \quad \mbox{with} \quad 
u_1 + u_2 = u_3 + u_4 = 0, \quad \mbox{and} \quad u_1 + u_2 + u_3 + u_4 = 0.
\]
Finally, let us notice that the cycles $u_2$ and $u_3$ are not boundary of any 3-chain. In more formal writing, we have: $u_2, u_3 \in (Z_2 - B_2) \subset C_2$.
\end{example}

\begin{example}[Cycles $\to$ boundaries] The assembly  \texttt{Lar.Struct([ tube, L.r(pi/2,0,0), tube, L.r(0,pi/2,0), tube ])}
of three instances of \texttt{cylinder()} with $n=16$ sides, default radius of length 1, height $h=2$, and $k=2$ decompositions in the axial direction are given below. The exploded unit 2-chains of the space arrangement, and the atoms corresponding to the 20 columns of $[\partial_3^+]$ and the 19 ones of $[\partial_3]$ are shown. In the center the outer boundary. Consider numbers of faces to split  ($288$ vs $54$) with two different data structures: boundary triangulations and LAR.

\vspace{4mm}

\begin{minipage}{0.3\textwidth}
   \centering
   \includegraphics[height=0.9\linewidth,width=\linewidth]{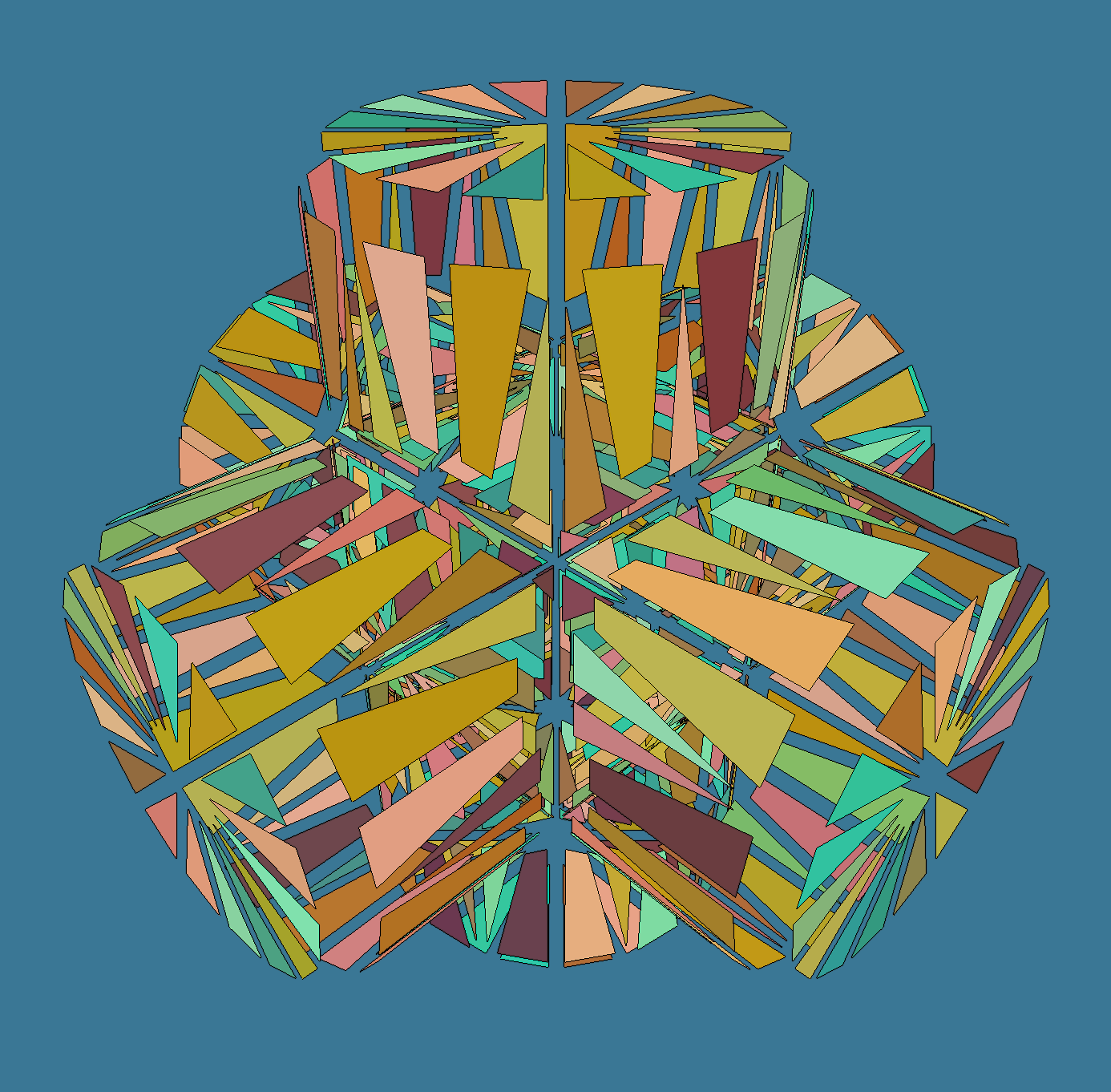} 
   \label{fig:ex2}
\end{minipage}
\begin{minipage}{0.3\textwidth}
   \centering
   \includegraphics[height=0.9\linewidth,width=\linewidth]{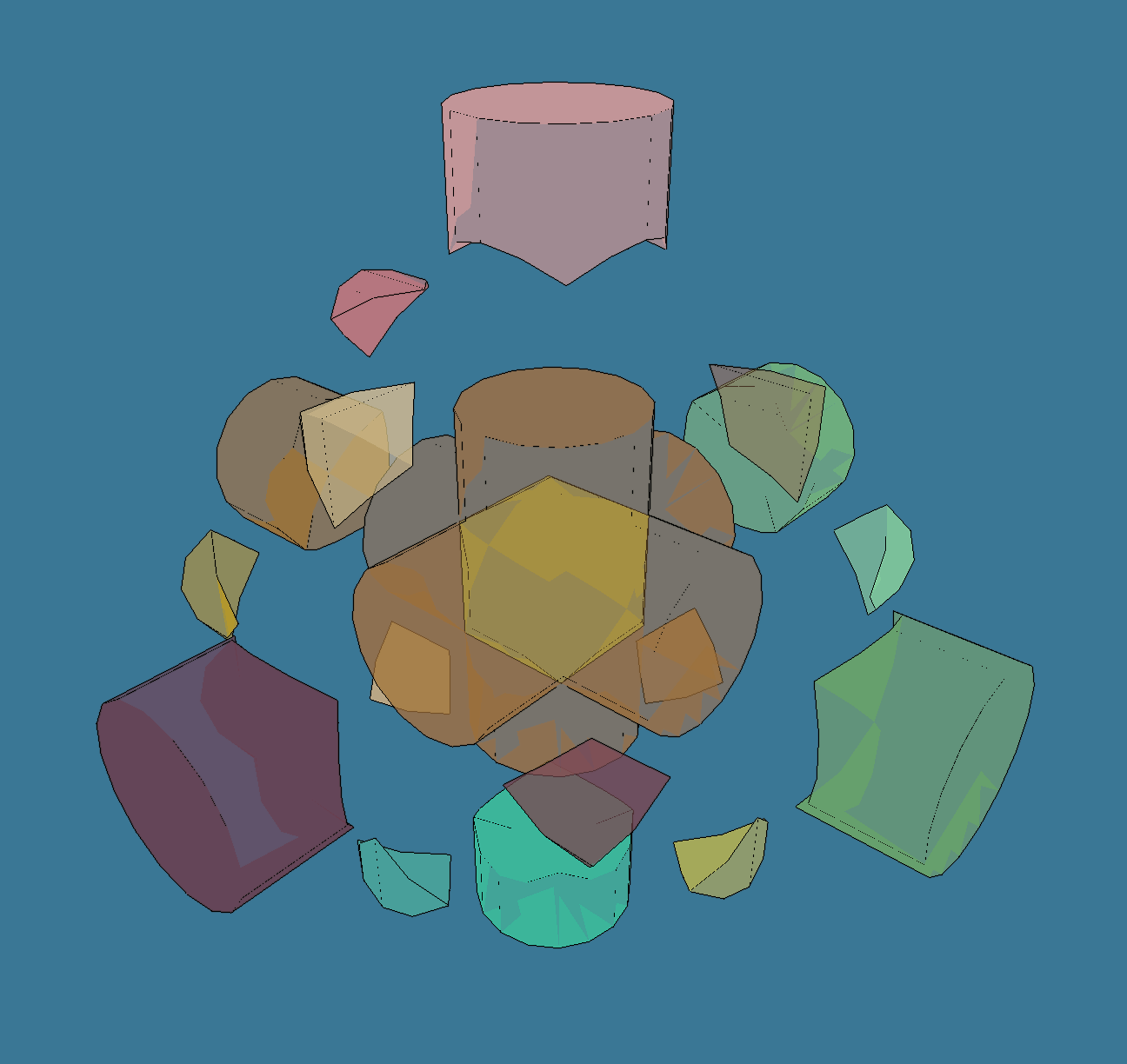} 
   \label{fig:ex2}
\end{minipage}
\begin{minipage}{0.3\textwidth}
   \centering
   \includegraphics[height=0.9\linewidth,width=\linewidth]{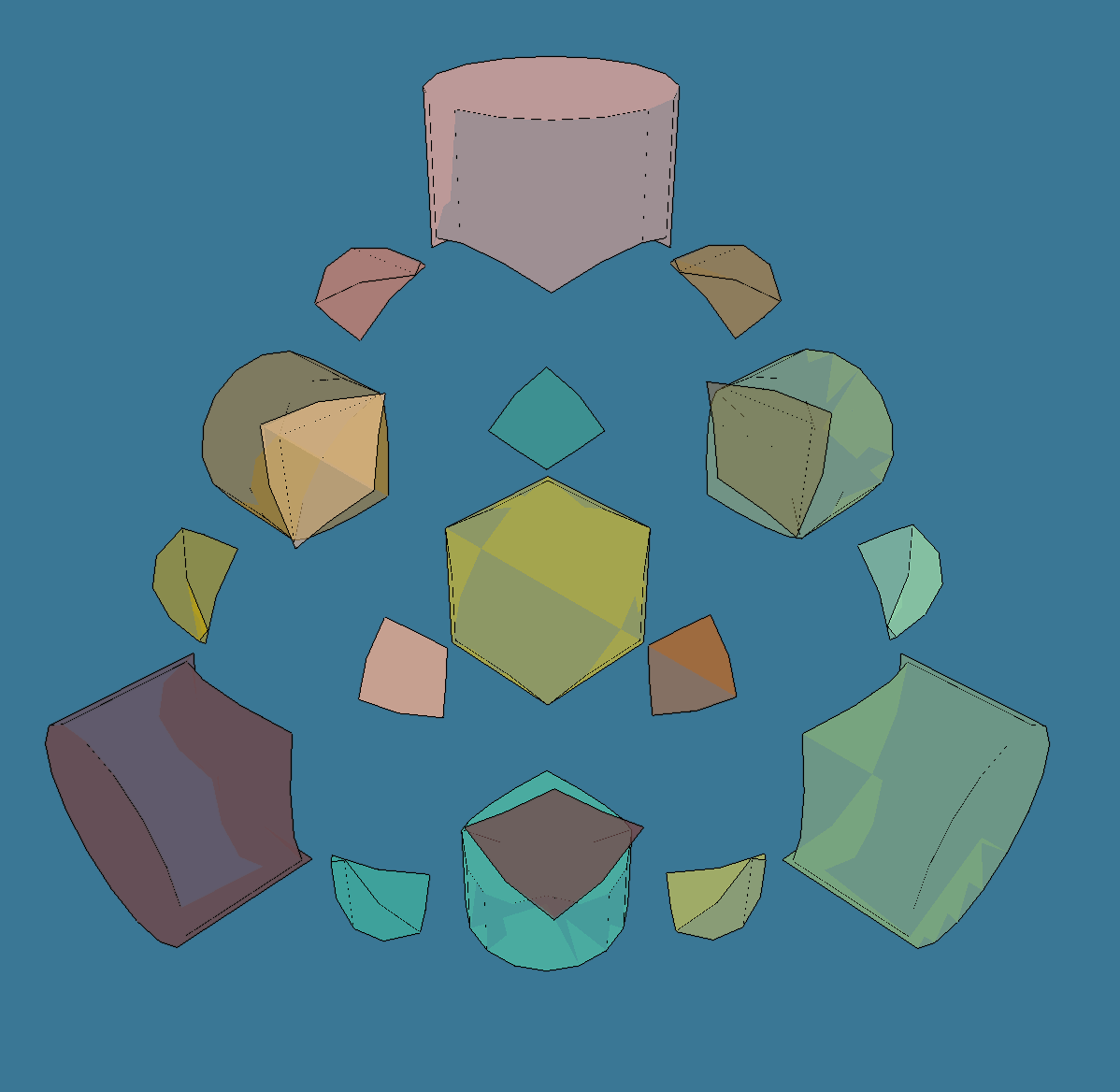} 
   \label{fig:ex3}
\end{minipage}

\begin{minipage}{0.3\textwidth}
   \centering
   \includegraphics[height=0.9\linewidth,width=\linewidth]{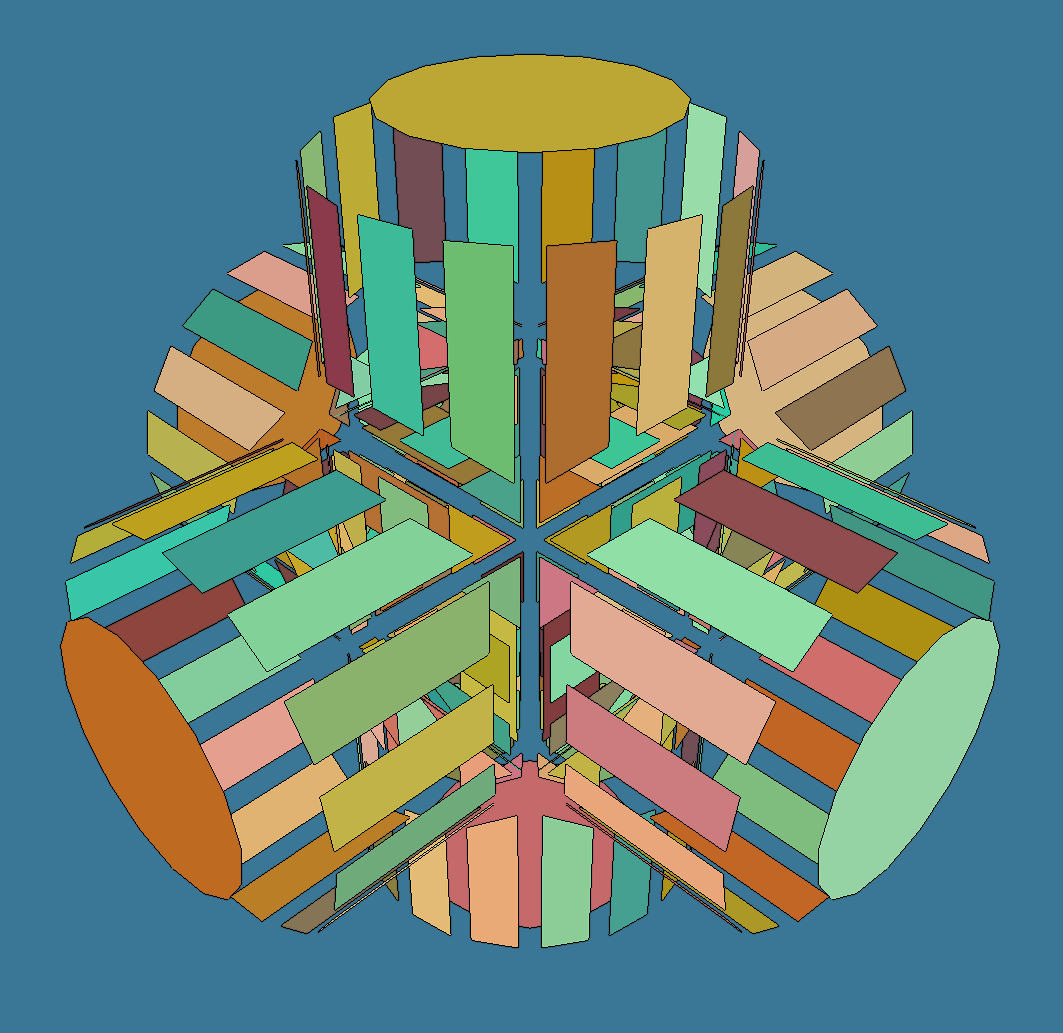} 
\end{minipage}
\begin{minipage}{0.3\textwidth}
   \centering
   \includegraphics[height=0.9\linewidth,width=\linewidth]{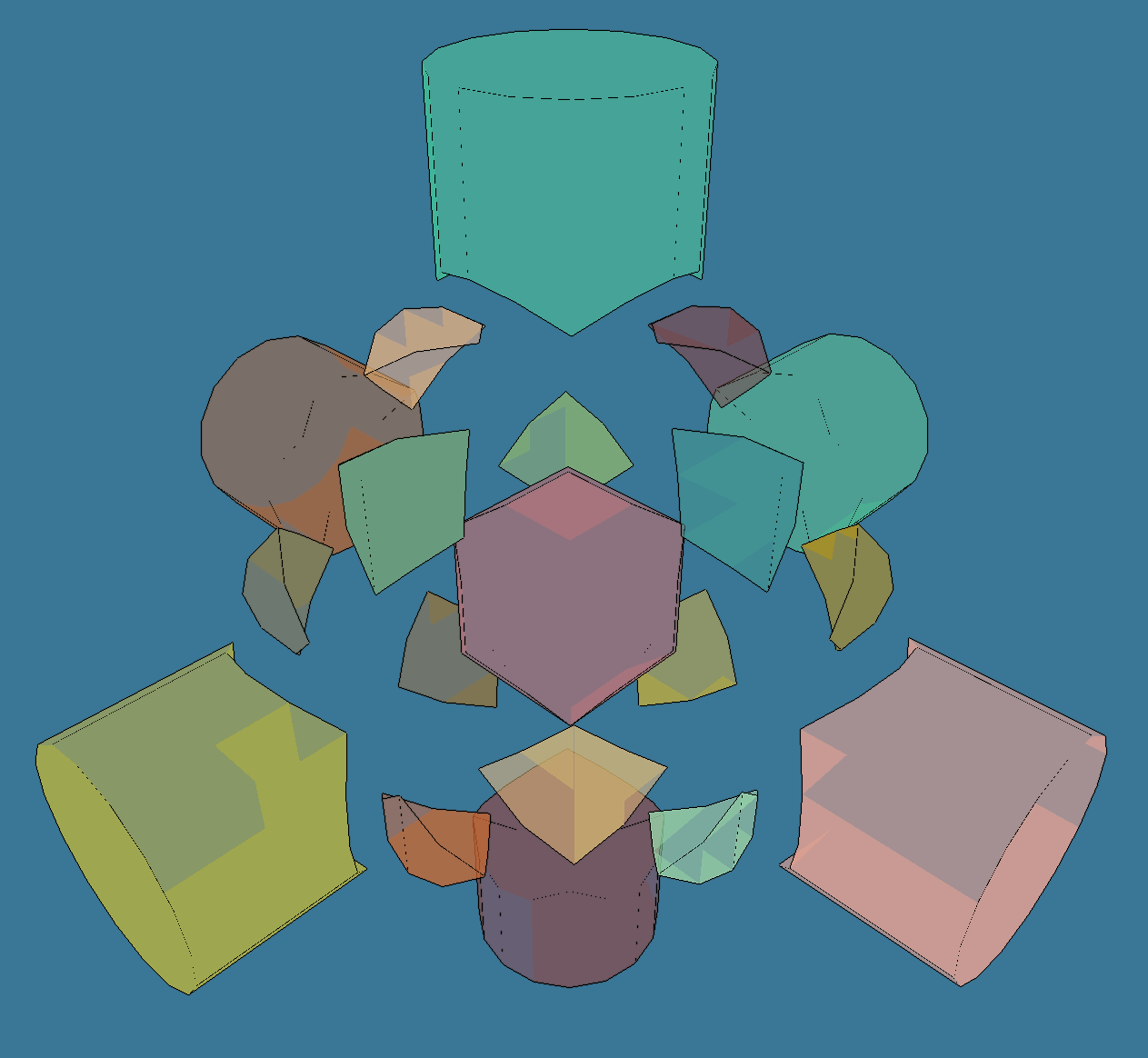} 
\end{minipage}
\hfill
\begin{minipage}{0.35\textwidth} \flushright
{\small \noindent
   Same 3-chain basis (with larger 2-faces), same atoms, and same assembly of three cylinders, but represented without boundary triangulations and with  more general LAR faces, i.e., with quads (quadrilaterals) on the lateral surface of the cylinders, and with two polygonal approximation of circle at the extremes. Compare the benefits in space and time  by number of faces in the basis: {\small $(16+2)\times 3 = 54 < 288 =(16\times 4 + 2 \times 16) \times 3$}, with smaller number of rows in $[\partial_3]$. }
\end{minipage} 
\end{example}

\begin{note}[Non intersecting shells]
In solid modeling, the word \emph{shell} is used to denote the connected boundary surfaces of a solid object~\citep{Paoluzzi:1989:BAO:70248.70249}. Here a shell come to be one of 2-cycles of the boundary of a maximally connected component of a 3-chain, or simply: one of boundary cycles of a maximal 3-component. The whole set of shells, including those internal to some unit 3-chain (atom), is obtained as 2-chain by matrix multiplication of $[\partial_3]$ times $[\textbf{1}]$, i.e., times the coordinate representation of the (whole) solid as a 3-chain.   
\end{note}

\begin{remark}[Correctness tests]
The number $N$ of columns of the $[\partial_3^+]$ matrix provides the dimension of the boundary space $B_3$, i.e., the number of irreducible elements of set $U_3$.
The boundary 2-cycle of the outer cell is obtained  by multiplication of $[\partial_3]$ times a column vector $[\mathbf{1}]$ with $N$ ones, equal to the sum of outer 2-cycles of component complexes. Of course, it is $[\partial_2][\partial_3][\mathbf{1}] = [\mathbf{0}]$, where the number of zeros equates the number of 1-cells of the complex.
\end{remark}

\begin{algorithm}[From cycles to boundaries] \label{alg:TGW3D}
Here we discuss how to reduce the matrix $[\partial_3^+]$ of irreducible 2-cycles, to the matrix $[\partial_3]$ of \emph{boundaries} of irreducible 3-chains (atoms) of the $\E^3$ arrangement induced by the input. 

By construction, each irreducible unit 3-chain $u\in U_3\subset C_3$ corresponds to a single connected PL 3-cell, possibly non-convex and non necessarily  path-contractible to a point.
In other words, the boundary of a 3-chain may possibly be non-connected, and made by one or more 2-cycles.

\begin{enumerate}
\item
Search, for each connected component of the 2-complex generated by congruence, the $[\partial_3^+]$  outer column;

\begin{enumerate}
\item
Repeat the  search and elimination for each matrix $[\partial_3^+]_k$ of a component. 

\begin{enumerate}
\item
In each $[\partial_3^+]_k$ look for the cycle which contains the highest number of non-zeros (i.e., the highest number of 2-cells), 

\item
Before elimination, check that the involved subset of vertices contains the extreme values (max and min) for each coordinate.

\item
If true, remove this cycle from the component matrix. In the very unlikely opposite case, the second longest cycle will be candidate, and so on. 
\end{enumerate}
\end{enumerate}

\item
All component matrices $[\partial_3]_k$ without the local outer cycle ($1\leq k\leq n$), are composed together columnwise is a single sparse block matrix 
$
[\partial_3] = [ [\partial_3]_1 \cdots [\partial_3]_k \cdots [\partial_3]_n ].
$
\end{enumerate}
\end{algorithm}

\subsection{Solid Algebra}\label{sec:algebra}

The second step of our approach to constructive solid geometry consists in generating a
representation of solid arguments as linear combinations of independent
3-chains, {i.e.,} of 3-cells of the space partition generated by the input. The
$U_3$ basis of 3-chains is represented by the columns of the
$\partial_3$ matrix. 
This set of 2-cycles can be seen as a collection of
point-sets, including the whole space and the empty set. In this sense, it generates both a
discrete topology of $\E^3$ and, via the Stone Representation
Theorem~\citep{key1501865m}, a finite Boolean algebra $\mathcal{B}$ over $C_3$ elements.

\begin{remark}
This second stage of the evaluation process of a functional form including solid
models and set operators is much simpler then the first stage, consisting
in executing a series of set-membership-classification (SMC) tests,  
computable in parallel, to build a representation of the input terms into the 
set algebra $\mathcal{B}$, using sparse arrays of bits.
\end{remark}

\subsubsection{Algebra of sets} 

Any finite collection of sets closed under
set-theoretic operations forms a Boolean algebra, {i.e.}~a complemented
distributive lattice, the \emph{join} and \emph{meet} operators being union
and intersection, and the \emph{complement} operator being set complement.
The bottom element is $\varnothing$  and the top element is the universal set
under consideration. By~\cite{key1501865m}, every Boolean algebra of $N$ atoms---and hence the
Boolean algebra of solid objects closed under regularized union and
intersection---is isomorphic to the Boolean algebra of sets over $\{ 0,1 \}^N$. 
Here, we look for the binary representation of solid objects terms 
in a solid geometry formula, {i.e.},~for
the coefficients of their components in basis $U_d \subset C_d$. 

\begin{note}[Julia representation]
If all chains in $U_d$ are equioriented, then such coefficients are simply drawn from
$\{ 0,1 \}$, and every coordinate vector for $C_d$ elements is a binary
sequence in $\{ 0,1 \}^N$, with $N={\#\,U_d}$. 
In Julia we implement this  
representation of algebraic terms using arrays of \texttt{BitArray} type, or sparse
arrays of type \texttt{Int8}, consuming few bytes per non-zero element.
\end{note}
\subsubsection{Boolean Atoms}\label{sec:atoms}


\begin{property}[Boolean atoms are unit 3-chains]
There is a natural transformation between $d$-chains defined on a space arrangement and the algebra generated by that arrangement. Unit $d$-chains correspond to atoms of the algebra; the $[c]$ coordinate representation (bit array) of any $d$-chain $c$ generates the  coordinate representation in boundary space $[\partial_d][c] = [b] \in B_{d-1} \subset C_{d-1}$.
\end{property}

\begin{conjecture}[Higher dimensional CSG]
The main result introduced by this paper is the representation of atoms of the CSG Boolean algebra generated by a partition of $\E^3$ space, with the basis $U_2$ of space $C_3$ represented by the columns of the matrix $[\partial_3]$ as possibly non-connected 2-cycles of faces (2-cells) of $\E^3 arrangement$. The same holds, of course with scaled indices, for two-dimensional CSG algebra, and probably should hold in higher dimensions.
\end{conjecture}

\subsubsection{Generate-and-test algorithm}\label{generate-and-test}

The representation of join-irreducible elements of $\mathcal{L}(\mathcal{S}) \cong \mathcal{A}(\mathcal{S}_{d-1})$ as discrete point-sets (see Section~\ref{sec:algebra}), is used to map the \emph{structure} of each term $X\in\mathcal{L}(\mathcal{S})$ to the \emph{set algebra} $\mathcal{B}\cong\mathcal{A}(\mathcal{S})$. Assume that:
(a)~$X\in\mathcal{L}(\mathcal{S})$;
(b)~a~partition of $\E^d$ into join-irreducible subsets $A_k\in \mathcal{L}(\mathcal{S})$ is known; 
(c)~a one-to-one mapping $A_k \mapsto p_k$ between atoms and internal points is given; 
(d)~a SMC oracle, {i.e.,}~a set-membership classification ~\citep{Tilove:80} 
test is available.

A naive approach to SMC, where each single point $p_k$ (in the interior of an atom $A_k\in\mathcal{L}(\mathcal{H})$) is tested  against \emph{all} input solid terms $X\in\mathcal{H}(\mathcal{S})$ may be computed in quadratic time $O(NM)$, where $N$ is the number of atoms, and $M$ is the number of input solid terms $X$. An efficient $O(N\log M)$ procedure is established here by using two ({i.e.}, $d-1$) one-dimensional interval-trees  for the decomposition $\{ A_k\}$ of $\E^3$, in order to execute the SMC test only against the terms in the subset $\mathcal{I}(p_k) \subseteq U_3$, whose containment boxes intersect a ray from the test point.

\begin{algorithm}[Generate-and-test] 
Therefore, the structure of term $X$ in the finite algebra $\mathcal{L}(\mathcal{S})$ can be computed using the \emph{generate-and-test} procedure. Such a SMC test is simple and does not involve the resolution of ``on-on ambiguities''~\cite{Tilove:80}, because of the choice of an \emph{internal} point $p_k$ in each algebra atom $A_k$.
{Set Membership Classification (SMC) via a point-polyhedron-containment test.}
\begin{algorithmic}[1]
\Procedure{Generate.And.Test}{$p_k$, $\mathcal{I}(p_k)$}{$(X)$} \Comment{Set Membership Computations}
	\State structure$(X)$ := $\emptyset$; $\quad[X] := [0]$   \Comment{Initialization of term $X$ in algebra of sets $\mathcal{A}(\mathcal{S}) \cong \mathcal{B}$}
    \ForAll{join-irreducible $A_i \in \mathcal{I}(p_k)$}
        \If{$\texttt{SetMembershipClassification}(p_k, A_i)$}  \Comment{if and only if $p_k \in \mathbf{i}A_i$}
            \State structure$(X)$ := structure$(X) \cup A_i$; $\quad[X][i] := 1$  \Comment{put 1 on $i$-th element of array $[X]$}
        \EndIf
    \EndFor
\EndProcedure  \Comment{Return the binary array $[X]$, coordinate vector for $X\in C_d$}
\end{algorithmic}
In our current implementation in 3D the SMC test is executed by intersecting a ray from $p_k$ with the planes containing the 2-cells of atoms in $\mathcal{I}(p_k)$, and testing for point-polygon-containment in these planes (via maps to the $z=0$ subspace). 
In~summary, we decompose $\E^d$ into join-irreducible elements $A_k$ of algebra $\mathcal{L}(\mathcal{S})$, and represent each $A_k$ with a point~$p_k$. 
By the Jordan curve theorem, an odd intersection number of the ray for $p_k$ with boundary 2-cells of $X$ produces an oracle answer about the query statement $p_k\in X$, and hence $A_k\subseteq X$. An even number gives the converse.
\end{algorithm}

\subsubsection{Binary representation of Boolean terms}

We contruct a representation of each term $X$ of a solid Boolean
expression, as a subset of $U_3$ (basis of 3-chains). Remember that, by construction, $U_3$ partitions both $\E^3$ and the input solid objects.

\begin{example}[Space arrangement from \texttt{assembly} tree]
Consider the \texttt{assembly} constructed by putting together three instances of
unit cube, suitably rotated and translated. The semantics of \texttt{Lar.Struct()} is 
similar to that of PHIGS structures~\citep{Kasper:1993:GPP:143741,Paoluzzi2003a}: The \texttt{Lar} constructor \texttt{cuboidGrid} of grids of cubes with ``shape''\footnote{
The \emph{shape} of a multidimensional array, in Julia, Python, and other computer languages, is a tuple or array with numbers of rows, columns, pages, etc. of data elements within the array. The \emph{length} of shape tells the array dimensions: 1=vector, 2=matrix, etc.
}
 $[m,n,p]$, returns (with \texttt{true} optional parameter) the whole collection of $p$-cells, $0\leq p\leq 3$ in arrays of arrays of vertex indices \texttt{VV,EV,FV,CV}.  Only \texttt{V,FV,EV} (vertices, faces, edges) are actually needed by the Boolean generation.

\begin{minipage}{0.6\textwidth}
\flushleft\footnotesize
\begin{lstlisting}[mathescape]
julia> m,n,p = 1,1,1;
    Lar = LinearAlgebraicRepresentation;
    V,(VV,EV,FV,CV) = Lar.cuboidGrid([m,n,p],true);
    cube = V,FV,EV;
julia> assembly = Lar.Struct([ cube,
    Lar.t(.3,.4,.25), Lar.r(pi/5,0,0), Lar.r(0,0,pi/12), cube, 
    Lar.t(-.2,.4,-.2), Lar.r(0,pi/5,0), Lar.r(0,pi/12,0), cube ]); 
julia> W, EV, FE, CF, boolmatrix = Lar.arrangement(assembly);
\end{lstlisting}
\end{minipage}
\begin{minipage}{0.36\textwidth}
	\centering
   \includegraphics[width=0.6\textwidth]{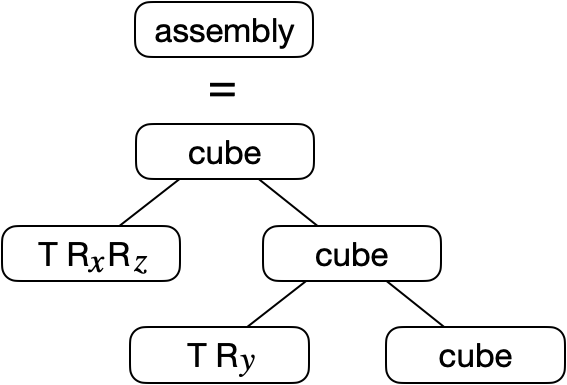} 
\end{minipage}
\vspace{1mm}

\noindent\normalsize \texttt{assembly} ${}={}$ \texttt{Lar.Struct()} is a linearization of DFS (Depth First Search) of the tree;
\texttt{Lar.arrangement()} function applied to \texttt{assembly}
returns the \texttt{(geometry,topology)} of 3D space
partition generated by it.  \texttt{Geometry} is given by
the embedding matrix \texttt{W} of all vertices (0-cells), and \texttt{topology} 
 by the  sparse matrices \texttt{CF}, \texttt{FE}, \texttt{EV},
{i.e.}, by~$\delta_2, \delta_1, \delta_0$, of  chain complex describing the   
$\mathcal{A}(\texttt{assembly})$ arrangement. See Definition~\ref{sec:geometriccomplex} (LAR Geometric Complex).

\label{ex:two} 
\end{example}

\begin{definition}[Bit vectors]
A subset $Y$ of $X$ can be identified with an indexed family of bits with index set $X$, with the bit indexed by $x \in X$ being 1 or 0 according to whether or not $x \in Y$.  
The Boolean algebra $\mathcal{B} = \mathscr{P}(X)$ of the power set of $X$ can be defined equivalently as the nonempty set of \emph{bit vectors}, all of the length $n = \# X$, with $\#\mathcal{B} = 2^n$.
\end{definition}

\begin{algorithm}[Transformation of 3-chain to bit-array]
To translate a solid CSG formula to machine language it is sufficient, once computed the $[\partial_3]$ matrix, and hence the $U_3$ basis,  for each unit 3-chain $u_k\in U_3$, to test for \emph{set-membership} a single \emph{internal} point $p_k\in u_k$, by
checking if $p_k \in X$. In the affirmative case the $k$-th bit of coordinate vector $[X]\in\{0,1\}^N \cong \mathscr{P}(U_3)$ is set to \texttt{true}.
\end{algorithm}

\begin{example}[Boolean matrix]\label{ex:abc}
\normalsize The array value of type \texttt{Bool} returned in the variable \texttt{boolmatrix} contains by column the results of efficient  
point-solid containment tests (SMC) for atomic 3-cells (rows), with respect to the terms of $\E^3$ partition: the outer 3-cell $\Omega$ and
each \texttt{C$_1$,C$_2$,C$_3$} \texttt{cube} instance (matrix columns), suitably mapped to world coordinates by \texttt{Struct} evaluation.

\begin{minipage}{0.6\textwidth}
\flushleft\footnotesize
\begin{lstlisting}[mathescape]
julia> Bmat = Matrix(boolmatrix) 
8x4 Array{Bool,2}:
 true  false  false  false 
false  false  false   true 
false   true   true  false
false   true   true   true 
false   true  false  false 
false  false   true  false
false   true  false   true 
false  false   true   true 
\end{lstlisting}
\end{minipage}
\end{example}

\begin{remark}[Boolean terms]
\normalsize Our objects \texttt{C$_1$,C$_2$,C$_3$} are extracted from \texttt{boolmatrix} columns into variables \texttt{A,B,C},
so describing how each one is partitioned by (ordered) 3-cells in $U_3$. The whole
space is given by $X = \Omega \cup \texttt{A} \cup \texttt{B} \cup \texttt{C}$.

\begin{minipage}{0.6\textwidth}
\footnotesize 
\begin{lstlisting}[mathescape]
julia> A,B,C = boolmatrix[:,2],boolmatrix[:,3],boolmatrix[:,4] 
(Bool[false, false, true, true, true, false, true, false], 
 Bool[false, false, true, true, false, true, false, true], 
 Bool[false, true, false, true, false, false, true, true]) 
\end{lstlisting}
\label{ex:twobis} 
\end{minipage}
\end{remark}

\subsubsection{Bitwise resolution of set algebra expressions}\normalsize

When the input initial solids have been mapped to arrays of Booleans, now called \texttt{A}, \texttt{B}, \texttt{C}, any expression of their finite Boolean algebra is evaluated by logical operators, that operate by comparing corresponding
bits of variables.  

\begin{definition}[Bitwise operators]
In particular, the Julia language offers bitwise
logical operators \emph{and} (\texttt{\&}), \emph{or} (\texttt{|}), 
\emph{xor} ($\texttt{$\veebar$}$), and
\emph{complement} (\texttt{!}), as well as a dot mechanism for applying elementwise
any function to arrays.  Hence, we can  write expressions like \texttt{(A .\& B)}
or \texttt{(A .| B)} that are bitwise evaluated and return the  result in a
new \texttt{BitArray} vector. We remark that these operators can be used also in
\emph{prefix} and \emph{variadic} form. Hence, bitwise operators can be applied 
at the same time to any finite number of variables.
\end{definition}

\begin{example}[Boolean formulas]
Some examples follow. The last expression is the intersection of the first
term \texttt{A} with complement of others terms \texttt{B} and \texttt{C}, with \texttt{A}, \texttt{B}, \texttt{C}
of Example~\ref{ex:abc}, giving the set difference $(\texttt{A}\!\setminus\!
\texttt{B})\!\setminus\! \texttt{C}$. 
The variable \texttt{AminBminC}  contains the result of the set difference denoted  \texttt{\&(A)}, $\overline{\texttt{B}}$, $\overline{\texttt{C}}$\texttt{)},  mapped into the model in Figure~\ref{fig:one-b}.

\begin{minipage}{0.6\textwidth}
\footnotesize 
\begin{lstlisting}[mathescape]
julia> AorB = A .| B; 
julia> AandB = A .& B;
julia> AxorB = A .? B; 
julia> AorBorC = A .| B .| C; 
julia> AorBorC = .|(A, B, C); 
julia> AandBandC = A .& B .& C; 
julia> AandBandC = .&(A, B, C);

julia> AminBminC = .&(A, .!B, .!C)
8-element BitArray{1}: 
[false, false, false, false, true, false, false, false] 
\end{lstlisting}
\end{minipage}
\end{example}

\begin{example}[Solid algebra 3D] \label{ex:rot-cubed}
Let us compute the arrangement of $\E^3$ produced by 8 concentric unit cubes randomly rotated about the origin. The input data set $\mathcal{S}$ is made by $6\times 8 = 48$ square 2-cells. The input configuration is close to the worst case $O(n^2)$ for Boolean operations, since every face is intersected by many other input faces.
 In the images below we see: (a) the fragmented faces after the 2-cell splitting; (b) the solid 3-cells assembled to give the solid union; (c) the explosed set of the atoms of the Boolean algebra associated to this arrangement; (d) one single (very complex) 3-cell evidenced. The central bigger atom is the intersection of all cubes, and is a 3-ball approximation.
 \vspace{2mm}

\begin{minipage}{0.24\textwidth}
\includegraphics[height=\linewidth,width=\linewidth]{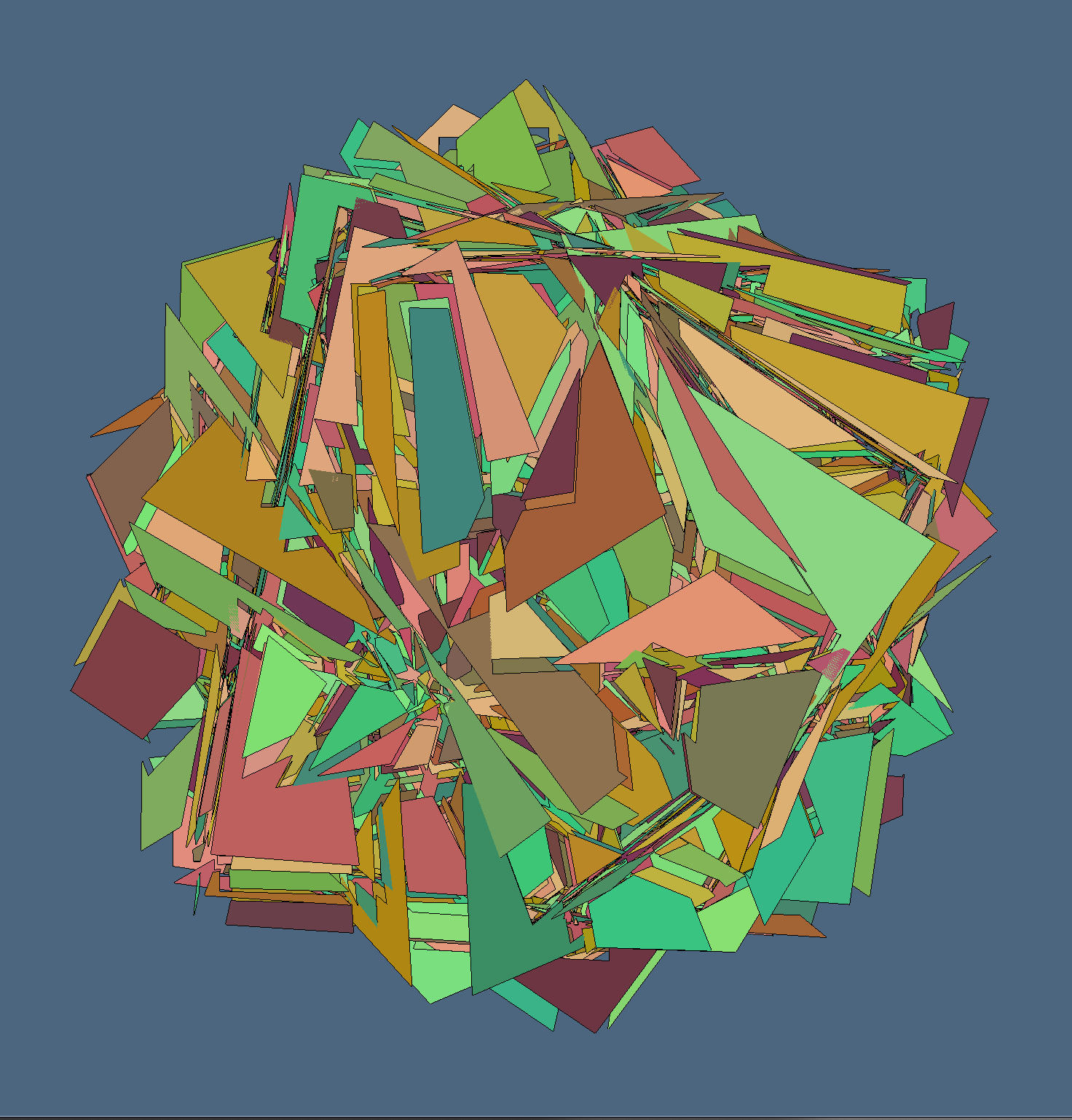}%
\end{minipage}
\begin{minipage}{0.24\textwidth}
\includegraphics[height=\linewidth,width=\linewidth]{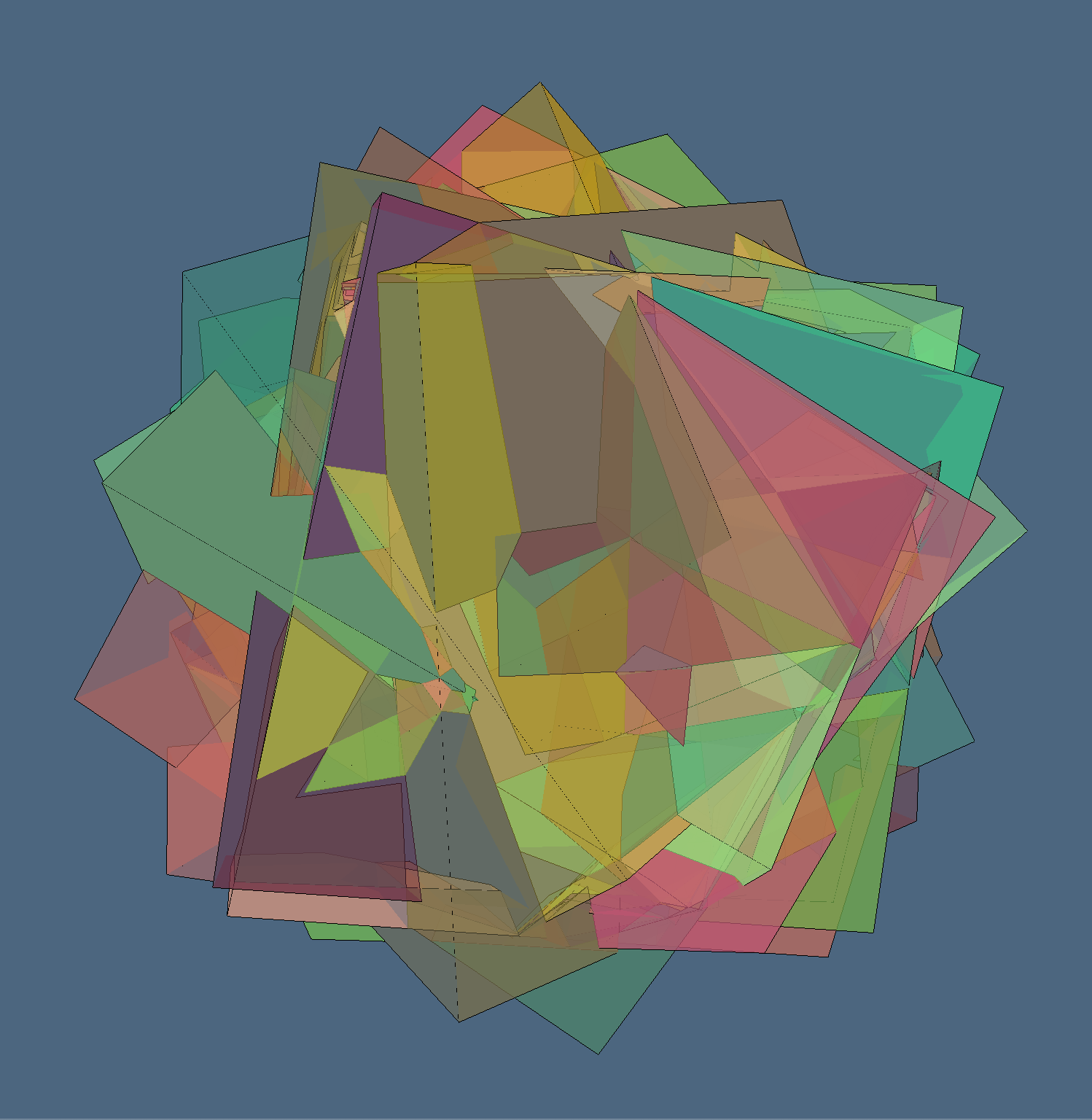}%
\end{minipage}
\begin{minipage}{0.24\textwidth}
\includegraphics[height=\linewidth,width=\linewidth]{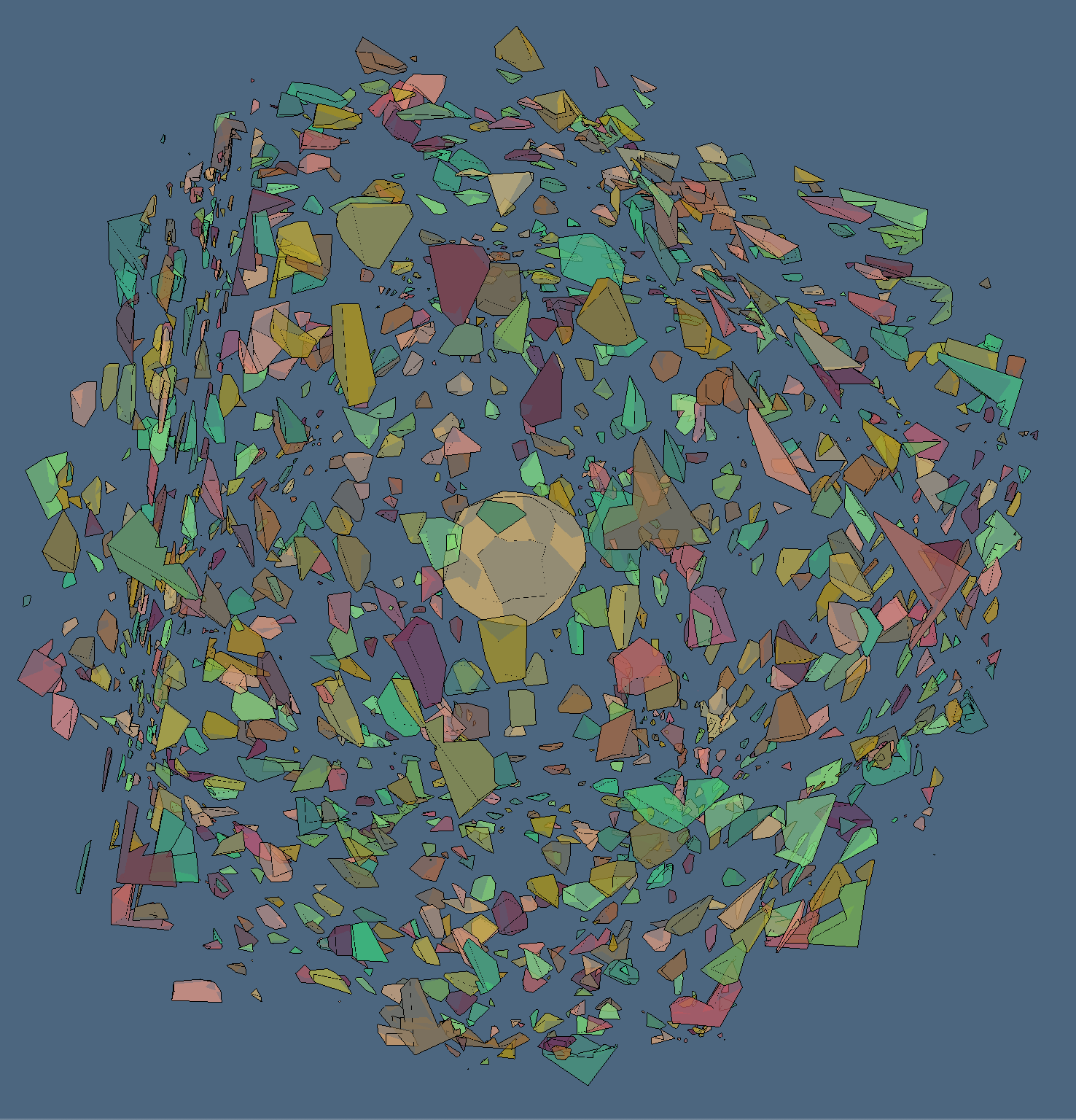}%
\end{minipage}
\begin{minipage}{0.24\textwidth}
\includegraphics[height=\linewidth,width=\linewidth]{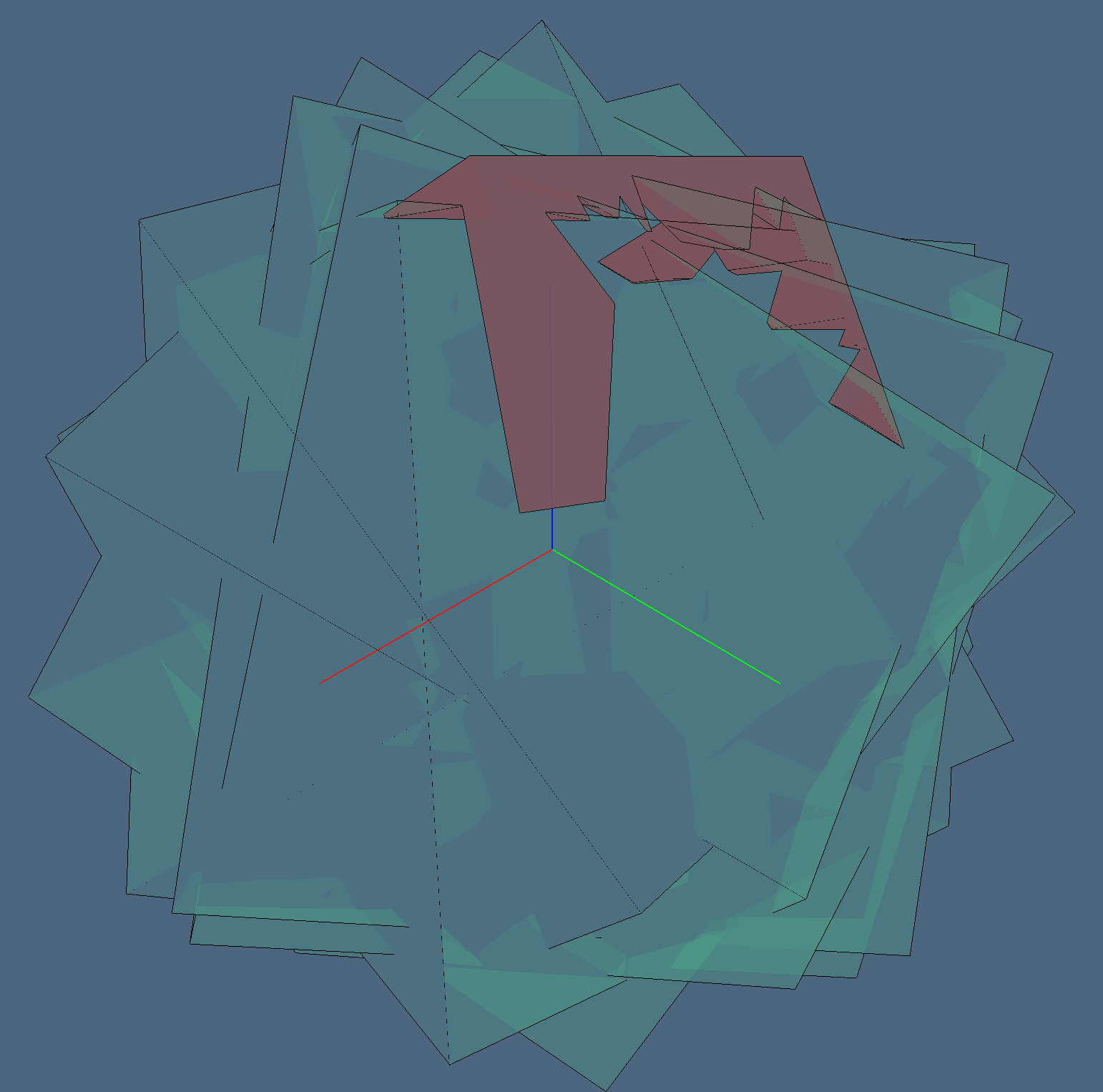}%
\end{minipage}

\end{example}

\begin{remark}[Topological robustness] 
The arrangement of $\E^3$ shown in Example~\ref{ex:rot-cubed} is a 3-complex with 2208 vertices, 5968 edges, 5360 faces, and 1600 solid cells. The Euler characteristic  $\chi$ is $\chi_0 - \chi_1 + \chi_2 - \chi_3 = 2208 - 5968 + 5360 - 1600 = 0$. This  count includes the outer (unbounded) 3-cell. We recall that Euclidean d-space is topologically equivalent to the d-sphere minus one point. the Euler characteristic of the $d$-sphere is $\chi = 1 + (-1)^d =$ 2 or 0, for either even or odd space dimension $d$.  It is worthwhile considering the complex shape of some 3-cells and 2-cells. These are handled by the LAR representation with the same simplicity than triangles or tetrahedra.
\end{remark}

\begin{example}[CSG example --- 2/2]\label{ex:forms}
\normalsize In this example we complete the computation of the CSG Boolean space induced by the space partition generated by the formula shown in Example~\ref{ex:explode}, and compute several other forms on the same Boolean algebra $\mathcal{B}$, by using the novel method proposed in this paper. 

Let us start by recalling that we have five input terms $\mathcal{S} = \{\texttt{X}_1,\texttt{X}_2,\texttt{X}_3,\texttt{Y},\texttt{Z}\}$, corresponding to three cylinder instances, a cube and a sphere.  The $\E^3$ arrangement they generate is $\mathcal{A}(S)$, shown below, together with several evaluated formulas of this Boolean algebra  (see Section~\ref{sec:DSL}). The value returned by a \texttt{@CSG} macro is of type \texttt{GeoComplex}, i.e., a geometric complex (see Definition~\ref{sec:geometriccomplex}).
It is worthwhile to remark that two resulting solids are stored as evaluated \texttt{GeoComplex} values within variables \texttt{A} and \texttt{B}, and that the symbols of such variables may be used in other \texttt{@CSG} macro expressions.
From top to bottom, and left to right, we display: \vspace{2mm}

{\small
\begin{minipage}{0.32\textwidth}\centering
	\texttt{@CSG (+,X$_1$,X$_2$,X$_3$,Y,Z)}; \\
	\texttt{A = @CSG (-,(*,Y,Z),X$_1$,X$_2$,X$_3$)};  \\
	\texttt{@CSG (+,A,X$_1$)}; 
\end{minipage}
\begin{minipage}{0.32\textwidth}\centering
	\texttt{@CSG (+,X$_1$,X$_2$)}; \\
	\texttt{@CSG (+,A,(*,B,Z))}; \\  
	\texttt{B = @CSG (+,X$_1$,X$_2$,X$_3$))}; 
\end{minipage}
\begin{minipage}{0.32\textwidth}\centering
	\texttt{@CSG X$_1$};  \\
	\texttt{@CSG (+,(*,Y,Z),A)};\\
	\texttt{@CSG (+ B,Y,Z)}
\end{minipage}}
\vspace{2mm}

\begin{minipage}{0.31\textwidth} 
   \includegraphics[height=\textwidth,width=\textwidth]{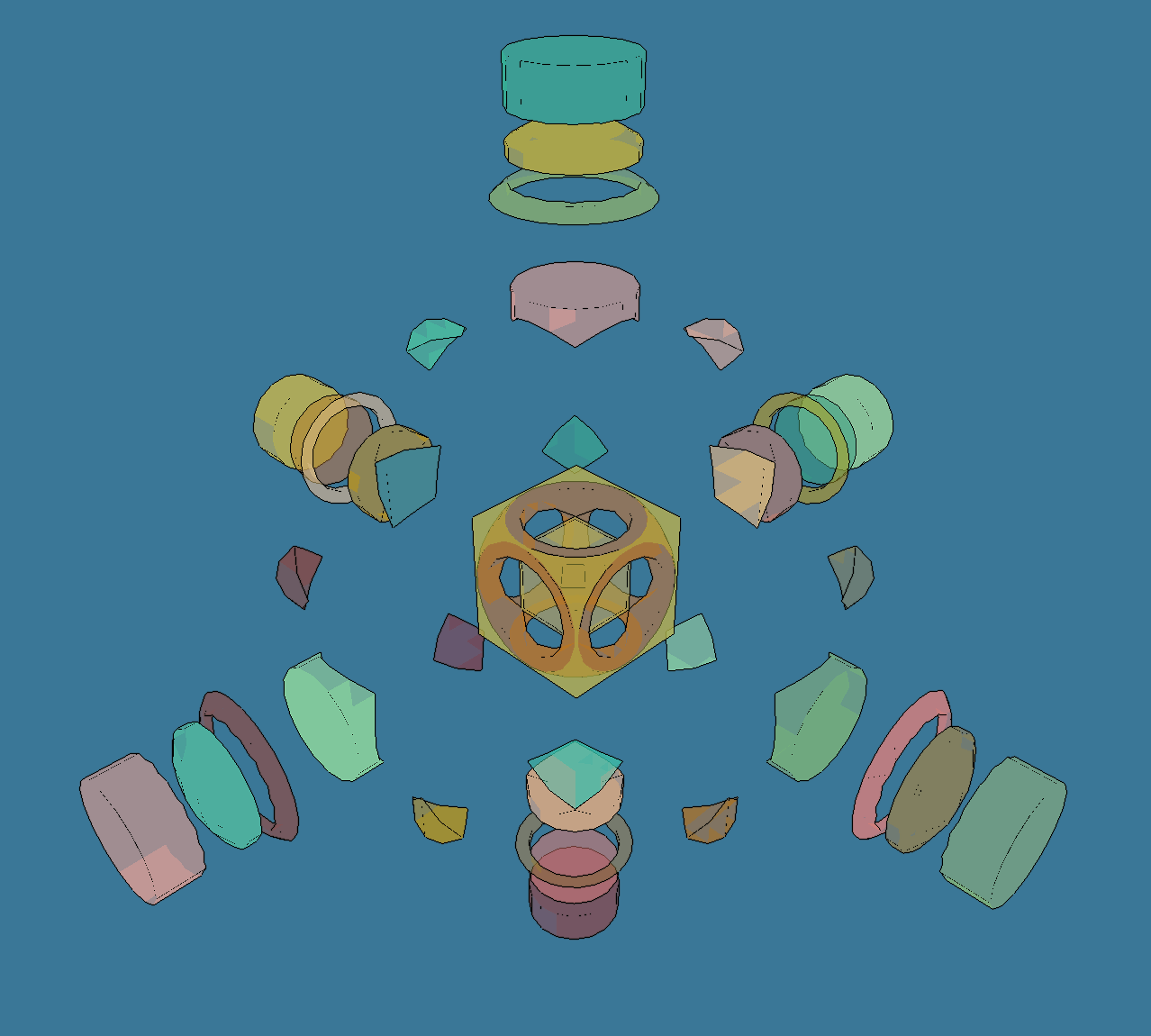} 
\end{minipage}
\begin{minipage}{0.31\textwidth}
   \includegraphics[height=\textwidth,width=\textwidth]{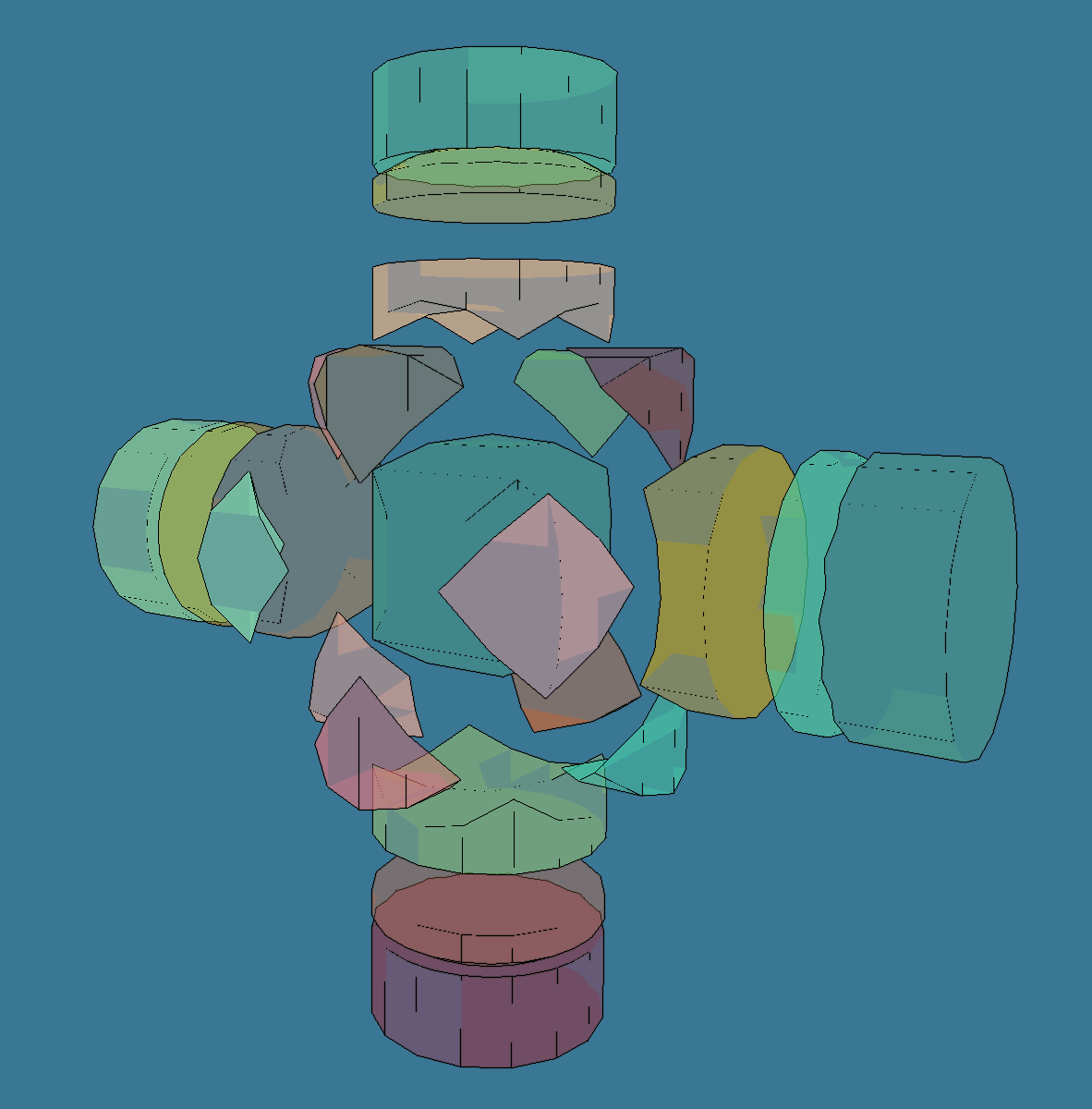} 
\end{minipage}
\begin{minipage}{0.31\textwidth}
   \includegraphics[height=\textwidth,width=\textwidth]{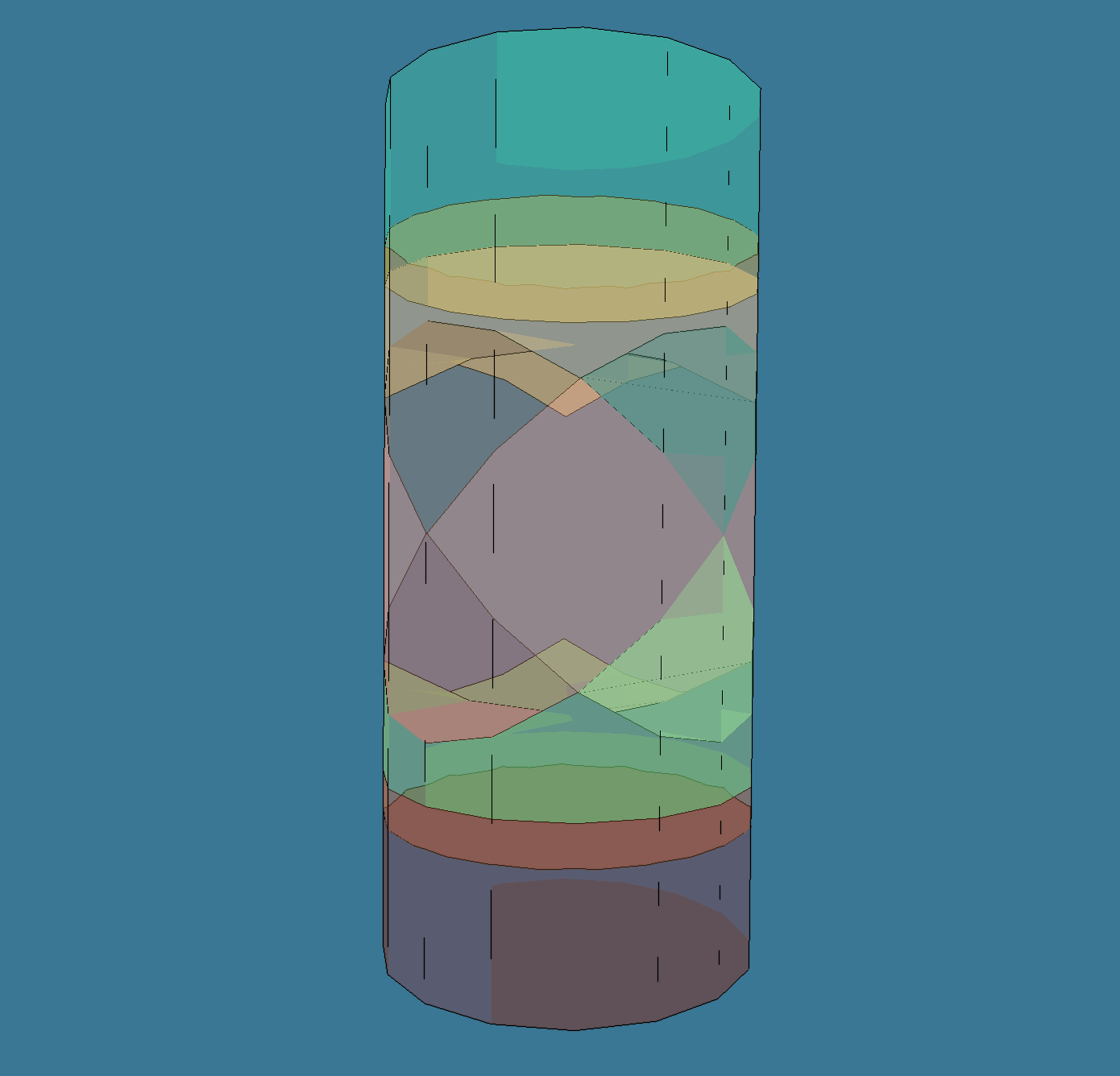} 
\end{minipage}

\begin{minipage}{0.31\textwidth} 
   \includegraphics[height=\textwidth,width=\textwidth]{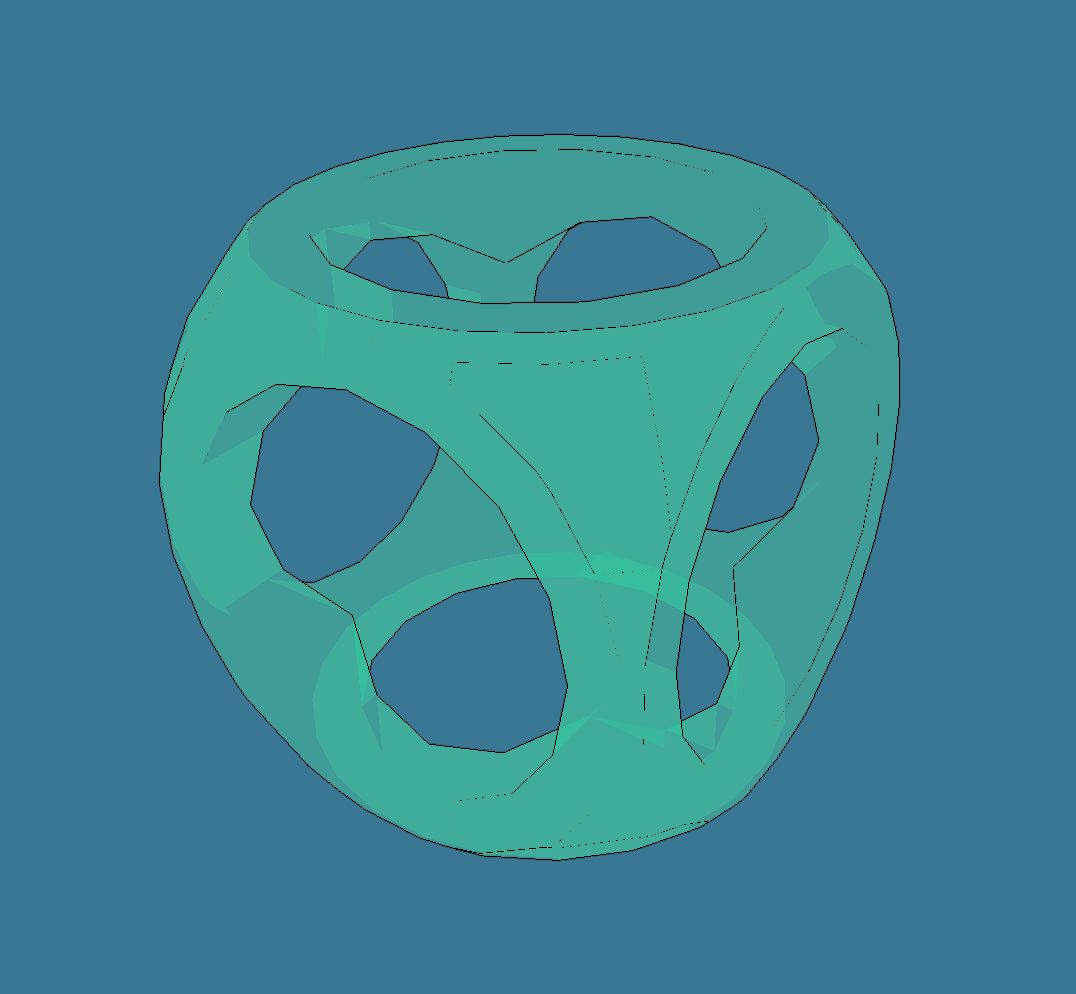} 
\end{minipage}
\begin{minipage}{0.31\textwidth}
   \includegraphics[height=\textwidth,width=\textwidth]{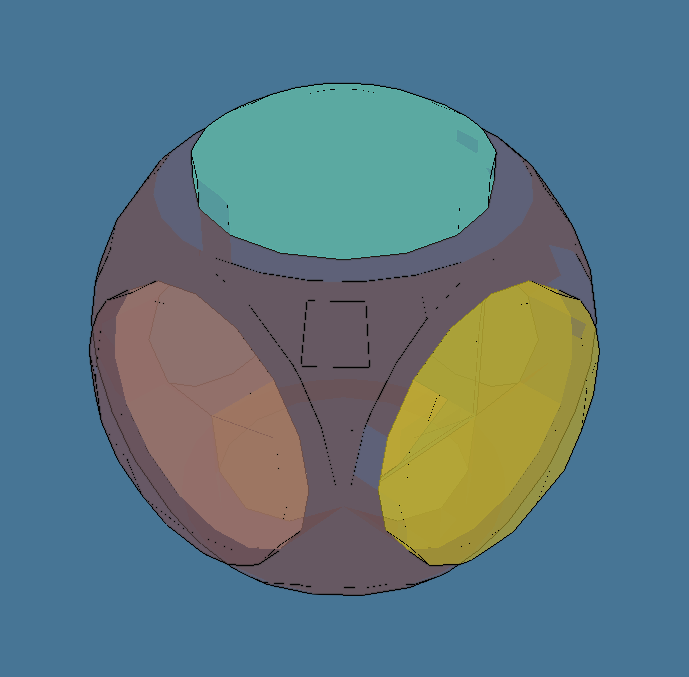} 
\end{minipage}
\begin{minipage}{0.31\textwidth}
   \includegraphics[height=\textwidth,width=\textwidth]{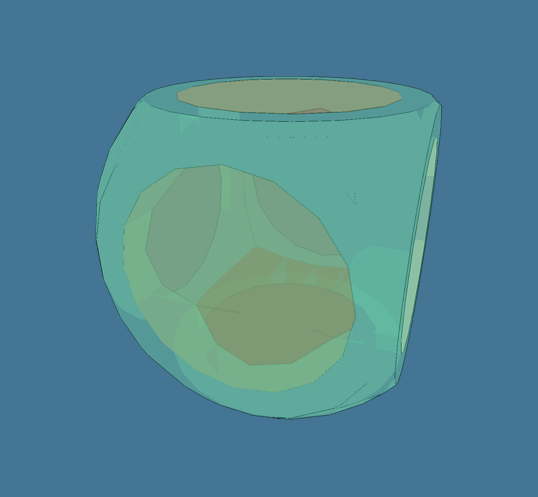} 
\end{minipage}

\begin{minipage}{0.31\textwidth} 
   \includegraphics[height=\textwidth,width=\textwidth]{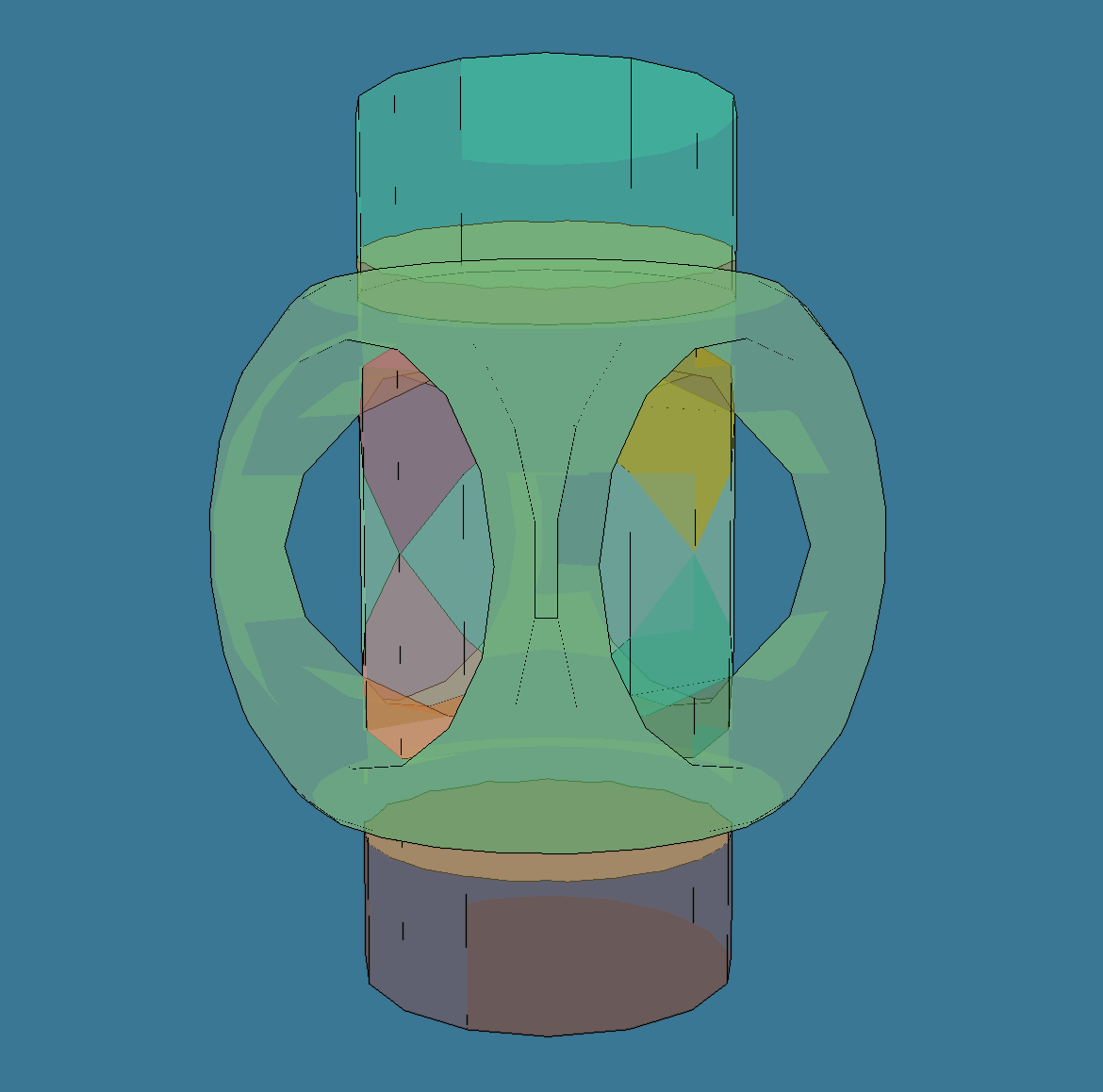} 
\end{minipage}
\begin{minipage}{0.31\textwidth}
   \includegraphics[height=\textwidth,width=\textwidth]{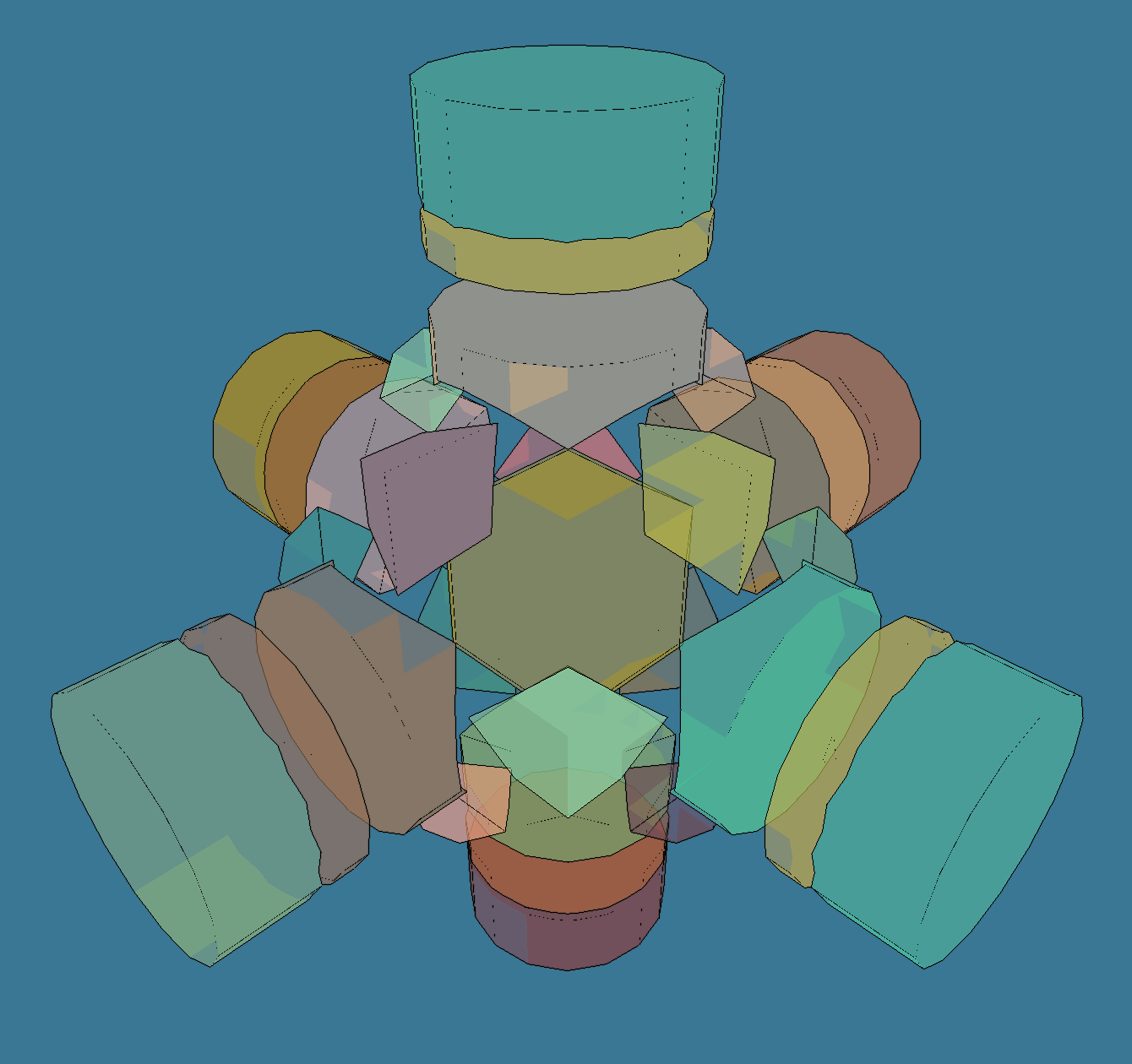} 
\end{minipage}
\begin{minipage}{0.31\textwidth}
   \includegraphics[height=\textwidth,width=\textwidth]{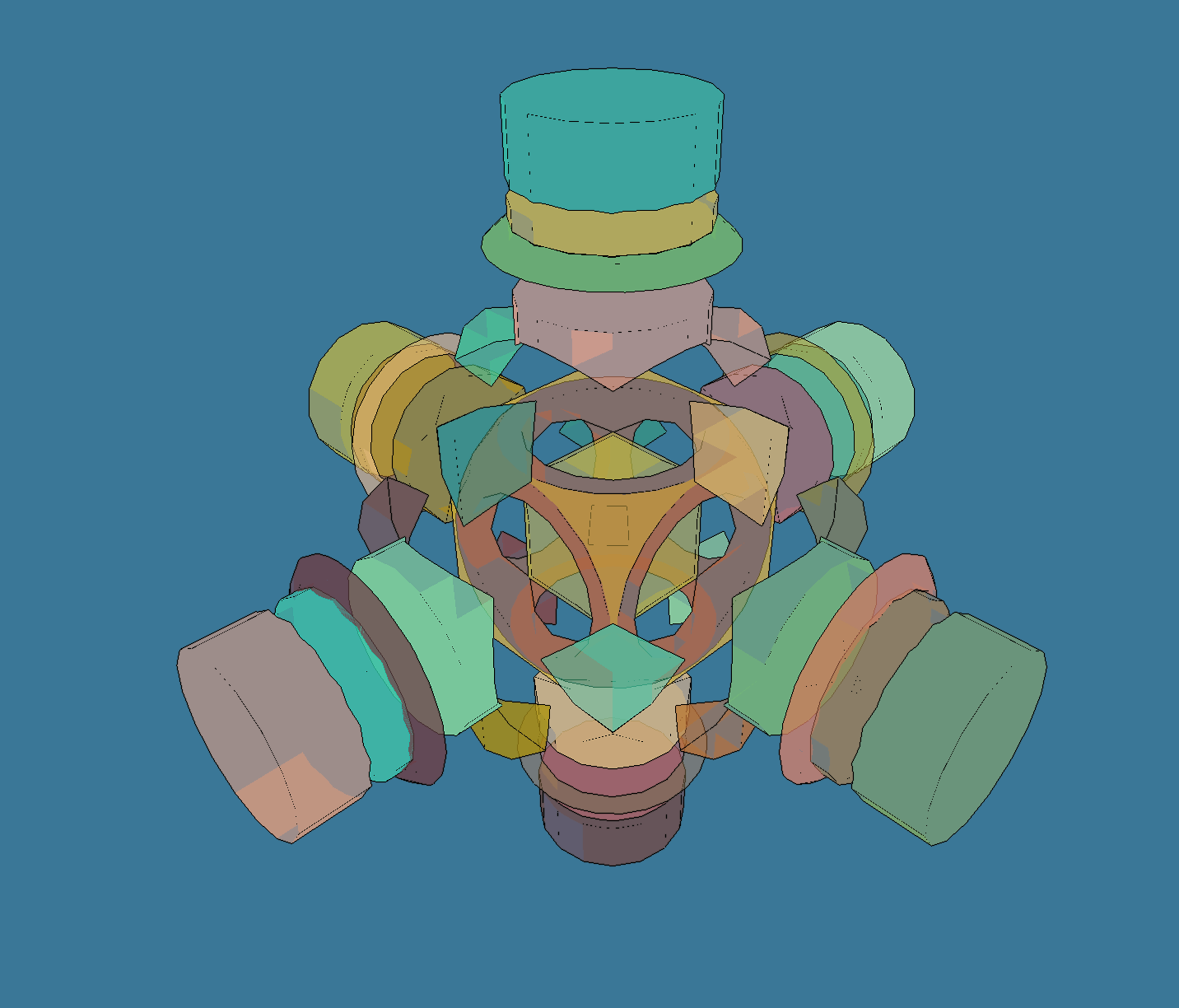} 
\end{minipage}
\vspace{2mm}

There are 40 atoms in the $\mathcal{B}$ algebra shown here, and $2^{40}\simeq 10^{12}$ terms with different \emph{structure}. Of course, the coordinate representation of each atom is a 40-element \texttt{bitArray} with only one bit to 1 (\texttt{true}), and all the other to 0 (\texttt{false}). Clearly, for computations of CSG formulas with bigger algebras, one may use sparse vectors.
\end{example}

\subsubsection{Boundary computation}

In most cases, the target geometric computational environment is able to display---more in general to handle---a~solid model only by using some boundary
representation, typically a triangulation. It is easy to get such a representation by multiplying the matrix of  3-boundary operator $\partial_3: C_3\to C_2$ times the coordinate vector 
in $C_3$ space of the solid expression, computed as a binary term of our set algebra. Once obtained in this way the signed coordinate vector of the solid object's boundary, i.e., the 2-chain of its oriented 2-cells (faces), these must be collected by columns into a sparse ``face matrix'', and translated to the corresponding matrix of oriented 1-cycles of edges, by right multiplications of $[\partial_2]$ times the face matrix. The generated boundary polygons will be finally triangulated and rendered by the graphics hardware, or exported to standard graphics file formats, or to other formats needed by applications.

\begin{figure}
\includegraphics[height=0.25\linewidth,width=0.25\linewidth]{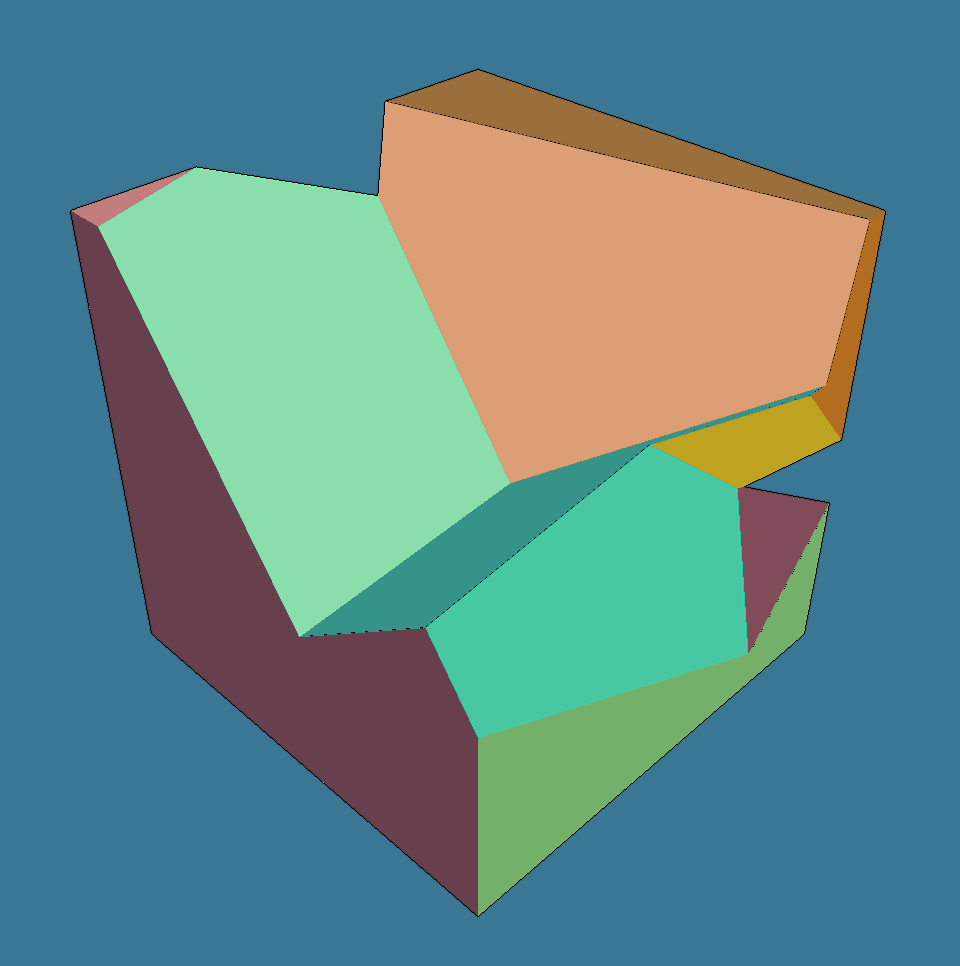}%
\includegraphics[height=0.25\linewidth,width=0.25\linewidth]{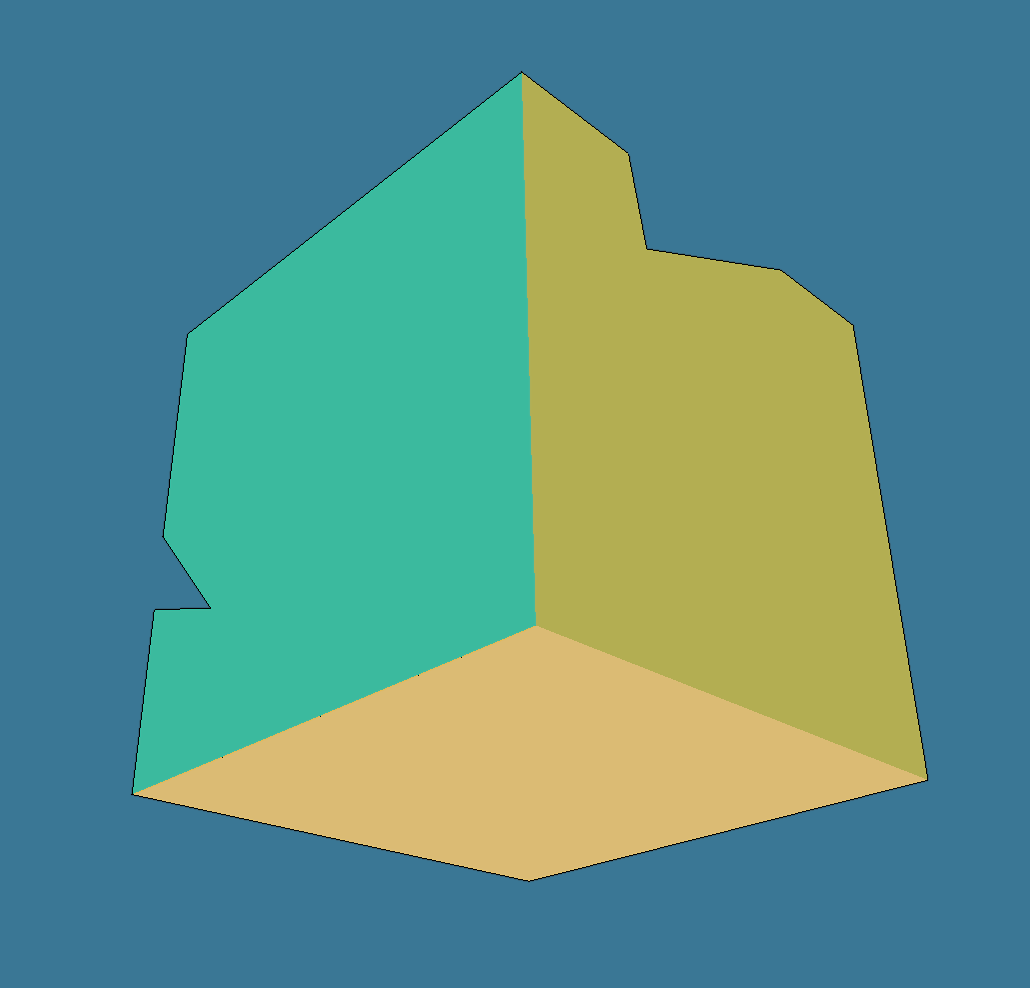}%
\includegraphics[height=0.25\linewidth,width=0.25\linewidth]{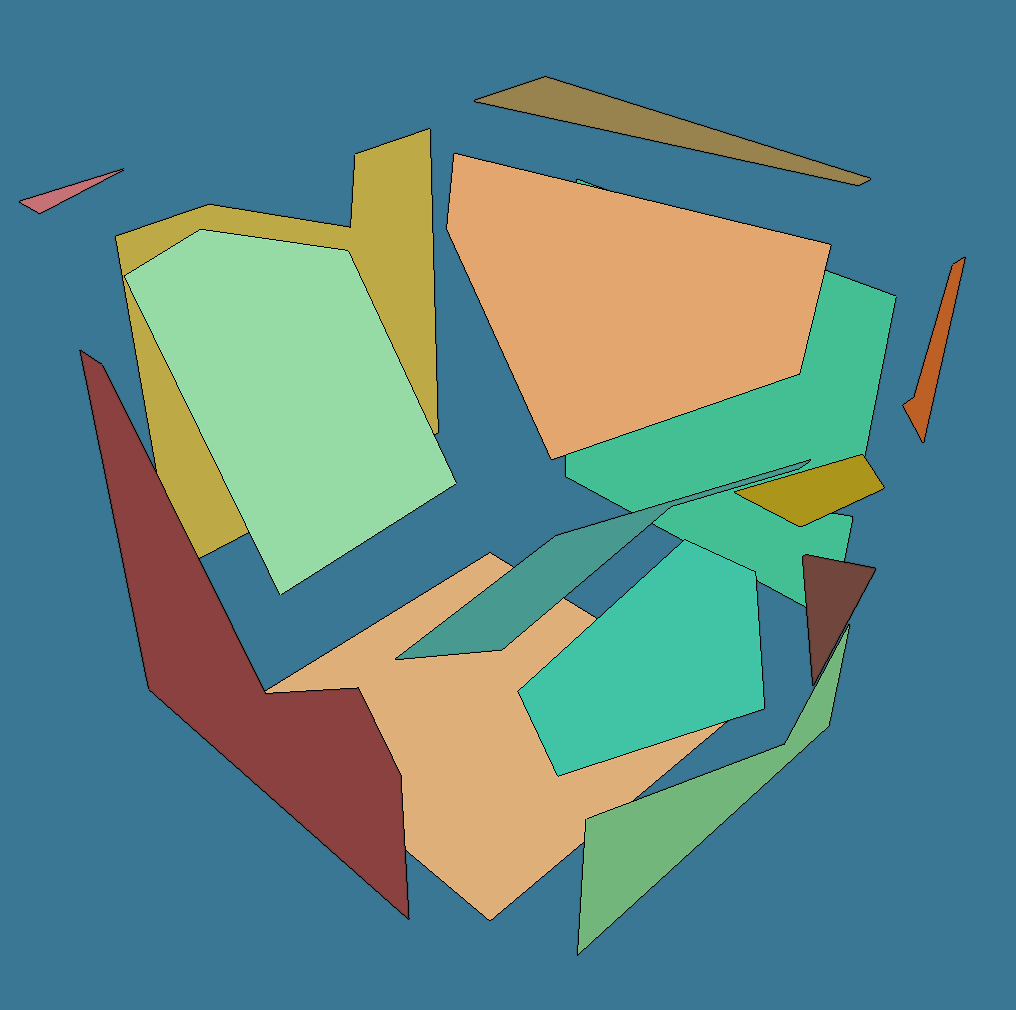}%
\includegraphics[height=0.25\linewidth,width=0.25\linewidth]{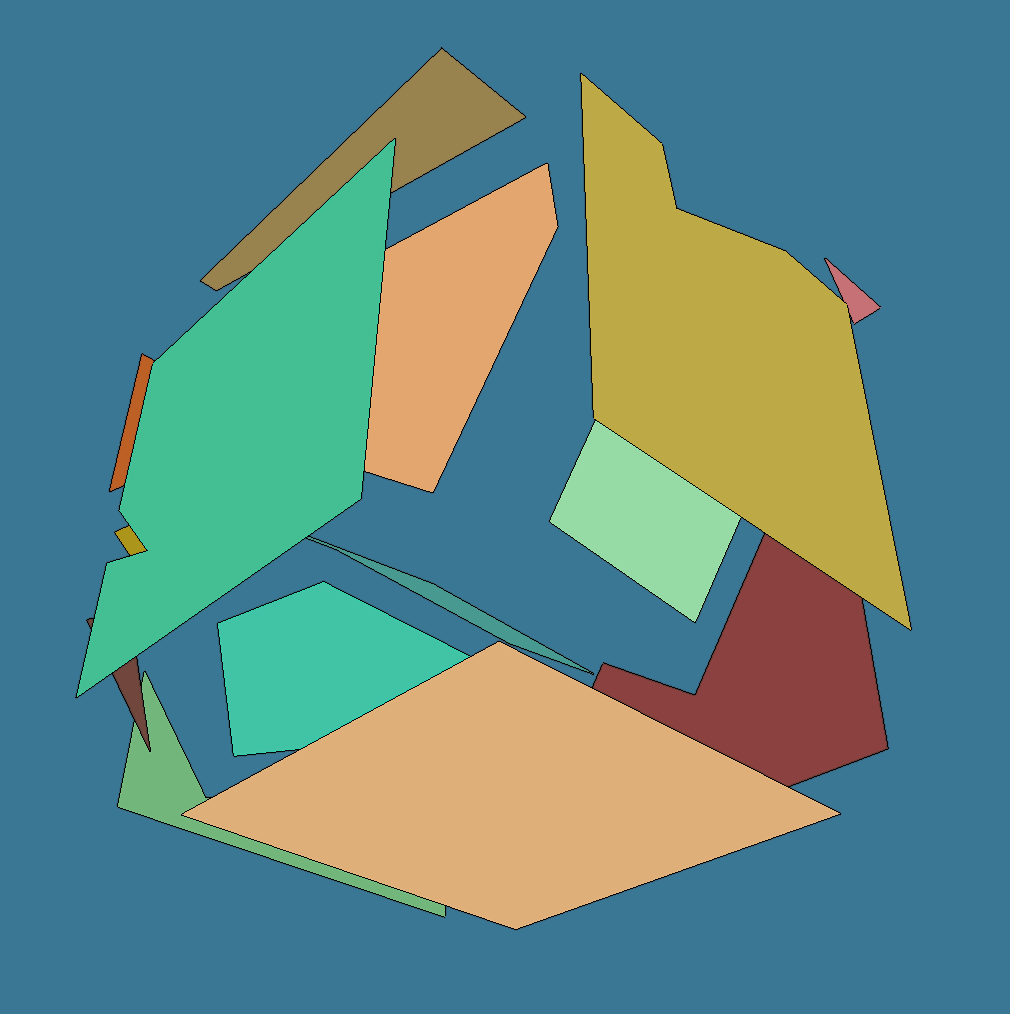}%

{\rm\footnotesize\hfill(a)\hfill\hfill(b)\hfill\hfill(c)\hfill\hfill(d)\hfill.} 
\caption{Boolean difference $(A \!\setminus\! B)
   \!\setminus\! C$ of three cubes, with 2-cells in different colors: (a) view from
   the front; (b) view from the back; (c) front with exploded 2-cells; (d)  back
   with exploded 2-cells. Note that 2-cells of the resulting boundary may be
   non-convex.} 
   
\label{fig:one-b} 
\end{figure}

\begin{definition}[LAR Geometric Complex]\label{sec:geometriccomplex}
It is wortwhile to remark that, in order to display a triangulation of  boundary faces  in their proper position in space, the whole information required (geometry $+$ topology) is contained within the LAR \emph{Geometric Complex} (GC):
\[
\mu: C_0\to\E^3,\ (\delta_2, \delta_1, \delta_0)
\qquad\equiv\qquad
\texttt{(geometry, topology) = (W, (CF, FE, EV))}
\]
A GC allows to transform the (possibly non connected) boundary 2-cycle of a Boolean result (see the example below) into a complete B-rep of the solid result. Note that ordered pairs of letters from \texttt{V,E,F,C}, correspond to the coboundary sequence  \emph{\emph{\texttt{V}}ertices$\to$\emph{\texttt{E}}dges$\to$\emph{\texttt{F}}aces$\to$\emph{\texttt{C}}ells} into the 
\emph{\emph{\texttt{C}}olumn$\to$\emph{\texttt{R}}ow} order of matrix maps of operators.
\end{definition}

\begin{remark}[Chain representation]
Let us recall that \emph{\texttt{CF}} (i.e.,~\emph{\texttt{F}}aces${}\to{}$\emph{\texttt{C}}ells) is the sparse matrix of coboundary
operator $\delta_2: C_2 \to C_3$, so that we have $[\partial_3] = [\delta_2]^t$,
which in Julia is \emph{\texttt{CF'}}.  The value of \emph{$\texttt{AminBminC = .\&(A,.!B,.!C)}$} given in Example~\ref{ex:boundary}, is the $C_2$ representation of the oriented boundary 2-cycle of solid difference, displayed in Figure~\ref{fig:one-b}.
\end{remark}

\begin{note}[From solid chains to B-reps]
Of course, the 14 non-zero elements in the \texttt{boundary} array (Example~\ref{ex:faces}) of the \texttt{AminBminC} variable, correspond to
the oriented boundary 2-cells of the solid result (see Example~\ref{ex:boundary}). Each of them is transformed into a possibly non connected 1-cycle by the \texttt{FE'} sparse matrix, {i.e.,}~by the 2-boundary matrix $[\partial_2] = [\delta_1]^t$. 
 Finally, every 1-cycle is transformed into one or more cyclic sequences of 0-cells, using
the \texttt{EV'} matrix, {i.e.}, by using $[\partial_1] = [\delta_0]^t$. The indices of 0-cells  are cyclically ordered, and used to generate sequences of 3D points via the embedding matrix \texttt{W}, which provides the vertex coordinates by column. 
This last step gives the ordered input for face triangulation using a CDT (Constrained Delaunay 
Triangulation) algorithm~\citep{SHEWCHUK200221} in 2D.
\end{note}

\begin{property}[Storage space of LAR Geometric Complex] 
The topology of a LAR 3-complex is fully represented by the operators $\delta_0, \delta_1, \delta_2$, i.e., by the sparse arrays 
\texttt{(EV,FE,CF)}, providing the incidences between vertices, edges and faces, for both B-reps and decompositive representations. If a boundary representation is used, LAR storage is 3/4 of the good ancient \emph{winged-edge} representation by~\cite{Baumgart:1972:WEP:891970}, often used as storage comparison for solid modeling, and very close ($3 \times 2E$, see~\cite{Woo:85}) to  \emph{half-edge}~\cite{Muller:78}, largely used in Computational Geometry.
\end{property}

\begin{example}\emph{Boundary of solid expression.}\label{ex:faces}
The variable \texttt{AminBminC} contains the logical representation of the Boolean expression $(\texttt{A} \!\setminus\! \texttt{B}) \!\setminus\! \texttt{C}$, specific to the particular algebra generated by terms in Example~\ref{ex:two}:

\begin{minipage}{0.9\textwidth}
\footnotesize 
\begin{lstlisting}[mathescape]
julia> difference = Int8.(AminBminC) 
8-element Array{Int8,1}:
[0,  0,  0,  0,  1,  0,  0,  0]

julia> boundary = CF' * difference 
47-element Array{Int8,1}: 
[1, 0, -1, 0, 0, 0, -1, 0, 1, 0, 1, 0, 0, 0, 1, -1, 0, -1, 0, -1, 0, 0, 1, 0, 0, 0, -1, 0, 0, 0, 0, 0, -1, 0, 0, 0, 0, 1, 0, 0, 0, 0, -1, 0, 0, 0, 0] 
\end{lstlisting}
\label{ex:boundary}
\end{minipage}

\normalsize
The value of the variable \texttt{AminBminC} is converted to the binary array \texttt{difference}---the coordinate representation of a 3-chain---by the \emph{vectorized} constructor \emph{``$\texttt{Int8}.$''}, in order to compute the \texttt{boundary} object of the Boolean expression which is shown in Figure~\ref{fig:one-b}, through multiplication times the boundary matrix \emph{$[\partial_3] \equiv \texttt{CF'}$}
Finally, we note that in chain notation it is possible to write the following expression for the oriented boundary of the ternary solid Boolean difference. The actual reduction to triangulated boundary is obtained using \texttt{FE'} and \texttt{EV'} sparse matrices, and a constrained Delaunay triangulation (CDT) algorithm.
\begin{equation}
\texttt{boundary} \mapsto f_{A\backslash B\backslash C} = f_{1} -f_{3}  -f_{7}  +f_{9}  +f_{11}  +f_{15}  -f_{16}  -f_{18}  -f_{20}  +f_{23}  -f_{27}  -f_{33}  +f_{38}  -f_{43}  \nonumber
\label{eq:faces}
\end{equation}
\end{example}

\subsection{Boolean DSL}\label{sec:DSL}
In this section we discuss the \emph{Domain Specific Language} (DSL) design about the introduced  Constructive Solid Geometry (CSG) algebra, as a Julia's small set of \emph{macros}, being currently under development.

\begin{definition}[Functional form]
Let $\langle \mathcal{A}; \otimes_1, \ldots, \otimes_k \rangle$ be an algebra generated by $\mathcal{H} = \{h_1,..., h_m\}$.
A syntactic expression constructed as a valid sequence of operations $\otimes_i$ on $n$ variables $x_i$ denoting elements of $\mathcal{A}$ is called a \emph{functional form} $\Phi$ over the algebra $\mathcal{A}$.
Note that $\Phi(x_1,..., x_n)$ is not a function (is not a set of ordered pairs) but \emph{defines} a function of $n$ arguments 
$
\phi : \mathcal{A}^n \to \mathcal{A}.
$ 
\end{definition}

\begin{remark}[Evaluation process]
We use capital Greek letters such $\Phi, \Pi$, etc. to denote forms, and denote functions over algebras by lower case Greek letters. The distinction between forms and functions is important.
Forms and functions over an algebra are formally related by an \emph{evaluation} process, assigning values to the variables in the form, and computing the resulting value of the expression. This relationship is expressed by writing
$
\phi(x_1,..., x_n) = |\Phi(x_1,..., x_n)|.
$
\end{remark}

\begin{property}[Evaluation process]
If $\mathcal{A}$ is a finite algebra, there is a finite number of distinct functions over $\mathcal{A}$, but an infinite number of distinct forms, e.g., strongly redundant.
If $S \in \mathcal{A}$ is an element of the algebra generated by $\mathcal{H}$, there exists a form $\Phi$ over $\mathcal{A}$ such that:
$
S = |\Phi(\mathcal{H})|.
$
We then say that $S$ is \emph{describable} in $\mathcal{A}$ by $\mathcal{H}$. In general, $\Phi(\mathcal{H})$ is not unique, but $|\Phi(\mathcal{H})|$ is unique by definition.
\end{property}

\begin{definition}[Domain Specific Language]
The \emph{Domain Specific Language} (DSL) proposed in this paper for our CSG algebra is very simple. Its well-formed formulas are made by Julia identifiers of variables, Boolean operation symbols, and round brackets. A CSG expression is coded as a Julia macro \texttt{@CSG}, to provide a simpler interface to create Julia objects with complicated structure. The list of object models associated to variables is extracted by traversing the \texttt{@CSG} expression. Each object must be located and oriented in world coordinates using the \texttt{Lar.Struct} datatype, allowing to combine hierarchical assemblies (cellular complexes in local coordinates) and affine transformations. The application of the function \texttt{Lar.struct2lar()} evaluates a geometric object of type \texttt{Lar.Struct}, generating the triple \texttt{(V,FV,EV)}, i.e., the LAR~\citep{Dicarlo:2014:TNL:2543138.2543294} \emph{minimal} models used by the \texttt{CSG.jl} package. 
\end{definition}

\begin{algorithm}[Evaluation process]
The evaluation of a 3D Boolean functional form \texttt{@CSG <expr>}, into an \emph{evaluated} LAR model \texttt{(V,(CF,FE,EV))}, made by a $3\times n$ array \texttt{V} and by a triple \texttt{(CF,FE,EV)} of sparse arrays, providing respectively the geometric embedding $C_0 \to \E^3$, and a chain complex $\delta_2, \delta_1, \delta_0$, is made by several tasks:

\begin{enumerate}
\item \emph{Evaluation of data and functional forms}. \textbf{Input}:  a single \texttt{@CSG} macro block, possibly delimited by \texttt{begin...end} clause, including solid terms and (possibly) component \texttt{@CSG} clauses. \textbf{Output}: parse tree of compound \texttt{@CSG} expression; ordered tuple of variable names associated to input objects in a single coord frame: \texttt{(A$_1$,$\ldots$,A$_m$)}.

\item \emph{Computation of space arrangement}.  \textbf{Input}: tuple of solid terms in world coordinates: \texttt{(A$_1$,$\ldots$,A$_m$)}. \textbf{Output}: sparse boundary matrices \texttt{(FC,EF,VE)} as $[\partial_3], [\partial_2], [\partial_1]$, and (new) vertex matrix \texttt{W}. The columns of $[\partial_3]$ are ono-to-one with the atoms generating the \texttt{@CSG} algebra.

\item \emph{Evaluation of Boolean terms}.  \textbf{Input}: the boundary representation of atoms of CSG algebra, generated by chain complex \texttt{(W,(FC,EF,VE))}, and the input \texttt{(A$_1$,$\ldots$,A$_m$)} terms. \textbf{Output}: logical $n\times m$ array \texttt{B = Array\{Bool,2\}} whose column $\texttt{B}_j$ gives the bit array representing the \texttt{A}$_j$ term of \texttt{@CSG} algebra input form.

\item \emph{Bitwise computing of Boolean function}.  \textbf{Input}: the logical array \texttt{B}; the parse tree of top-level \texttt{@CSG} expression. \textbf{Output}: a single \texttt{BitArray} of length $n$, as the number of Boolean atoms. The  1$s$  in this array denote the atomic 3-chains whose disjoint union produce the Boolean result, i.e., the value of evaluated \texttt{@CSG} form.

\item \emph{Boundary evaluation of output 3-chain}.  \textbf{Input}: the 3-chain (binary) coordinate representation of the Boolean result; the chain complex \texttt{W,(FC,EF,VE)}. \textbf{Output}: the boundary representation of the Boolean result, possibly exported in a standard vector graphics format, like, e.g., the  file formats \texttt{.OBY}, \texttt{.PLY}, or \texttt{.DAE} (Digital Asset Exchange --- \texttt{COLLADA XML}). Our software prototype may currently export in 2D and 3D \texttt{.OBY}.

\end{enumerate}

\end{algorithm}

\begin{definition}[Boolean form]
Clause (expression formed from a finite collection of literals) that matches other clauses and/or \texttt{LAR} models, by matching Boolean combinations of other clauses. It is built using one or more \texttt{@CSG} macro, each one with a typed ``Struct'' occurrence. 
The semantics of a \texttt{Struct}, natively including only ``aggregation'' and ``affine transformation'', is enriched with $n$-ary union, intersection, difference, and complement symbols. 
Let $\texttt{A}, \texttt{B}, \texttt{C}, \dots$ be either \texttt{LAR} models or \texttt{Struct} literals or values. We have:
\[
\begin{aligned}
\texttt{:(+, A,B,C,...)} & \texttt{::= Struct}\left( \texttt{:union}, \texttt{[A,B,C,...]} \right)\\
\texttt{:(*, A,B,C,...)} & \texttt{::= Struct}\left( \texttt{:intersect}, \texttt{[A,B,C,...]} \right)\\
\texttt{:(-, A,B,C,...)} & \texttt{::= Struct}\left( \texttt{:diff}, \texttt{[A,B,C,...]} \right)
\end{aligned}
\]
\end{definition}

\section{Computational examples}\label{sec:examples}

\begin{figure} 
\includegraphics[height=0.25\linewidth,width=0.25\linewidth]{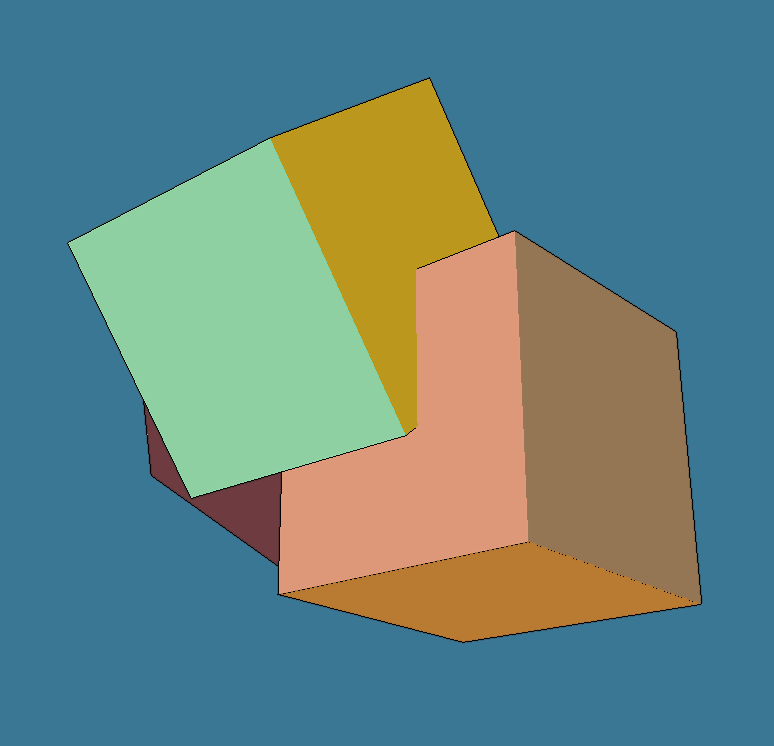}%
\includegraphics[height=0.25\linewidth,width=0.25\linewidth]{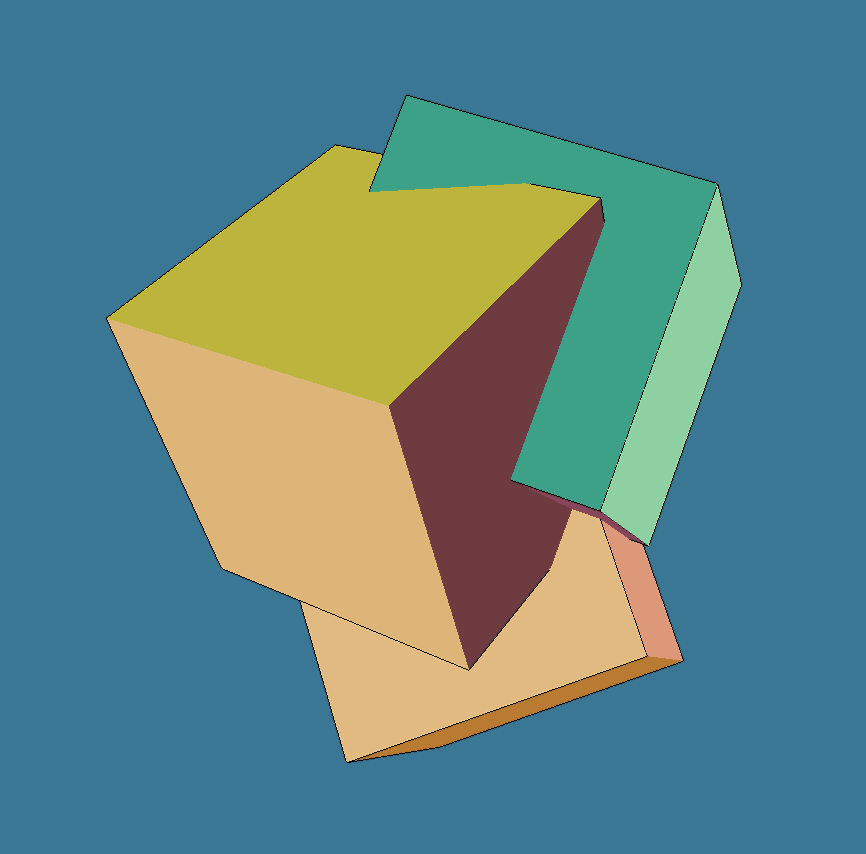}%
\includegraphics[height=0.25\linewidth,width=0.25\linewidth]{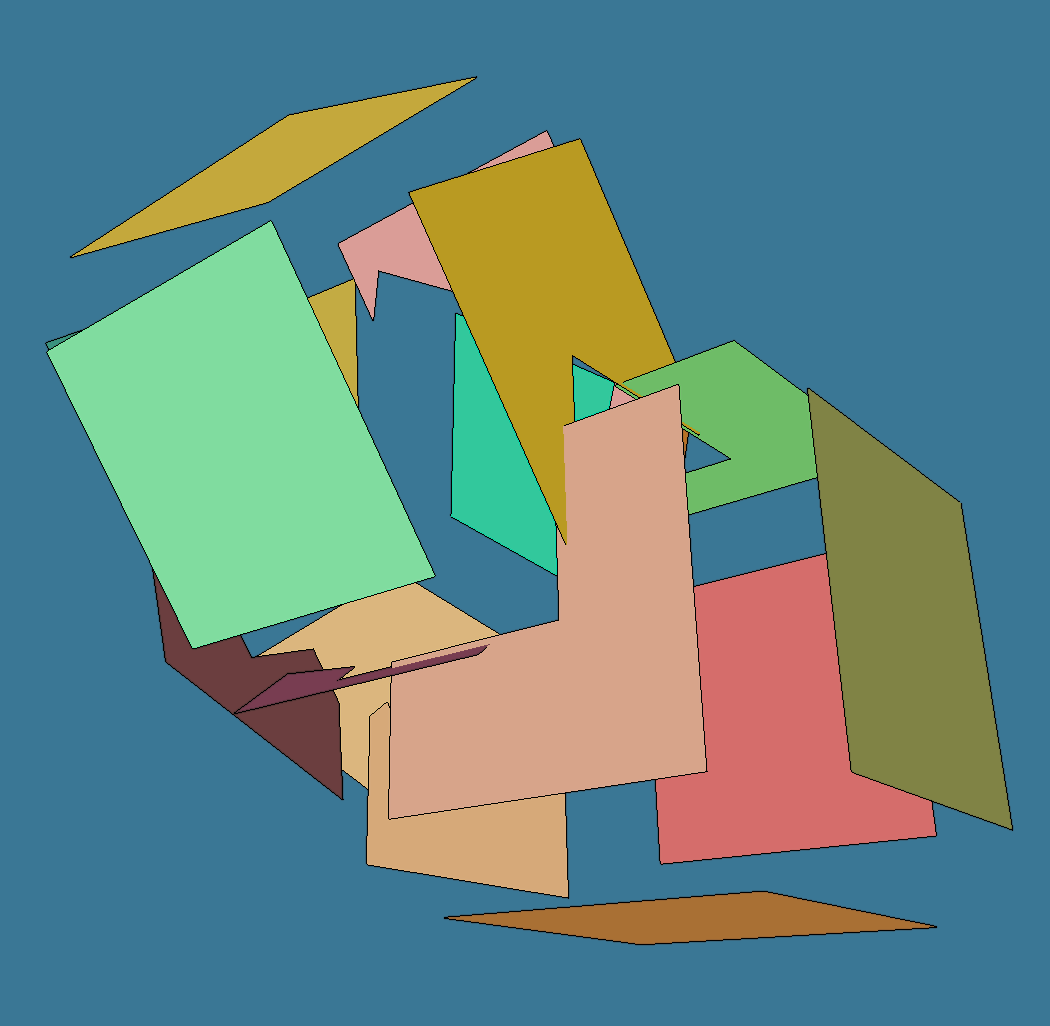}%
\includegraphics[height=0.25\linewidth,width=0.25\linewidth]{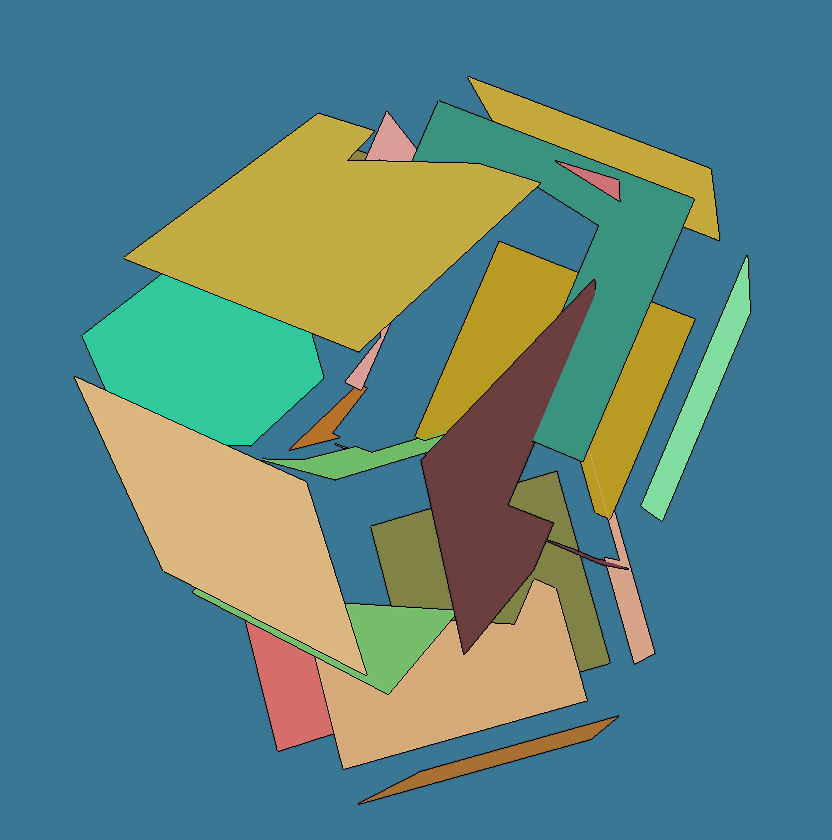}%

{\rm\footnotesize\hfill(a)\hfill\hfill(b)\hfill\hfill(c)\hfill\hfill(d)\hfill.} 
\caption{Boolean union $A \cup B
   \cup C$ of three cubes (from \texttt{assembly} of Example~\ref{ex:two}), with 2-cells in different colors: (a) view from
   the front; (b) view from the back; (c) front with exploded 2-cells; (d)  back
   with exploded 2-cells. For clarity sake, only the boundary 2-cells are displayed. The space 
   partition generated other 2-cells in the interior, of course.}
\label{fig:union} 
\end{figure}

This section provides the reader with some examples of solid modeling \emph{programming style} with Julia and its sparse and non-sparse arrays. We believe that giving  a look at a few simple concrete examples is useful for understanding our computational approach and its possible developments. 
Sections~\ref{seq:union} and~\ref{sec:five} aim at showing the sequence of spaces and transformations that finally produce a solid model when evaluating a Boolean expression through  a  function application  \texttt{Lar.bool3d(assembly)}. We show in Section~\ref{sec:checks} that the exactness property ($\partial^2 = 0$) of any chain complex can be used to check the accuracy of calculations.  The \emph{Euler characteristic} of the \texttt{union} solid model generated in Example~\ref{ex:two} is stepwise computed in Section~\ref{sec:euler}, where we use the chain maps $\partial_3, \partial_2, \partial_1$ to obtain the sets (and numbers) of \emph{faces}, \emph{edges} and~\emph{vertices} belonging to the boundary of the \texttt{union} solid, which is a closed 2-manifold of genus zero (see Figure~\ref{fig:union}).
As we told before, a specific API and a mini DSL for CSG expressions are being planned.

\subsection{Variadic union}\label{seq:union}
\normalsize  
In order to compute the \texttt{union} of three affinely transformed instances of the unit \texttt{cube}, we consider the \texttt{assembly} expression given in Example~\ref{ex:two}. First we get the $\E^3$ space partition and the term structure generated by \texttt{assembly} through the function \texttt{Lar.bool3d}; 
then we combine the logical arrays \texttt{A}, \texttt{B}, and \texttt{C}, building the
value of \texttt{BitArray} type for the \texttt{union} variable, that stores the logical representation of the specific 3-chain. Let us remark that the bitwise ``\texttt{or}'' operator (``\texttt{$|$}'') is applied in a vectorized way to arrays, by inserting a dot character:

\begin{minipage}{0.9\textwidth}
\small 
\begin{lstlisting}[mathescape]
julia> W, (EV, FE, CF), boolmatrix = Lar.bool3d(assembly);  
julia> A,B,C = boolmatrix[:,2],boolmatrix[:,3],boolmatrix[:,4] 
julia> union = .|(A, B, C);
julia> @show union;

union = Bool[false, true, true, true, true, true, true, true]
\end{lstlisting}
\end{minipage}

\normalsize  Finally,  the  boundary 2-cycle  \texttt{faces} is generated by multiplication of the sparse matrix $[\partial_3]$ ({i.e.}, \texttt{CF'})  times the binary converted \texttt{union}. The mapping \texttt{Bool $\to\{0,1\}$} is applied via a vectorized application of the  \texttt{Int8} constructor:

\begin{minipage}{0.9\textwidth}
\small 
\begin{lstlisting}[mathescape]
julia> faces = CF' * Int8.(union);
julia> @show faces;

faces = [1, 0, -1, 0, 0, 0, -1, 0, 1, 0, 1, 0, 0, 0, 1, -1, 0, -1, 0, 0, 1, -1, 0, -1, 1, 
0, 0, 0, 1, 1, 0, -1, 0, 0, 1, 0, -1, 0, 0, 0, -1, 1, 0, 1, 0, 0, -1]
\end{lstlisting}
\end{minipage}

\normalsize\noindent
With $f_k\in U_2$, where $U_2$ is the basis of chain space $C_2$ generated by $\mathcal{A}\,(\texttt{assembly})$,  we may write in chain notation: 
\begin{equation}
\begin{split}\small 
\texttt{faces} \mapsto f_{A\cup B\cup C} =& f_{1} -f_{3} -f_{7} +f_{9} +f_{11} +f_{15} -f_{16} -f_{18} +f_{21} -f_{22} -f_{24} \\
&+f_{25} +f_{29} +f_{30} -f_{32} +f_{35} -f_{37} -f_{41} +f_{42} +f_{44} -f_{47} \label{eq:faces}
\end{split}
\end{equation}
The boundary representation, i.e., the subset of boundary's oriented faces, is given by the 2-cycle in Eqation~\ref{eq:faces}. They are transformed into a boundary triangulation, needed for graphic display, by proper use of $[\partial_2] = \texttt{FE'}$ and $[\partial_1] = \texttt{EV'}$.

\subsection{Meaning of $\partial$ matrices}\label{sec:five}

By definition, the matrix $[\partial_2] = \texttt{FE'} = \texttt{EF}$ contains by columns the $U_2$ basis expressed as an ordered sequence of 1-cycle vectors. Signed values provide cyclic ordering of 1-cycles of edges, easily extendible to higher dimensions\footnote{It is often stated that cyclic ordering of polygon sides is not extendable to higher dimensions. On the contrary, this can be done, using d-cycles.}. With $e_h \in U_1$, we have:

\begin{minipage}{0.9\textwidth}
\small 
\begin{lstlisting}[mathescape]
FE'[:, 1] $\mapsto$ f$_1$ =  e$_1$ -e$_2$ +e$_4$ -e$_5$ +e$_6$ -e$_7$ +e$_8$ 
FE'[:, 2] $\mapsto$ f$_2$ =  e$_3$ +e$_7$ -e$_8$  
... 	...		... 
FE'[:,47] $\mapsto$ f$_{47}$ = e$_{64}$ -e$_{69}$ +e$_{77}$ -e$_{81}$ -e$_{86}$ +e$_{88}$ 
\end{lstlisting}
\end{minipage}

\normalsize
Each 1-cell $e_h$ is mapped to an ordered pair of 0-cells $v_j \in U_0$, via the boundary matrix $[\partial_1] = [\delta_0]^t =  \texttt{EV'}$:

\begin{minipage}{0.9\textwidth}
\small 
\begin{lstlisting}[mathescape]
EV'[:, 1] $\mapsto$ e$_1$ =  v$_2$ - v$_1$ 
EV'[:, 2] $\mapsto$ e$_2$ =  v$_6$ - v$_3$  
... 	...		... 
EV'[:,49] $\mapsto$ e$_{49}$ =  v$_{29}$ - v$_{26}$  
\end{lstlisting}
\end{minipage}

The geometric embedding in $\E^3$ of the $A\cup B\cup C$ model generated in Section~\ref{seq:union} is provided by the coordinate array $\texttt{W} \in \R^3_{49}$,  with 3 rows and 43 columns. The   coordinate array \texttt{V} embedding the initial \texttt{assembly} model, consisting of three non (yet) intersected cubes, has instead  dimension $3\times 24$.

\subsection{Correctness checks}\label{sec:checks}
\normalsize

A computational approach based on chain complexes offers unique tools for checking the accuracy of calculations, that are correct by construction, since both $\partial^2=0$ and $\delta^2=0$ hold, when applied to any chain.   In words, \emph{every boundary is a cycle,} or equivalently: the \emph{chain complex is exact}. 
Also, we know that the boundary of every solid is a possibly non-connected closed surface, hence a 2-cycle.
This holds if and only if the construction of $[\partial]$ matrices is done correctly.
In the following example, we start from the chain of boundary \texttt{faces} of Section~\ref{seq:union}.

\begin{minipage}{0.9\textwidth}
\footnotesize 
\begin{lstlisting}[mathescape]
julia> pairs = [(f,sign) for (f,sign) in enumerate(copCF' * Int8.(union)) if sign != 0];
julia> faces = map(prod, pairs);
julia> @show faces;
faces = [1,-3,-7,9,11,15,-16,-18,21,-22,-24,25,29,30,-32,35,-37,-41,42,44,-47]
\end{lstlisting}
\end{minipage}

\normalsize We obtain the following 2-chain as boundary 2-cycle. 
The cardinality of the \texttt{faces} array is $\chi_2=21$.
\small
\begin{equation}
\begin{split}
\texttt{\mbox{faces}} \mapsto f_{A\cup B\cup C} =& f_{1 } -f_{3 } -f_{7 } +f_{9 } +f_{11 } +f_{15 } -f_{16 } -f_{18 } +f_{21 } -f_{22 } -f_{24 } +f_{25 } +f_{29 } +f_{30 }\\ 
&-f_{32 } +f_{35 } -f_{37 } -f_{41 } +f_{42 } +f_{44 } -f_{47} \label{eq:faces}
\end{split}
\end{equation}
\normalsize
The edge subset on the boundary of the \texttt{union} solid is computed by: (a) transforming the \texttt{faces} array of signed indices of 2-cells into the \texttt{COORD} representation (\texttt{rows,cols,vals}) of a sparse matrix~\citep{coosparse}; (b) holding a single boundary face (2-cell) per column in \texttt{facemat}; (c) multiplying this matrix times the $[\partial_2]$ operator, thus obtaining the new sparse matrix \texttt{edges4face}, holding a face 1-cycle per column; and, finally, (d)~extracting only the positive instance of boundary edges belonging to the boundary faces. The number of boundary edges $\chi_1=57$ is one half of the non-zero terms in the sparse vector \texttt{edges4face}. The other half has the opposite sign, so that the total sum is zero:

\begin{minipage}{0.9\textwidth}
\footnotesize 
\begin{lstlisting}[mathescape]
julia> nonzeros = hcat([[abs.(face),k,sign(face)] for (k,face) in enumerate(faces)]...);
julia> facemat = sparse([nonzeros[k,:] for k=1:size(nonzeros,1)]...);
julia> edges4face = copFE' * facemat;
julia> rows,cols,vals = findnz(edges4face);
\end{lstlisting}
\end{minipage}

\begin{minipage}{0.96\textwidth}
\small 
\begin{lstlisting}[mathescape]
julia> edges = [e*sign for (e,sign) in zip(rows,vals) if sign==1];
julia> @show edges;

edges = [1,4,6,8,10,14,18,20,9,23,26,27,2,34,30,36,38,5,29,24,43,15,42,28,47,51,53,54,21,52,57,
		 48,61,62,64,50,59,66,55,58,40,63,68,77,79,80,35,78,83,72,88,7,74,87,69,81,86]
\end{lstlisting}
\end{minipage}

We remark again that, without filtering out the terms of negative \texttt{sign}, we would get an \texttt{edges} array of signed indices summing to zero, according to the constraint $\partial^2=0$. This attests to the exactness of calculations.

\subsection{Euler characteristic}\label{sec:euler}

 Euler characteristic of a solid of genus $g$  in $\E^3$  is defined as 
\[
\chi(g) = \chi_0 - \chi_1 + \chi_2 = 2 - 2g,
\]
where $\chi_0, \chi_1, \chi_2$ are, respectively, equal to the number of vertices, edges and faces on the boundary of the solid. In our case, 
the number of boundary faces is $\chi_2 = 21$, according to Eq.~(\ref{eq:faces}). 
The \texttt{edges} indices given in Section~\ref{sec:checks} determine the 1-chain $e_{A\cup B\cup C}$ (or, more precisely, the non-independent 1-cycle generated by the independent 2-cycles of boundary faces). 
The cardinality of the \texttt{edges} array provides $\chi_1=57$. Note that, in order to counting the edges, we consider only the  positive instances of 1-cells, since they appear in pairs (positive and negative) in a closed and coherently oriented
cellular 2-complex.
\begin{align}\footnotesize
\begin{split}
\texttt{edges} \mapsto e_{A\cup B\cup C} &= e_{1 } +e_{4 } +e_{6 } +e_{8 } +e_{10 } +e_{14 } +e_{18 } +e_{20 } +e_{9 } +e_{23 } +e_{26 } +e_{27 } +e_{2 } +e_{34 } +e_{30 } +e_{36 } +e_{38 } +e_{5 } +e_{29 } 
+e_{24 } \\
&+e_{43 } +e_{15 } +e_{42 } +e_{28 } +e_{47 } +e_{51 } +e_{53 } +e_{54 } +e_{21 } +e_{52 } +e_{57 } +e_{48 } +e_{61 } +e_{62 } +e_{64 } +e_{50 } +e_{59 } +e_{66 } +e_{55 } \\
&+e_{68 } +e_{77 } +e_{79 } +e_{58 } +e_{40 } +e_{63 } +e_{80 } +e_{35 } +e_{78 } +e_{83 } +e_{72 } +e_{88 } +e_{7 } +e_{74 } +e_{87 } +e_{69 } +e_{81 } +e_{86} 
\end{split}
\label{eq:edges}
\end{align}

\normalsize The final script computes the 0-chain $v_{A\cup B\cup C}$ of vertices  on the boundary of \texttt{union} solid, with cardinality $\chi_0 = 38$.

\begin{minipage}{0.8\textwidth}
\small 
\begin{lstlisting}[mathescape]
julia> nonzeros = hcat([[abs(e),k,sign(e)] for (k,e) in enumerate(edges)]...);
julia> edgemat = sparse([nonzeros[k,:] for k=1:size(nonzeros,1)]...);
julia> verts = sort(collect(Set(findnz(copEV' * edgemat)[1])));
julia> @show verts;
verts = [1,2,3,4,5,6,7,8,9,10,12,15,17,18,19,20,21,24,25,26,27,30,31,32,33,34,
		 35,36,37,38,39,41,44,45,46,47,48,49]
\end{lstlisting}
\end{minipage}

In chain notation:
\begin{equation}\footnotesize
\begin{split}
\texttt{verts} \mapsto v_{A\cup B\cup C} &= v_{1 } +v_{2 } +v_{3 } +v_{4 } +v_{5 } +v_{6 } +v_{7 } +v_{8 } +v_{9 } +v_{10 } +v_{12 } +v_{15 } +v_{17 } +v_{18 } +v_{19 } +v_{20 } +v_{21 } +v_{24 } +v_{25 } +v_{26 } \\
&+v_{27 } +v_{30 } +v_{31 } +v_{32 } +v_{33 } +v_{34 } +v_{35 } +v_{36 } 
+v_{37} +v_{38 } +v_{39 } +v_{41 } +v_{44 } +v_{45 } +v_{46 } +v_{47 } +v_{48 } +v_{49}\\[2mm]
\end{split}
\label{eq:verts}
\end{equation}

\normalsize
The generated \texttt{union} solid model has topological genus $g=0$ (see Figure~\ref{fig:union}). Therefore, the test of correctness provided by checking the Euler characteristic via Eqs.~(\ref{eq:faces}), (\ref{eq:edges}), and (\ref{eq:verts}), gives the correct answer:
\[
\chi(\texttt{union}) = \chi_0 - \chi_1 + \chi_2 = 38 - 57 + 21 = 2
\]

\begin{figure} 
\includegraphics[height=0.25\linewidth,width=0.25\linewidth]{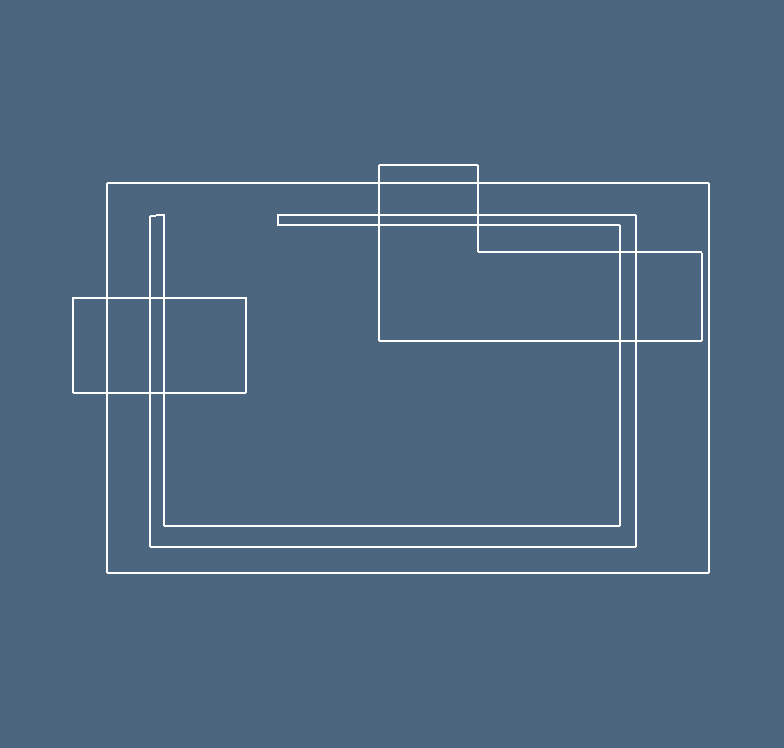}%
\includegraphics[height=0.25\linewidth,width=0.25\linewidth]{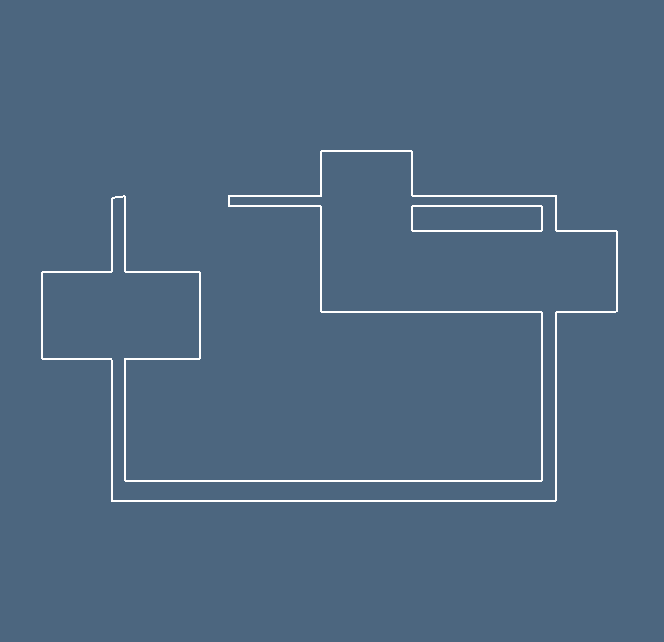}%
\includegraphics[height=0.25\linewidth,width=0.25\linewidth]{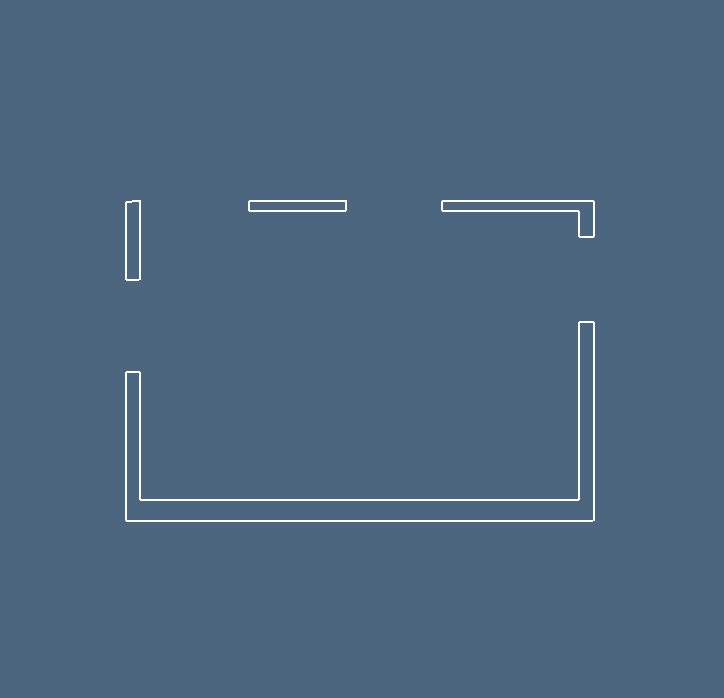}%
\includegraphics[height=0.25\linewidth,width=0.25\linewidth]{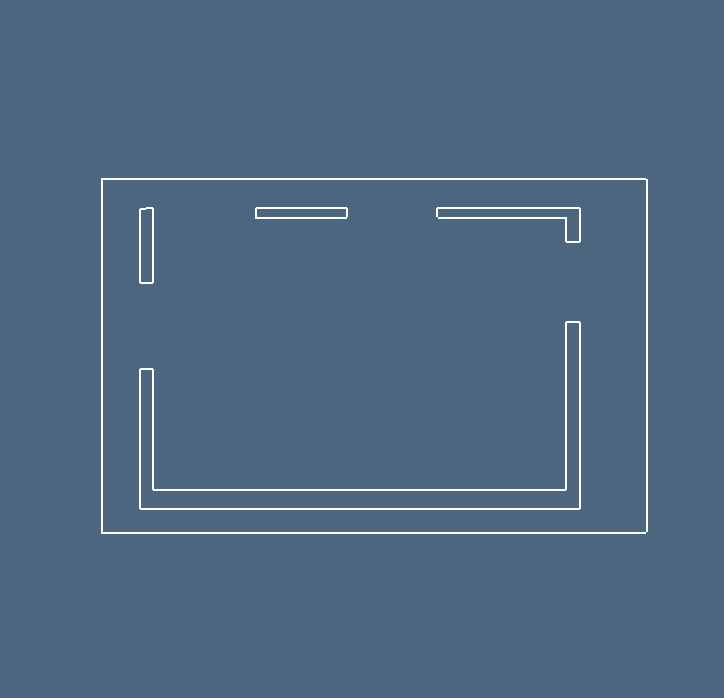}%

\vspace{-9mm}
\includegraphics[height=0.25\linewidth,width=0.25\linewidth]{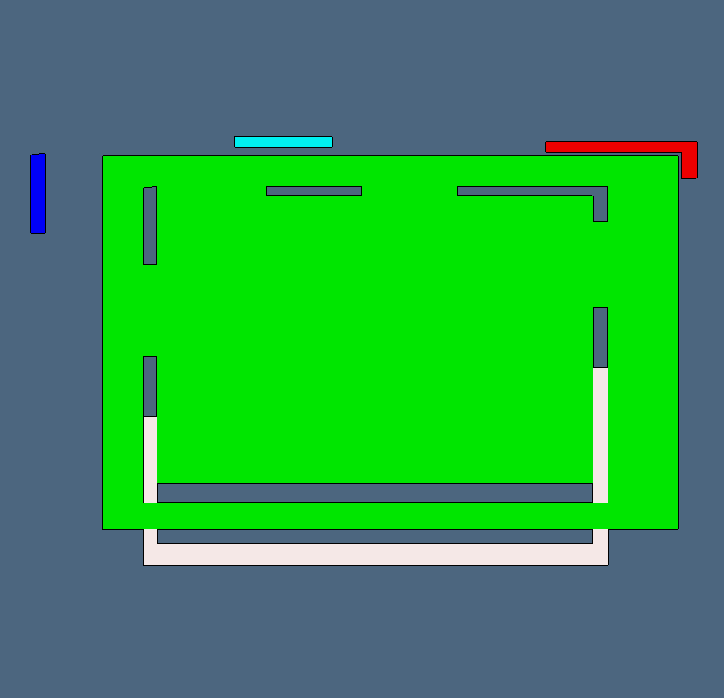}%
\includegraphics[height=0.25\linewidth,width=0.25\linewidth]{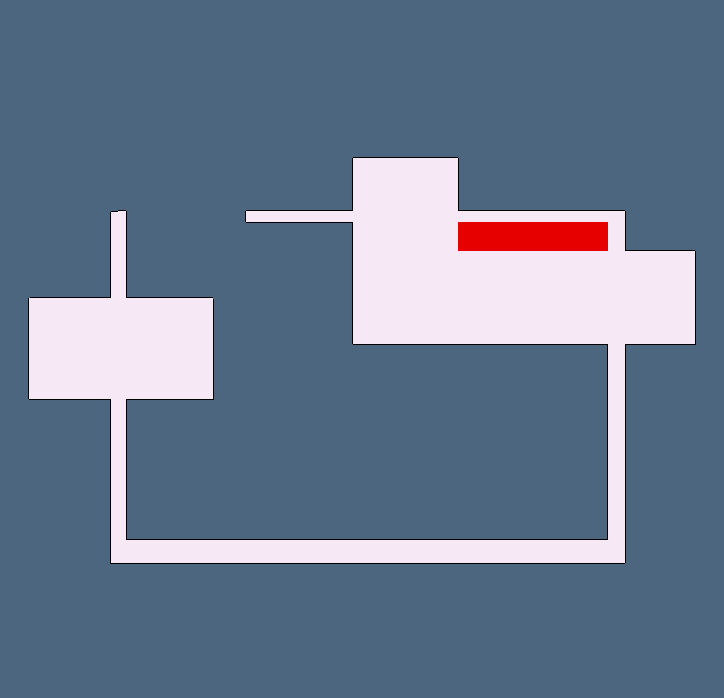}%
\includegraphics[height=0.25\linewidth,width=0.25\linewidth]{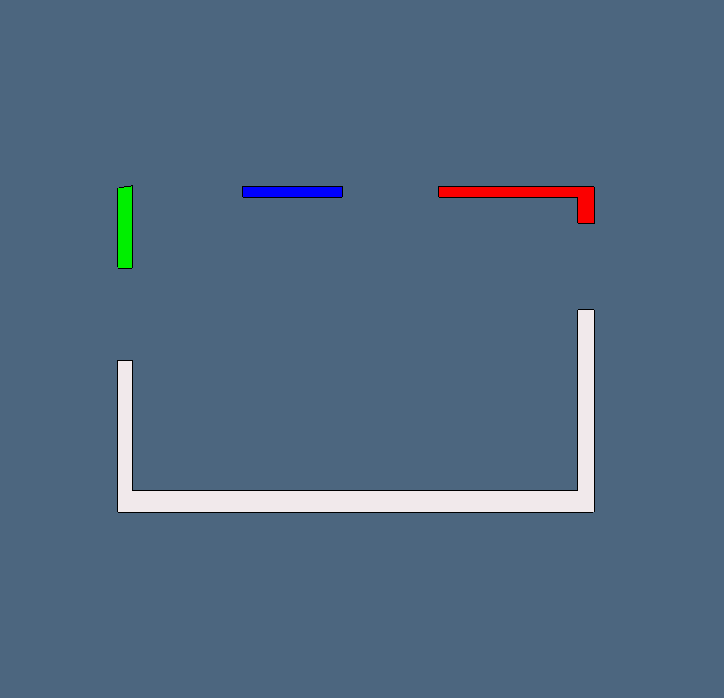}%
\includegraphics[height=0.25\linewidth,width=0.25\linewidth]{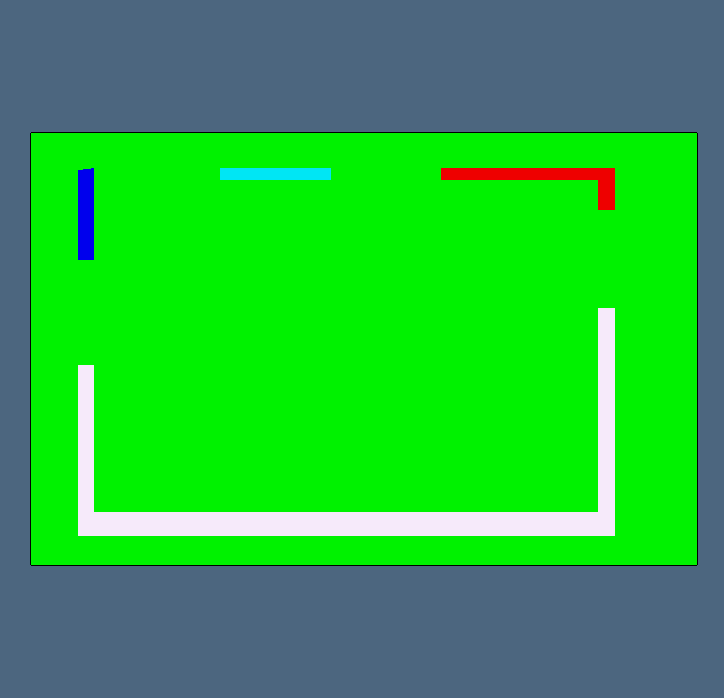}%

{\rm\footnotesize\hfill(a,b)\hfill\hfill(c,d)\hfill\hfill(e,f)\hfill\hfill(h,g)\hfill.} 
\caption{Some examples of variadic Booleans, obtained by applying the Boolean operators ``\texttt{.|}'', ``\texttt{.-}'',  ``\texttt{.!}'', and ``\texttt{.\&}'' to the  \texttt{assembly} variable,  with 1-cells (B-reps) imported from \texttt{SVG} files. (a,b)  start-to-end: from polygon input via \texttt{SVG} to (exploded) difference of outer box and interior walls; (c,d) boundary of union and union of some shapes; (e,f) difference 2-complex and its boundary; (g,h) boundary of outer box minus the previous 2-complex, and the corresponding 2-complex.}
\label{fig:union-2d} 
\end{figure}

\section{Result Discussion}\label{sec:discussion}
%
The approach put forth in this paper is novel in several ways. 
First of all, we introduced here the first method of solid computational geometry, at authors' knowledge, to assemble the terms of a CSG form using the atoms of a Boolean algebra, identified with the basis 3-chains of a partition of the embedding space, in turn given by the columns of the $[\partial_3]$ matrix in chain space. 
Furthermore, we initiated to set up the structure of Boolean form's terms as binary arrays, aka coordinate vectors in $C_3$ space,  and combining them by native bitwise operators.
Moreover, with the only exception of $k$d-trees and interval-trees for acceleration of geometric queries, our approach does not use standard methods of computational geometry---in particular, of solid modeling subfield---which are based on highly specialized data structures, that are often very complex and require the implementation of very complicated algorithms. 

We use instead basic tools and methods of linear algebra and algebraic topology, {i.e.},~(sparse) matrices of linear operators and semiring matrix multiplication~\citep{DBLP:journals/corr/KepnerABBFGHKLM16}, plus filtering. Hence, accuracy of topological computations is guaranteed by contruction, since operator matrices and chain bases, as
well as any chain, satisfy the (graded) constraints $\partial^2 = 0$ and $\delta^2 = 0$, which are easy to check.
Furthermore, we have shown that  space and time complexities are comparable with those of traditional methods, and it is reasonable to expect that our approach might be extended to generic dimensions and/or implemented on highly parallel computational engines, using pretty standard and highly general computing libraries, e.g., GraphBLAS and GPU kernels.  In this case a
good speed-up is expected, given the high level of possible parallelization of our algorithms. 

%
In this paper we have also given a new characterization of the computational process for evaluating CSG expressions of any depth and complexity. Traditional evaluation methods require a post-order DFS traversal of the expression tree, and the sequential computation   of each Boolean operation encountered on nodes, until the root operation is evaluated. It is well known that such a process lacks in robustness and accumulates numerical errors, that ultimately modify the local topology and make the applications  stop on error. Hence, intermediate boundary representations need to be generated and carefully curated, before continuing the traversal.  

With our approach, the evaluation of a CSG expression of any complexity is done with a different computational process. A hierarchical structure of macros is used both to parse the expression tree and to apply affine transformations to solid primitives, in order to scale, rotate and translate them in their final (``world'') positions and attitudes. All their 2-cells are thus accumulated in a single collection, and each of them is \emph{independently} fragmented, generating a collection of local topologies that are merged by \emph{boundary congruence}, using a \emph{single} round-off  operation on vertices. The global space partition is so generated, and all input solids are classified with respect to 3-cells of this partition, i.e., the \emph{atomic objects} of a  Boolean algebra, with a single point-membership test per atom. Finally,  \emph{any Boolean form} of arbitrary complexity, with the same variables, can be quickly evaluated by bitwise vectorized logical operations.

Last but not least, all distinction is removed between manifold and non-manifold geometries, both in 2D and 3D, allowing for mixing B-reps and cellular decompositions of the interior of elementary solids. B-reps may be always  generated  for efficiency purposes---via boundary operator multiplication---in order to strongly reduce the number of 2-cells to be decomposed (possibly in parallel), but this is not strictly required. This flexibility about input data types could be useful, in particular, for combining outer surfaces with regular solid grids, and is consistent with the generation of internal structures to make 3D printed models more reliable and easy to produce.

%
\subsection*{Acknowledgements} 
Several people have contributed or are currently contributing to implement different parts and algorithms of this project, including Gianmaria DelMonte, Francesco Furiani, Giulio Martella, Elia Onofri, and Giorgio Scorzelli. Antonio Bottaro believed in this project and funded its development as responsible of R\&D of Sogei (Italian Ministry of Economy and Finance) and CEO of the Geoweb subsidiary.
 

\section{Conclusions and future prospects}\label{sec:conclusion}
%
We have introduced a novel approach to computation of Boolean operations between solid models. In particular, we have shown that any Boolean expression between solid models may be evaluated using the finite algebra associated with the set of independent generators of the chain space produced by the arrangement of Euclidean space generated by a collection of geometric objects. 
The computational approach to geometric design introduced by this paper is~quite different from traditional methods of geometric modeling and computational geometry.

Two prototypes of open-source implementation were written in Julia language~\citep{BEKS14}, mostly using sparse arrays. At the present time, we can only  evaluate CSG formulas  with closed regular cells. 
A novel implementation for closure algebras is planned, using the parallel and distribute features of Julia, that provides best-in-class support to linear algebra and matrix computations, including the new  GraphBLAS~\citep{DBLP:journals/corr/KepnerABBFGHKLM16} standard.

We  think that, by introducing linear methods in solid geometry, we have in some sense only scratched the surface of the new body of knowledge being currently investigated by machine learning methods for image understanding, that also use tensors and linear algebra. Next generation ML methods need formal and abstract techniques for building and merging solid models derived from partial images. We have already done some early experiments of our topological algebraic method in 3D medical imaging, by combining boundary and coboundary operators with filters, aiming at discovering and tracing complex interior structures. We hope that the techniques introduced by this paper may provide some basic tools for many applications in this and other fields.

\bibliographystyle{cas-model2-names}

\bibliography{cagd-2019}

\end{document}